\documentclass[12pt]{article}
\usepackage{graphicx}
\usepackage{jheppub}
\usepackage{amsmath, amsthm, amssymb,latexsym,epstopdf,epsfig}
\makeatletter
\def\@fpheader{\relax}
\makeatother 

\newcommand{\be}{\begin{equation}}
\newcommand{\ee}{\end{equation}}
\newcommand{\beq}{\begin{equation}}
\newcommand{\eeq}{\end{equation}}
\newcommand{\ba}{\begin{eqnarray}}
\newcommand{\ea}{\end{eqnarray}}
\newcommand{\bea}{\begin{eqnarray}}
\newcommand{\eea}{\end{eqnarray}}
\newcommand{\nn}{\nonumber}

\newcommand{\NO}{\nonumber}

\def\co{{\cal O}}

\def \r {\rho}
\def\om{\omega}
\newcommand{\volf}{\mathrm {Vol_{(4)}}}
\def\Ffive{F_{(5)}}
\def\Fthree{F_{(3)}}
\def\Btwo{B_{(2)}}
\def\Hthree{H_{(3)}}
\def\Xifour{\Xi_{(4)}}
\def\eps{\epsilon}
\def\MM{\mathcal{M}}
\def\bpscp{\bold{c_+}}
\def\bpscm{\bold{c_-}}
\newcommand{\ansatz}{ans\"atz\ }

\preprint{UTTG-07-12}	
\title{On the Stability of Non-Extremal Conifold Backgrounds with Sources}

\author[a,b]{Elena C\'aceres}
\author[b]{and Steve Young}
\affiliation[a]{Facultad de Ciencias, Universidad de Colima, Mexico}
\affiliation[b]{ Theory Group, Department of Physics,\\
University of Texas at Austin, Austin, TX 78727, USA.}
 
\emailAdd{elenac@zippy.ph.utexas.edu}
\emailAdd{scyoung@zippy.ph.utexas.edu}

\abstract{We present finite temperature  solutions describing  $N_c$ $D5$  branes wrapped on the $S^2$ of the resolved conifold  in the presence of  $N_f$ flavor branes  sources and their backreaction {\em i.e}   $N_f/N_c \sim 1$. In these solutions  the dilaton does not blow up at infinity but stabilizes to a finite value. Thus, we can use them   to  generate new ones with $D5$ and   $D3$ charge. The resulting  backgrounds are  non-extremal versions  of  the ``flavored"  resolved deformed conifold. It is tempting to interpret  these solutions  as  gravity  duals of finite temperature field theories exhibiting non-trivial phenomena as Seiberg dualities, Higgsing and confinement. However, a first necessary step in this direction is to investigate their  stability. We study the  specific heat of these new flavored backgrounds and find that  they are thermodynamically unstable.  Our results on the stability also  apply to some of the  non-extremal backgrounds with   Klebanov-Strassler asymptotics found in the literature.

		 }

\begin{document}

%--------+---------+---------+---------+---------+---------+---------+

	\maketitle
\setcounter{page}{1}
%--------+---------+---------+---------+---------+---------+---------+

\newpage
 \section{Introduction}

 The gauge/gravity duality \cite{Maldacena:1997re},\cite{Gubser:1998bc},\cite{Witten:1998qj} provides a completely new framework for the understanding  of strongly coupled field theories. It  has  opened a new area of study at the interface of  field theory  and string theory. The applications  are numerous and include formal developments as well as  phenomenological topics. One of the goals is to understand aspects of  QCD, at zero and finite temperature,  through the use of a suitable string dual. In this spirit, the two landmarks for non-conformal  backgrounds with $\mathcal{N}=1$ supersymmetry --- the deformed conifold model of Klebanov and Strassler (KS) \cite{Klebanov:2000hb} and the wrapped branes model of Maldacena and N\'u\~nez (MN) \cite{TOWARDS_SYM} --- have been thoroughly studied and the connections between them well understood. Both backgrounds, MN and  and KS, are special cases of a general \ansatz proposed by Papadopoulos and Tseytlin in \cite{PT}.  In \cite{BUTTI}, a  family of solutions interpolating between KS and MN was found by   imposing  supersymmetry conditions on the \ansatz of \cite{PT}. In the field theory side this family of solutions was shown to correspond  to different vacuum expectation values of a baryonic operator. Varying the string coupling constant of this family of SUGRA solutions smoothly interpolates between KS and MN. 
 In order to incorporate dynamical flavor in the fundamental representation in these backgrounds we have to consider  the backreaction of flavor branes. This difficult problem can be  tackled using  a smearing technique and has been extensively explored in the literature \cite{Casero:2006pt,Benini:2006hh,Caceres:2007mu,Bigazzi:2008zt,Nunez:2010sf,Benini:2007gx,Bigazzi:2008qq}.

 In \cite{Maldacena:2009mw} the authors presented  a particular solution of wrapped $D5$  branes that  interpolates between the deformed conifold with flux and the resolved conifold with branes. Applying U-dualities  to this solution adds $D3$ charge and we recover the Klebanov-Strassler baryonic branch. Thus, the  chain of dualities takes us from a background with only dilaton and three form flux $H_{(3)}$, to one with dilaton, $H_{(3)}, \ F_{(3)}$ and $F_{(5)}$. This procedure --also refer to as {\it rotation} \footnote{As shown in \cite{WARPED} it is  a rotation in the space of Killing spinors.}-- requires that we start with a $D5$ brane background whose gravity modes and field theory modes are not decoupled. Unlike MN and the flavored deformations of MN \cite{Casero:2006pt}, this type of solutions have a stabilized dilaton. In this framework, the decoupling limit is taken after the dualities have been applied so it is only the backgrounds with dilaton, $H_{(3)}, \ F_{(3)}$ and $F_{(5)}$ that  will have an interpretation as field theory duals \cite{Maldacena:2009mw}.
 This  solution generating technique  can also be used to study backgrounds with flavor and backreaction. In \cite{WARPED} the authors presented a new flavored wrapped brane solution with stabilized dilaton. Using this new solution as a seed for the  {\it rotation} procedure,  a flavored generalization of the KS baryonic branch can be obtained. The dual field theory exhibits Seiberg dualities and Higgsing and  was conjectured to be a {\it mesonic} branch of KS.  The $SU(N_f/2)$ flavor symmetry arises only in the IR, at the bottom of the cascade after Seiberg dualities and Higgsing have taken place; the {\it quarks} are really bi-fundamentals but since one of the gauge groups is very weakly coupled it can be thought of as flavor. This scenario was further developed in \cite{Conde:2011aa},\cite{Conde:2011ab},\cite{Elander:2011mh}.

The {\it rotation}  procedure  can also  be applied to non-supersymmetric \cite{Bennett:2011va} or non-extremal backgrounds \cite{HEATING}. Thus, it  provides an alternative way of generating non-extremal deformations of KS which is a fascinating subject with rich new  physics \cite{Buchel:2010wp}. In particular, KS black holes were studied in   \cite{Aharony:2007vg},\cite{Mahato:2007zm}. 

In this paper we present solutions that are non-extremal deformations of the flavored warped deformed conifold of  \cite{WARPED}, {\it i.e}  we present new non-extremal flavored  solutions that through a chain of dualities  take us to a non-extremal cascading theory. We also  study their stability.  The backgrounds presented here 
%presented   generalize  the results of \cite{HEATING} in two ways: we consider the entire baryonic branch {\it i.e}
contain the finite-temperature solutions of  \cite{HEATING} as a special case ($N_f=0$, $a(r)=0$) \footnote{ $a(r)$ is a function appearing in the metric whose infrared behavior is related to the parameter $\xi$ used in \cite{BUTTI} to parametrize the   baryonic branch. For small values of $r$,   $a(r)\sim 1 - \xi r^2 + \mathcal{ O}(r^4)$ \cite{BUTTI}.}.
%. As pointed out in \cite{WARPED}, the backreaction  of the  flavor branes  modifies the asymptotics and   the UV is KS  only when the flavor is turned off, $N_f\rightarrow 0$. Thus,  our solutions 
On the other hand, in  our solutions  the non-extremal parameter enters in the warp factor,  $F_{(3)},\ H_{(3)}$ and $F_{(5)}$ and thus, they   are not (for $N_f=0$)  of the type studied in    \cite{Aharony:2007vg},\cite{Mahato:2007zm}.   

We study the specific heat of  these new backgrounds and find it to be negative
both before and after the rotation. In \cite{GTV} the authors showed that non-extremal MN is unstable and in \cite{Buchel:tachyon}  this instability was  related  to the existence of a  tachyonic quasinormal mode .  Even though in this work we consider different wrapped D5 solutions than the ones studied in \cite{GTV},\cite{Buchel:tachyon}, it is possible that the same effect is present here. Also in \cite{GTV} the authors  argued the existence of a  phase transition above which the theory is in a high temperature thermodinamically unstable phase. 
It is possible  that the U-dualities map the phase transition of \cite{GTV} to a phase transition in a cascading non-extremal background  \cite{Gubser:2001ri},\cite{Buchel:2010wp},
\cite{Buchel:tachyon},\cite{Aharony:2007vg},\cite{Mahato:2007zm}. We leave these issues for future study.

The remainder of the  paper is organized as follows. In  Section 2  we review the general ansatz for wrapped fivebranes branes at zero temperature with and without flavor. This ansatz is very general and  different types of solutions have been explored in the literature. In \cite{Casero:2006pt}  solutions with flavor and backreaction  were obtained using a smearing procedure. The solutions of \cite{Casero:2006pt}  have the standard linear dilaton behavior. In \cite{HoyosBadajoz:2008fw} and \cite{WARPED} it was shown that there exist  also flavored  solutions with a dilaton that goes to a constant in the UV; these are the type of backgrounds to which the U-dualities can be applied and the ones that are relevant in the present work. In  Section 2.2 we present the equations of motion for a non-extremal deformation of the flavored backgrounds of \cite{WARPED}. In Section 3 we obtain a numerical solution to the equations of Section 2.2. 
 Once these solutions are under control we add $D3$ charge through the U-dualities and arrive at a non-extremal flavored deformed resolved conifold. 
 In order to numerically solve the equations of motion  we first have to obtain a consistent UV expansion. We find a UV expansion with 11 free parameters that completely determine the UV behavior at any order. We then integrate back and match to the required regular horizon behavior. Having obtained the non-extremal solution we proceed in Section 4 to apply the chain of dualities to obtain solutions with nontrivial dialton, $F_{(3)},\  H_{(3)} {\rm and}  F_{(5)}$. We present the numerical results as well as the  UV  behavior. 
 In section 5 we study the thermodynamic stability of the seed background (wrapped fivebranes) and of the background generated by U-duality. We find that they are both unstable. Our solutions contain as a special case the ones presented in \cite{HEATING} thus we expect our results to hold for those backgrounds as well.

 \section{Flavored backgrounds}

 \subsection{Review of flavored wrapped D5 backgrounds, $T=0$ }
In a seminal work, Casero, Nu{\~n}ez and Paredes \cite{Casero:2006pt} presented a framework to  incorporate the backreaction of the flavor D5-branes. The geometries obtained  depend on the ratio $N_f/N_c$ which can be  of order one even in the $N_c \rightarrow \infty$ limit. These spaces  are singular at the origin, but the singularity is a ``good" one in the sense of the criterion in \cite{Maldacena:2000mw} which means that the metric component $g_{tt}$ remains finite in the limit $r \rightarrow 0$. Everywhere else the geometry is smooth and the curvature small as long as $g_sN_c\gg1$. These backgrounds are conjectured to be dual to ${\mathcal N}=1$ SQCD in four dimensions with a large number of flavors, up to the same caveats concerning the decoupling of the KK modes that were already present in the discussion of the original Maldacena-Nu\~nez  background without flavor.\\
The general strategy is the following \cite{Casero:2006pt}: One introduces a deformation of the Maldacena-Nu\~nez background due to the presence of flavor D5-branes, derives the corresponding BPS equations (see Appendix B of \cite{Casero:2006pt}), and  attempts to solve them. The flavor D5-branes are taken to extend along the $(x^0,x^1,x^2,x^3,\psi,r)$ directions and are smeared over the $(\theta,\varphi,\tilde{\theta},\tilde{\varphi})$ directions. These branes can be shown to preserve the same supersymmetry as the background for arbitrary values of the angles $\theta, \varphi, \tilde{\theta}, \tilde{\varphi}$ \cite{Nunez:2003cf}. Moreover, the smeared flavor branes will be sources for the RR 3-form, resulting in RR fluxes in the deformed background that can be observed as a ``violation" of the original Bianchi identity. \\
The  \ansatz for the deformation of the MN  background ($H_{(3)}=0,\, F_{(5)}=0$)
is,
\begin{align}\label{eq:defmetansatz}
	ds_{10}^2 &= \alpha' g_s N_c e^{\phi(r)/2} \left[\vphantom{\frac{e^{2g}}{4}}\frac{1}{\alpha' g_s N_c} dx_{1,3}^2+dr^2
+ e^{2h(r)} (d\theta^2 + \sin^2 \theta d\varphi^2) + \right.\nonumber\\
& \quad \left. +\frac{e^{2g(r)}}{4} \left( (\omega_1 +a(r)d\theta)^2 +(\omega_2 -a(r) \sin \theta d \varphi)^2 \right) + \frac{e^{2k(r)}}{4}(\omega_3 + \cos \theta d \varphi)^2 \right],
\end{align}
The RR 3-form field strength reads
\begin{align}\label{eq:F3o}
F_{(3)}& = \frac{\alpha' g_s N_c}{4} \left[ -(\omega_1 +b(r) d\theta)\wedge (\omega_2 - b(r) \sin \theta d\varphi) \wedge (\omega_3 +\cos \theta d \varphi)  \right.\nonumber\\
&\left. + b' dr \wedge (-d \theta \wedge \omega_1 + \sin \theta d\varphi \wedge \omega_2) + (1-b(r)^2) \sin \theta d \theta \wedge d\varphi \wedge \omega_3\right],
\end{align}
and automatically satisfies the Bianchi identity $dF_{(3)}=0$. The left-invariant one-forms $\om_a$ on $S^3$ are
\begin{align}
\omega_1 &= \cos\psi d\tilde\theta\,+\,\sin\psi\sin\tilde\theta
d\tilde\varphi,\nonumber\\
\omega_2&=-\sin\psi d\tilde\theta\,+\,\cos\psi\sin\tilde\theta d\tilde\varphi,\nonumber\\
\omega_3&=d\psi\,+\,\cos\tilde\theta d\tilde\varphi.
\end{align}
%We also introduce the standard notation,
%\begin{align}
%	e_1&= d\theta, \ \ \ \ e_2=\sin\theta
%d\varphi,  \ \ \ \ \tilde\om_3= \om_3 + \cos \theta d\varphi
%\end{align}
We also introduce the new radial coordinate $\rho$,
\begin{equation}
d \rho = e^{-k(r)} dr,
\end{equation}
and the standard notation
\begin{equation}
	e_1= d\theta, \ \ \ \ e_2=\sin\theta
d\varphi\nonumber
\end{equation}
\begin{equation}
\tilde\om_1= \om_1 +a(\r)e_1,\quad \tilde\om_2= \om_2 -a(\r)e_2,\quad \tilde\om_3= \om_3 + \cos \theta d\varphi.
\end{equation}
%The metric then becomes
%\begin{align}\label{eq:defmetansatzrho}
%	ds_{10}^2 &= e^{\phi(\r)/2} \left[\vphantom{\frac{e^{2g(\r)}}{4}}dx_{1,3}^2+ e^{2 k(\r)}d\r^2
%		+ e^{2h(\r)} ({e_1}^2 + {e_2}^2) + \right.\nonumber\\
%& \quad \left. +\frac{e^{2g(\r)}}{4} \left( (\om_1 +a(\r)e_1)^2 +(\om_2 -a(\r)e_2)^2 \right) + \frac{e^{2k(\r)}}{4} (\tilde\om_3)^2 \right],
%\end{align}
\\
\noindent
The metric then becomes
\begin{align}\label{eq:defmetansatzrho}
	ds_{10}^2 &= e^{\phi(\r)/2} \left[\vphantom{\frac{e^{2g(\r)}}{4}}dx_{1,3}^2+ e^{2 k(\r)}d\r^2
		+ e^{2h(\r)} ({e_1}^2 + {e_2}^2) + \right.\nonumber\\
& \quad \left. +\frac{e^{2g(\r)}}{4} \left( \tilde\om_1^2 +\tilde\om_2^2 \right) + \frac{e^{2k(\r)}}{4} (\tilde\om_3)^2 \right],
\end{align}
where we have set $\alpha' =g_s =1$ and $N_c$ has been absorbed into $e^{2h}, e^{2g}, e^{2k}$ and $d\r^2$. 

The more familiar  MN background is obtained from \ref{eq:defmetansatz} and \ref{eq:F3o} with  
\begin{align}
&a(r)=\frac{2r}{\sinh 2r}, \qquad b(r)=0, \qquad  k(r)=g(r)=1, \\
&e^{2 h(r)}=r \coth 2r - \frac{r^2}{\sinh^2 2r} -\frac{1}{4}\qquad e^{-2\phi}=2  e^{-2\phi_0} \frac{ e^{h(r)}}{\sinh 2r} &
\end{align}
Note that even in the $N_f=0$ case,  the system allows for other solutions -albeit with singular behavior in the IR- one of them is the solution with $a(r)=0$ often referred to as ''abelian"\footnote {The abelian solution is characterized by  $a(r)=b(r)=0$, $k(r)=g(r)=1$, $e^{2h(r)}=r$ and $e^{-2\phi}= 4 e^{-2\phi_0} \sqrt{r} e^{-2r} $},  others were studied in  \cite{HoyosBadajoz:2008fw}. If we consider $N_f\ne0$ then there is a plethora of possible solutions. Among them there are solutions where  the dilaton goes to a constant  as $r \rightarrow \infty$. Unlike MN that has a linear dilaton, the stabilized dilaton solutions do not correspond to a near horizon limit,  gravity and field theory modes are coupled. However,  as explained in \cite{Maldacena:2009mw} these are precisely the type of solutions we need to use as ''seed" solutions for the rotation procedure to be used in Section \ref{section_rotating},  Appendix \ref{appendix-UDuality}.  Note also that the \ansatz for  $F_{(3)}$, (\ref{eq:F3o}), guarantees that $F_{(3)}$ is a closed form and thus it  does not include any backreation yet.

To include  flavor branes and its backreaction, the action must be augmented by the DBI and Wess-Zumino actions for the flavor D5-branes. The complete action then reads
\begin{equation}
S=S_{\text{grav}}+ S_{\text{sources}},
\end{equation}
where, in Einstein frame, we have
\begin{equation}\label{eq:Sgrav}
S_{\text{grav}}= \frac{1}{2\kappa_{10}^2}\int d^{10}x \sqrt{-g_{10}} \left( R -\frac{1}{2} (\partial_{\mu}\phi)(\partial^{\mu} \phi) - \frac{1}{12}e^{\phi}F_{(3)}^2 \right),
\end{equation}
and
\begin{equation}\label{eq:Sflavor}
S_{\text{sources}}= T_5 \sum^{N_f} \left( - \int_{\MM_6}d^6x e^{\phi/2} \sqrt{-g_{6}} + \int_{\MM_6} P[C_{(6)}]\right).
\end{equation}
where	$T_5 =\frac{1}{(2\pi )^5}$ and $2 \kappa_{10}^2 = (2 \pi)^7$. One of the effects of smearing the $N_f \rightarrow \infty$ flavor branes along the two transverse 2-spheres is that there will be no dependence on the angular coordinates $(\theta,\varphi,\tilde{\theta},\tilde{\varphi})$ of the functions $f(\rho), h(\rho), g(\rho)$ and $k(\rho)$ that determine our metric \ansatz (\ref{eq:defmetansatzrho}) --- significantly simplifying the computations. After the smearing we can write
\begin{equation}\label{eq:SflavorSmeared}
	S_{\text{sources}} =\frac{T_5 N_f}{(4 \pi)^2}\left( -\int d^{10}x \sin \theta \sin \tilde{\theta} e^{\phi/2} \sqrt{-g_{6}} + \int \text{Vol}{\mathcal Y}_{(4)} \wedge C_{(6)}\right),
\end{equation}
with the definition $\text{Vol}({\mathcal Y}_4) = \sin \theta \sin\tilde \theta d \theta \wedge d \varphi \wedge d \tilde{\theta} \wedge d \tilde{\varphi}$. Once the smeared flavor D5-branes are incorporated into the background, the Bianchi identity for $F_{(3)}$ (which is identical to the EOM for $C_{(6)}$) gets modified to
\begin{equation}
d F_{(3)}= \frac{N_f}{4} \sin \theta \sin \tilde{\theta} d \theta \wedge d \varphi \wedge d \tilde{\theta} \wedge d \tilde{\varphi}= \frac{N_f}{4} \om_1
\wedge\om_2\wedge e_1\wedge e_2. 
\end{equation}
as a result of adding a Wess-Zumino term. This can be solved by adding the following term to the original $F_{(3)}$ in (\ref{eq:F3o}),
\begin{equation}
F_{(3)}^{\,\text{sources}}= -\frac{N_f}{4} e_1\wedge e_2 \wedge \om_3
\end{equation}
The full RR 3-form field strength now reads
\begin{align}
F_{(3)}& =  \frac{N_c}{4} \Bigg[ -(\omega_1 + b\, e_1)\wedge (\omega_2 - b e_2) \wedge \tilde\om_3 \nonumber\\
&+ b' d\r \wedge (-e_1 \wedge \omega_1 + e_2 \wedge\omega_2) + \left( 1-b^2-\frac{N_f}{N_c} \right) e_1 \wedge e_2 \wedge \omega_3\Bigg] \equiv \frac{ N_c}{4}f_{(3)},\nonumber\\
\label{eq:F3ansatz}
\end{align}
where the only modification is the appearance of the term involving $N_f/N_c$ in the second line.\\
The first order BPS equations for this \ansatz can be solved for $b(\r), h(\r), g(\r)$ leaving a system of three coupled differential equations for $a(\rho),k(\rho)$ and $\phi(\r)$. The details can be found in \cite{Casero:2006pt,HoyosBadajoz:2008fw,Casero:2007jj}.

In the present work we are  interested in non-extremal generalizations of a particular type of  flavored backgrounds: the ones with stabilized dilaton presented in \cite{WARPED}. The non-zero temperature breaks supersymmetry and thus we cannot resort to  the BPS equations nor to the master equation formalism developed in \cite{HoyosBadajoz:2008fw}. We are bound to solve the second order equations of motion (EOMs) for  $b(\r), h(\r), g(\r), a(\rho),k(\rho)$ and $\phi(\r)$ that will be obtained in the next section.

\subsection{Non-extremal flavored backgrounds}
\label{subsection-non_extremal_flavored_backgrounds}
One of our goals is to find non-extremal deformations of the flavored backgrounds presented in  the previous section. To that aim we consider the following \ansatz for the metric,
\bea
\label{eq:baseMetric}
 ds^2_{IIB}&=& e^{\phi/2}\Big[-e^{-8x} dt^2 + dx_1^2 + dx_2^2 + dx_3^2
\Big] +
e^{\phi/2} ds_{6}^2,\nonumber\\
ds_6^2&=&
\Big[e^{8x}e^{2k}d\r^2+\frac{e^{2k}}{4}(\omega_3 +
\cos\theta d\varphi)^2 +e^{2h}(d\theta^2+\sin^2\theta
d\varphi^2)\nonumber\\
&+&\frac{e^{2g}}{4}\left((\omega_1+a(\rho)d\theta)^2
+ (\omega_2-a(\rho)\sin\theta d\varphi)^2\right)   \Big] \nonumber\\
\eea
and the RR 3-form gauge field
\be\label{eq:F3}
F_{(3)}= \frac{N_c}{4} f_{(3)}
\ee
\noindent where $f_{(3)}$ is defined in (\ref{eq:F3ansatz}).
%\bea
%F_{(3)}&=& \frac{N_c}{4} \Bigg[ -(\tilde\omega^1 + b\, d\theta)\wedge (\tilde\omega^2 - b \sin \theta d\varphi) \wedge (\tilde\omega^3 +\cos \theta d \varphi) \nonumber\\
%&+& b' dr \wedge (-d \theta \wedge \tilde\omega^1 + \sin \theta d\varphi \wedge \tilde\omega^2) + \left( 1-b^2-\frac{N_f}{N_c} \right) \sin \theta d \theta \wedge d\varphi \wedge \tilde\omega^3\Bigg] \equiv \frac{\alpha' N_c}{4}w_3.\nonumber\\
%\label{F3ansatz}
%\eea
%Here $\tilde\omega_i$ are the left-invariant forms of $SU(2)$
%\bea\lab{su2}
%&&\tilde{\omega}_1= \cos\psi d\tilde\theta\,+\,\sin\psi\sin\tilde\theta
%d\tilde\varphi,\nonumber\\
%&&\tilde{\omega}_2=-\sin\psi d\tilde\theta\,+\,\cos\psi\sin\tilde\theta d\tilde\varphi,\nonumber\\
%&&\tilde{\omega}_3=d\psi\,+\,\cos\tilde\theta d\tilde\varphi.
%

%\eea
This \ansatz is general enough to account for the effect of the backreaction of a large number $N_f$ of flavor D5 branes, smeared along the $\theta, \varphi, \tilde \theta, \tilde \varphi$ directions, and spanning both the Minkowski and $\r$ directions. This gives a source contribution to the $F_{(3)}$ Bianchi identity, which we define as the smearing form $\Xifour$:
\be
\Xifour \equiv d F_{(3)}= \frac{N_f}{4}\om_1\wedge \om_2 \wedge e_1 \wedge e_2.
\ee 
\noindent
 The factor $e^{8x(\r)}$ is a non-extremal deformation that  accounts for the appearance of a horizon.  The total Lagrangian for the bulk fields and smeared flavor branes is
\begin{equation}
S=S_{\text{grav}}+ S_{\text{sources}},
\end{equation}
where $S_{\text{grav}}$ and $S_{\text{sources}}$ are given in (\ref{eq:Sgrav}) and (\ref{eq:SflavorSmeared}) respectively.
%where, in Einstein frame, we have

%	\begin{align}
%		S_{IIB} = &\int \sqrt{-g} \left( R - \frac{1}{2} \partial_{\mu} \phi \partial^{\mu} \phi \right) 
%		&- \frac{1}{2} \int e^{\phi} F_{(3)} \wedge *F_{(3)} 	\end{align}
%S_{\text{grav}}= \frac{1}{2\kappa_{(10)}^2}\int d^{10}x \sqrt{-g_{(10)}} \left( R -\frac{1}{2} (\partial_{\mu} \phi) (\partial^{\mu} \phi) - \frac{1}{12}e^{\phi}F_{(3)}^2 \right),
%
%\end{align}
%and
%\begin{equation}
%S_{\text{flavor}} =\frac{T_5 N_f}{4 \pi^2}\left( -\int d^{10}x \sin \theta \sin \tilde{\theta} e^{\phi/2} \sqrt{-g_{(6)}} + \int (\sin \theta \sin\tilde \theta d \theta \wedge d \varphi \wedge d \tilde{\theta} \wedge d \tilde{\varphi}) \wedge C_{(6)}\right).
%\label{Sflavbef}
%\end{equation}
%Here, the D5-brane tension and 10d Newton's constan

Before proceeding to present the EOMs note that the Wess-Zumino term in the flavor action does not involve the metric nor the dilaton so it will not enter   Einstein's equations. Its effect is to   change the equation for  $C_{(6)} $ which was $d*F_{(7)} \equiv dF_{(3)} =0$ and now in the presence of sources is $
 d F_{(3)}=\Xifour.$
The   \ansatz (\ref{eq:baseMetric}) assumes all the angular dependence is fixed by the symmetries of the background and the only non-trivial dependence is  on the radial variable $\rho$ %\footnote{Note also  that  we are dealing with backgrounds with $F_5=0$ and $H_3=0$ and with a constant smearing form.}
. 
Following \cite{GTV} we  derive a one dimensional Lagrangian from which we can obtain  the EOMs for all fields. We introduce  the ans\"atze (\ref{eq:baseMetric} - \ref{eq:F3}) in the action,  integrate  over the angular variables and drop the overall volume factor,
\be 
S_1= \int (T-U) dr 
\ee
\noindent 
where the explicit form of $T$ and $U$ and the details of the derivation are given in Appendix \ref{appendix-EOMs}. Note that this action should be supplemented with the constraint  coming from reparametrization invariance, $T+U=0$.

%a one dimensional Lagrangian from which we can derive the EOMs for all the fields. 
Finally, defining $s\equiv \frac{N_f}{N_c}$ the EOMs read (seeAppendix \ref{appendix-EOMs} for details),
\bea\label{eq:eomfirst}
&&-\frac{1}{8} e^{-4 h+8 x} s^2+\frac{1}{4} e^{-2 g-4 h+8 x} s \left(-2 e^{2 (h+k)}+e^{2 g} \left(1-2 b a+a^2\right)\right)\nonumber\\
&&+\frac{1}{8} e^{-4 (g+h)} \left(-16 e^{4 h+8 x}-e^{4 g+8 x} \left(1-2 b a+a^2\right)^2+2 e^{2 (g+h)} \left(-4 e^{8 x} (b-a)^2-b'^2\right)\right)\nonumber\\
&&+2 \left(g'+h'-4 x'\right) \phi '+2 \phi '^2+\phi ''= 0
\eea
\bea
&&2 g' x'+2 h' x'-8 x'^2+2 x' \phi '+x'' = 0
\eea
\bea
&&\frac{1}{8} e^{-4 (g+h-2 x)} \bigg[e^{4 g}+16 e^{4 h}-e^{4 (g+k)}-16 e^{4 (h+k)}-2 e^{4 g} s+e^{4 g} s^2\nonumber\\
&&+2 e^{2 g} \left(4 e^{2 h}+4 e^{4 g+2 h}+e^{2 g+4 k}-4 e^{2 h+4 k}-e^{2 g} (-1+s)\right) a^2-e^{4 g} \left(-1+e^{4 k}\right) a^4\nonumber\\
&&-4 e^{2 g} b a \left(4 e^{2 h}-e^{2 g} (-1+s)+e^{2 g} a^2\right)+b^2 \left(8 e^{2 (g+h)}+4 e^{4 g} a^2\right)\bigg]\nonumber\\
&&+2 k' \left(g'+h'-4 x'+\phi '\right)+k''= 0
\eea
\bea
&&\frac{1}{8} e^{-2 (g+2 h)} \bigg[-8 e^{2 (g+h+k+4 x)}+e^{2 g+8 x}+e^{2 g+4 k+8 x}+4 e^{2 (h+k+4 x)} s-2 e^{2 g+8 x} s\nonumber\\
&& +e^{2 g+8 x} s^2+2 e^{8 x} \left(2 e^{2 h}+2 e^{4 g+2 h}-4 e^{2 (g+h+k)}-e^{2 g+4 k}+2 e^{2 h+4 k}-e^{2 g} (-1+s)\right) a^2\nonumber\\
&&+e^{2 g+8 x} \left(1+e^{4 k}\right) a^4+4 e^{8 x} b^2 \left(e^{2 h}+e^{2 g} a^2\right)-4 e^{8 x} b a \left(2 e^{2 h}-e^{2 g} (-1+s)+e^{2 g} a^2\right)\nonumber\\
&&+e^{2 h} b'^2+e^{4 g+2 h} a'^2\bigg]+2 h' \left(g'-4 x'+\phi '\right)+2 h'^2+h''=0
\eea
\bea
&&\frac{1}{8} e^{-2 (2 g+h)} \bigg[-32 e^{2 (g+h+k+4 x)}+16 e^{2 h+8 x}+16 e^{2 h+4 k+8 x}+4 e^{2 (g+k+4 x)} s\nonumber\\
&&+4 e^{2 g+8 x} b^2-8 e^{2 g+8 x} b a-4 e^{2 g+8 x} \left(-1+e^{4 g}-e^{4 k}\right) a^2+e^{2 g} b'^2-e^{6 g} a'^2\bigg]\nonumber\\
&&+2 g' \left(h'-4 x'+\phi '\right)+ 2 g'^2 + g''=0
\eea
\bea
&&e^{-4 g-2 h+8 x} \bigg[-2 e^{2 g} b^2 a+b \left(4 e^{2 h}-e^{2 g} (-1+s)+3 e^{2 g} a^2\right)\nonumber\\
&&-a \left(4 e^{2 h}(1+e^{4 g}+e^{4k})-8 e^{2 (g+h+k)}-e^{2 g+4 k}-e^{2 g} (-1+s)+e^{2 g} \left(1+e^{4 k}\right) a^2\right)\bigg]\nonumber\\
&&+2 a' \left(2 g'-4 x'+\phi '\right)+a''=0
\eea
\bea\label{eq:eomlast}
&&-4 e^{8 x} b+4 e^{8 x} a+e^{2 g-2 h+8 x} a \left(1-s-2 b a+a^2\right)+b' \left(-8 x'+2 \phi '\right)+b''=0.\nonumber\\
\eea\\
\noindent These equations will be solved together with the constraint coming from reparametrization invariance,
%\bea
%\label{constrainteqn}
%&&\frac{1}{256} e^{-2 (g+h+4 x-\phi )} \bigg[e^{4 (g+k+2 x)}+16 e^{4 (h+k+2 x)}-16 e^{2 (2 g+h+k+4 x)}-64 e^{2 (g+2 h+k+4 x)}\nonumber\\
%&&+e^{4 g+8 x}+16 e^{4 h+8 x}+8 e^{2 (g+h+k+4 x)} s-2 e^{4 g+8 x} s+e^{4 g+8 x} s^2\nonumber\\
%&&+2 e^{2 g+8 x} \left(4 e^{2 h}+4 e^{4 g+2 h}-8 e^{2 (g+h+k)}-e^{2 g+4 k}+4 e^{2 h+4 k}-e^{2 g} (-1+s)\right) a^2\nonumber\\
%&&+e^{4 g+8 x} \left(1+e^{4 k}\right) a^4+4 e^{2 g+8 x} b^2 \left(2 e^{2 h}+e^{2 g} a^2\right)\nonumber\\
%&&-4 e^{2 g+8 x} b a \left(4 e^{2 h}-e^{2 g} (-1+s)+e^{2 g} a^2\right)+16 e^{4 (g+h)} g'^2+64 e^{4 (g+h)} g' h'\nonumber\\
%^&&+16 e^{4 (g+h)} h'^2+32 e^{4 (g+h)} g' k'+32 e^{4 (g+h)} h' k'-2 e^{2 (g+h)} b'^2-2 e^{6 g+2 h} a'^2\nonumber\\
%&&-128 e^{4 (g+h)} g' x'-128 e^{4 (g+h)} h' x'-64 e^{4 (g+h)} k' x'\bigg]\nonumber\\
%&&+\frac{1}{8} e^{2 (g+h-4 x+\phi )} \left(2 g'+2 h'+k'-4 x'\right) \phi '+\frac{1}{8} e^{2 (g+h-4 x+\phi )} \phi '^2=0
%\eea\\
\bea
\label{constrainteqn}
&&\frac{1}{256} e^{-4 (g+h)} \bigg[(-e^{4 (g+k+2 x)}-16 e^{4 (h+k+2 x)}+16 e^{2 (2 g+h+k+4 x)}+64 e^{2 (g+2 h+k+4 x)}\nonumber\\
&&-e^{4 g+8 x}-16 e^{4 h+8 x}-8 e^{2 (g+h+k+4 x)} s+2 e^{4 g+8 x} s-e^{4 g+8 x} s^2\nonumber\\
&&-2 e^{2 g+8 x} \left(4 e^{2 h}+4 e^{4 g+2 h}-8 e^{2 (g+h+k)}-e^{2 g+4 k}+4 e^{2 h+4 k}-e^{2 g} (-1+s)\right) a^2\nonumber\\
&&-e^{4 g+8 x} \left(1+e^{4 k}\right) a^4+4 e^{2 g+8 x} a \left(4 e^{2 h}-e^{2 g} (-1+s)+e^{2 g} a^2\right) b\nonumber\\
&&-4 e^{2 g+8 x} \left(2 e^{2 h}+e^{2 g} a^2\right) b^2+2 e^{6 g+2 h} a'^2+2 e^{2 (g+h)} b'^2 \bigg] \nonumber\\
&&-\frac{1}{16} \left(g'^2+h'^2+2 h' (k'-4 x')+2 g' (2 h'+k'-4 x')-4 k' x'\right)\nonumber\\
&&+\frac{1}{8} (-2 g'-2 h'-k'+4 x') \phi'-\frac{1}{8} \phi'^2 = 0.\nonumber\\
\eea
\noindent
%The details of the derivation are given in Appendix \ref{appendix-EOMs}.

\section{New non-extremal flavored solutions with stabilized dilaton}
\label{section-new_unrotated_solutions}
In this section, we numerically solve the Einstein equations of motion (\ref{eq:eomfirst})-(\ref{constrainteqn}). Our method combines the virtues of the approaches developed in \cite{Aharony:2007vg} and \cite{Mahato:2007zm, PandoZayas:2006sa}. Following \cite{Aharony:2007vg} we first find a set of parameters that  completely determines the UV behavior of the  solutions to any order. In our case there turn out to be eleven free parameters.  The IR behavior is determined by regularity at the horizon and involves seven free parameters.
We then solve numerically the equations of motion and the constraint  using these expansions as boundary conditions. 
To derive the EOMs we follow the framework in \cite{GTV} (see Appendix \ref{appendix-EOMs} for details). 

Two points are worth mentioning
\begin{itemize}
	\item The solutions we are looking for are non-extremal  deformations of the supersymmetric flavored solutions studied in \cite{WARPED}. Therefore, we fix some  UV parameters so that the UV asymptotics reduce to the  supersymmetric solutions when the non-extremal parmeter  goes to zero.
	\item  An important feature of our solutions is a dilaton whose value is ``stabilized" --- \emph{i.e.} it asymptotes to a constant value in the UV. This property is required if the rotation procedure is to produce new solutions where the gravity modes decouple. Recall that, as discussed in \cite{Maldacena:2009mw}, \cite{WARPED},  in the $N_f=0, \ T=0$ case we have  a two-parameter family  of solutions. MN corresponds to the particular case when  both parameters are taken to  zero --or equivalently when the decoupling limit is taken-- while we are interested in solutions where  one of the parameters is infinity  and the other one is finite. Thus in the $N_f\rightarrow 0,\  T \rightarrow 0$ limit our solutions are not expected to reduce to MN. \footnote{ Using the notation of \cite{WARPED} let $\beta$ and $c$ be the parameters that characterize the family of solutions. We have the following cases: 1)when $\beta=c= 0$ we get  MN \ 2) when  $\beta\rightarrow \infty, c \rightarrow \infty$  we obtain  KS, \ 3)when  $\beta\rightarrow \infty$ and c is a free but finite parameter  we get the KS baryonic branch}
\end{itemize}
		\subsection{UV expansions}\label{FTUVexpansions}

To determine the UV behavior of the metric and gauge functions for a finite-temperature deformation of the flavored solutions of \cite{WARPED}, we begin by taking a UV series expansion of the form
\begin{alignat}{2}\label{UVexp}
e^{2h} =& \sum_{i=0}^{\infty} \sum_{j=0}^{i} h_{i,j} \; \rho^j \; e^{4(1-i) \rho /3}\quad\quad\quad\quad& e^{4\phi} =& \sum_{i=1}^{\infty} \sum_{j=0}^{i} f_{i,j} \; \rho^j \; e^{4(1-i) \rho /3}\nonumber\\
e^{2g} =& \sum_{i=0}^{\infty} \sum_{j=0}^{i} g_{i,j} \; \rho^j \; e^{4(1-i) \rho /3}& e^{8x}=&\sum_{i=1}^{\infty} \sum_{j=0}^{i} x_{i,j} \; \rho^j \; e^{2(1-i) \rho /3}\nonumber\\\
e^{2k} =& \sum_{i=0}^{\infty} \sum_{j=0}^{i} k_{i,j} \; \rho^j \; e^{4(1-i) \rho /3}& a =& \sum_{i=1}^{\infty} \sum_{j=0}^{i} a_{i,j} \; \rho^j \; e^{2(1-i) \rho /3}\nonumber\\
&\quad& b =& \sum_{i=1}^{\infty} \sum_{j=0}^{i} b_{i,j} \; \rho^j \; e^{2(1-i) \rho /3},
\end{alignat}
and requiring that it satisfies the Einstein equations given in Appendix \ref{appendix-EOMs}.  For an expansion of this form, we find that all of the coefficients --- up to arbitrary order --- are determined in terms of a set of eleven free parameters. Details of the relations between the expansion coefficients are given in Appendix \ref{appendix-UV asymptotics FT}. 
We next set a number of the free parameters to agree with the supersymmetric UV asymptotics of \cite{WARPED} (summarized in Appendix \ref{appendix-UV asymptotics SUSY}):

\begin{eqnarray}\label{indep_const}
 f_{1,0}=1,& f_{3,0}=3/4 c_+^2 - 3s/4 c_+^2 + 51 s^2/8c_+^2,& h_{1,0}=-1/4+13 s/32,\NO\\
 k_{0,0}=2 c_+/3,& k_{3,0}=\frac{-512 c_+^3 + 8 c_{-} +3s(32 - 32 s + 17 s^2)}{384 c_+^2}& h_{1,1}=1/2 -s/4,\NO\\
a_{2,0}=0,& a_{4,0}=2, &b_{4,0}=0
 \end{eqnarray}

After doing this, the asymptotics reduce to those of \cite{WARPED} in the limit where the parameter $x_{5,0}$ goes to zero.
In order to agree with the convention used in \cite{HEATING}, we will refer to the parameter $x_{5,0}$ as $C_2$ in what follows.\footnote{Note however, that the sign of our $C_2$ is flipped with respect to theirs: we expand $e^{8x}$, whereas they expand $e^{-8x}$.} With the choice of parameters in equation \eqref{indep_const}, the UV asymptotics of our metric and gauge functions are given by
\begin{align}\label{eq:UVexpansion}
e^{8x}& = 1 + C_2 e^{-8 \rho/3} - \frac{C_2 e^{-4 \rho} s}{2 c_+} + \mathcal{O}(e^{-8 \rho /3})\NO\\
e^{2k}&=\frac{2}{3} c_+ e^{4 \rho /3}+\frac{s}{2}-\frac{e^{-4 \rho /3} \left(400-400 s+91 s^2-160 (-2+s)^2 \rho +64 (-2+s)^2 \rho ^2\right)}{96 c_+} + \mathcal{O}(e^{-8 \rho /3})\NO\\
e^{2h}&=\frac{1}{4} c_+ e^{4 \rho /3}+\frac{1}{32} (-8+13 s-8 (-2+s) \rho )\NO\\
&+\frac{1}{256 c_+}e^{-4 \rho /3} \left(208-208 s+43 s^2-64 (-2+s)^2 \rho +64 (-2+s)^2 \rho ^2\right)+ \mathcal{O}(e^{-8 \rho /3})\NO\\
e^{2g}&=c_+ e^{4 \rho /3}+\bigg(1+\frac{5 s}{8}\bigg)+(-2+s) \rho\NO\\
&+\frac{1}{64 c_+}e^{-4 \rho /3} \left(208-208 s+43 s^2-64 (-2+s)^2 \rho +64 (-2+s)^2 \rho ^2\right) + \mathcal{O}(e^{-8 \rho /3})\NO\\
e^{4\phi}&= 1-\frac{3 e^{-4 \rho/3} s}{c_+}+\frac{e^{-8 \rho/3} \left(6+s^2 (51-12 \rho)-48 \rho+6 s (-1+8 \rho)\right)}{8 c_+^2} + \mathcal{O}(e^{-4 \rho})\NO\\
a&=2 e^{-2\rho} + \mathcal{O}(e^{-10 \rho/3})\;\;\;\;\;\;\;\;\;\;b=2 e^{-2 \rho}(2-s)\rho +  \mathcal{O}(e^{-14 \rho/3}).
\end{align}
All of these expressions in fact depend on $C_2$, but only at higher order than we have shown here. The parameter $c_-$ also enters at higher order. In our numerical solutions, we will set $c_-$ equal to 0 for simplicity. This choice of UV behavior leaves us with three free parameters: $c_+, C_2, \text{and } s$.%{********{add notes on the $Q_0$ setting parameter and a compact summary of the type \textbf{N} solutions --- no longer than the summary in WARPED. This is important beacause a) our FT sols reduce to them in the UV when $C_2 \to 0$, b) we use their UV asymptotics exactly, with no free parameters set, when calculating the ADM energy.}}.p

\subsection{IR  asymptotics}
\label{nearhorasymp}

Near the horizon, $\r = \r_{h}$, we expand the equations of motion in a power series up to fifth order: \footnote{The horizon expansion could be taken to higher order to improve accuracy. However, due to the presence of $s$ the expressions  grow unwieldy in complexity. We find  that a  fifth order expansion  provides enough accuracy to obtain well behaved  numerical solutions.}
\begin{align}\label{eq:exp-hor}
 e^{-8x(\r)} &= x_1 (\r-\r_h) + x_2 (\r-\r_h)^2+ x_3 (\r-\r_h)^3+ x_4(\r-\r_h)^4+ x_5(\r-\r_h)^5+\cdots\nonumber\\
e^{2 h(\r)} &= h_0 + h_1 (\r-\r_h) + h_2 (\r-\r_h)^2+ h_3 (\r-\r_h)^3+ h_4 (\r-\r_h)^4+ h_5(\r-\r_h)^5+\cdots\nonumber\\
&\cdots \nonumber\\
e^{4\phi(\r)} &= f_0 + f_1 (\r-\r_h) + f_2 (\r-\r_h)^2+ f_3 (\r-\r_h)^3+ f_4 (\r-\r_h)^4+ f_5(\r-\r_h)^5+\cdots\nonumber\\
a(\r)& = a_0 + a_1 (\r-\r_h) + a_2 (\r-\r_h)^2+ a_3(\r-\r_h)^3+ a_4 (\r-\r_h)^4+a_5 (\r-\r_h)^5+\cdots\nonumber\\
b(\r)& = b_0 + b_1 (\r-\r_h) + b_2 (\r-\r_h)^2+  b_3(\r-\r_h)^3+ b_4 (\r-\r_h)^4+b_5 (\r-\r_h)^5+\cdots
\end{align}

\noindent
Demanding that these expansions satisfy the equations of  motion, we derive expressions for $x_2 ...,h_1 ..., g_1 ..., k_1 ..., f_1 ..., a_1 ..., b_1 ...$ in terms of $x_1,h_0,g_0,k_0,a_0,b_0$.\footnote{The higher order coefficients do not depend on $f_0$. This corresponds to the invariance of the EOMs under constant shifts in the dilaton.} Along with $f_0$, this gives seven independent parameters coming from the horizon expansion.  The expressions for the dependent coefficients are quite  cumbersome and we will not present their explicit form here;  a  Mathematica  file  is available from  the authors upon request.

\subsection{Numerics}
\label{section-numerics}

 In this section we outline the numerical procedure used to find solutions to the equations of motion \eqref{eq:eomfirst}-\eqref{eq:eomlast}  that have a horizon at finite  $\r=\r_h$ and obey boundary conditions \eqref{eq:UVexpansion}. Due to the large dimensional  parameter space (3 free parameters in the UV and 7 in the IR) this is a difficult numerical problem. The numerical method we use  does not pretend to solve the problem in all generality and is not a tool to thoroughly study the full parameter space. However,  within certain limitations to be elucidated below, it allows us to construct numerical solutions with enough accuracy to determine the temperature, energy  and specific heat of these backgrounds. More details of the numerics are given in Appendix \ref{appendix-numerics}.   
 
Our strategy is to start in the UV where there are fewer free parameters. We  pick a set of values for  $c_+$ and $C_2$,\footnote{For simplicity, we set  $c_-$ equal to zero.} which  determine the UV behavior of the background. 

With these UV boundary conditions, we integrate back from $\r_\infty$ and look for values of $c_+$ and $C_2$ which produce solutions with horizon behavior. We will refer to these solutions as the \textit{UV-shot} solutions.
The value of $\r_\infty$ is chosen such that the dilaton reaches its asymptotic value, $e^{4 \phi}|_{\r_\infty}\sim 1$. We next require that this numerical solution matches the horizon expansion \eqref{eq:exp-hor}  and its derivatives up to fifth order. Doing this for a given choice of $(c_+,C_2)$ determines the free parameters $(x_1,h_0,g_0,k_0,f_0,a_0,b_0)$. A black hole solution will not exist for every choice of  $(c_+,C_2)$; for some values of  $(c,C_2)$ there will not be a horizon and for others there may be naked singularities outside the horizon.

%The numerical solutions should not only satisfy the equations of motion derived from the one dimensional effective action, $T-U$,  but also the constraint $T+U$ which arises from  reparametrization invariance. 
%Since we are using a finite value $\r_\infty$ for the UV boundary, and our UV asymptotic expansion is taken only up to finite order, such a solution will still have some remaining inaccuracy, which we can quantify by evaluating the constraint equation $T+U$.

%The constraint should equal zero over the entire interval. For finite temperature solutions,  the divergence of $e^{8x}$ will always cause it to differ from zero near the horizon. This behavior was also seen in the numerical solutions in \cite{HEATING}. In addition to this,  non-zero values of $s$ contribute to the oscillatory behavior close to the horizon. 
%In order to control this behavior and get well behaved solutions we had to drastically change  the parameters of the numerical routine 

In order to match the solution to the horizon expansion, we define  a ``mismatch" function evaluated at a point $\r_0$ close to the horizon, $\r_0=\r_h+ \eps$. 
\begin{align}\label{eq:mismatch1}
m(\r_0)=&[x_{sh}(\r_0) -x_{num}(\r_0)]^2 + [x_{sh}'(\r_0) -x'_{num}(\r_0)]^2+[g_{sh}(\r_0) -g_{num}(\r_0)]^2 + \nonumber\\
&[g_{sh}'(\r_0) -g'_{num}(\r_0)]^2+[h_{sh}(\r_0) -h_{num}(\r_0)]^2 + [h_{sh}'(\r_0) -h'_{num}(\r_0)]^2 +\nonumber\\
&[k_{sh}(\r_0) -k_{num}(\r_0)]^2 + [k_{sh}'(\r_0) -k'_{num}(\r_0)]^2 %+ [\phi_{sh}(\r_0) -\phi_{num}(\r_0)]^2 +\nonumber\\
%&[\phi_{sh}'(\r_0) -\phi'_{num}(\r_0)]^2.
\end{align} 
The subscript $num$ denotes the UV-shot numerical solution and $sh$  denotes the series expansion \eqref{eq:exp-hor}. In \cite{HEATING} the authors used a similar technique to find abelian, $a=b=0$, solutions. Unlike \cite{HEATING}, here we are dealing with non-abelian solutions and due to numerical inaccuracies near the horizon, including $a$, $b$, and $\phi$ in the mismatch function does not allow for an accurate matching. Our  strategy to find the horizon parameters  will be a three step procedure. First we  will minimize a mismatch function $m$ (that does not contain $a$,  $b$, and $\phi$) and  determine $(x_1, h_0, g_0, k_0)$. Then we will tune  $a_0 $ and $b_0$ to eliminate unwanted behavior in $b$. Finally, we will use the invariance of the EOMs under constant shifts of the dilaton to set $f_0$ so that the horizon-shot and UV-shot dilaton agree at the UV boundary. Keep in mind that we want to perform  the matching as close to the horizon as possible and obtain  stable solutions.
We use the word ``stable" loosely to mean horizon shot solutions that agree with the UV shot ones.
This is not an easy task because close to the horizon the UV-shot solutions always display numerical inaccuracies.  %In \cite{HEATING} the authors dealt with this problem  by using a very high order expansion. In our case the presence of  non-zero $a$, $b$, and $s$ significantly adds to the difficulty.
To overcome this problem we do the following:
we first match at a larger value of $\eps = 0.7$,  then we  use the resulting values of the matched horizon parameters as seeds for a new match performed at a lower $\eps=.15$. In the last step we also constrain the allowed variance of $(x_1,g_0,h_0,k_0)$ around the seed values until the resulting $\eps=.15$ match gives horizon shot solutions which agree with the UV shot solutions. The end result is $m(\r_h + .15) < 10^{-7}$ and a set of parameters highly tuned  to give stable solutions.  

\begin{figure}[h]
\begin{minipage}{.495\textwidth}
%\captionsetup{width=0.8\textwidth}
 \begin{center}
 \includegraphics[width=7.9cm]{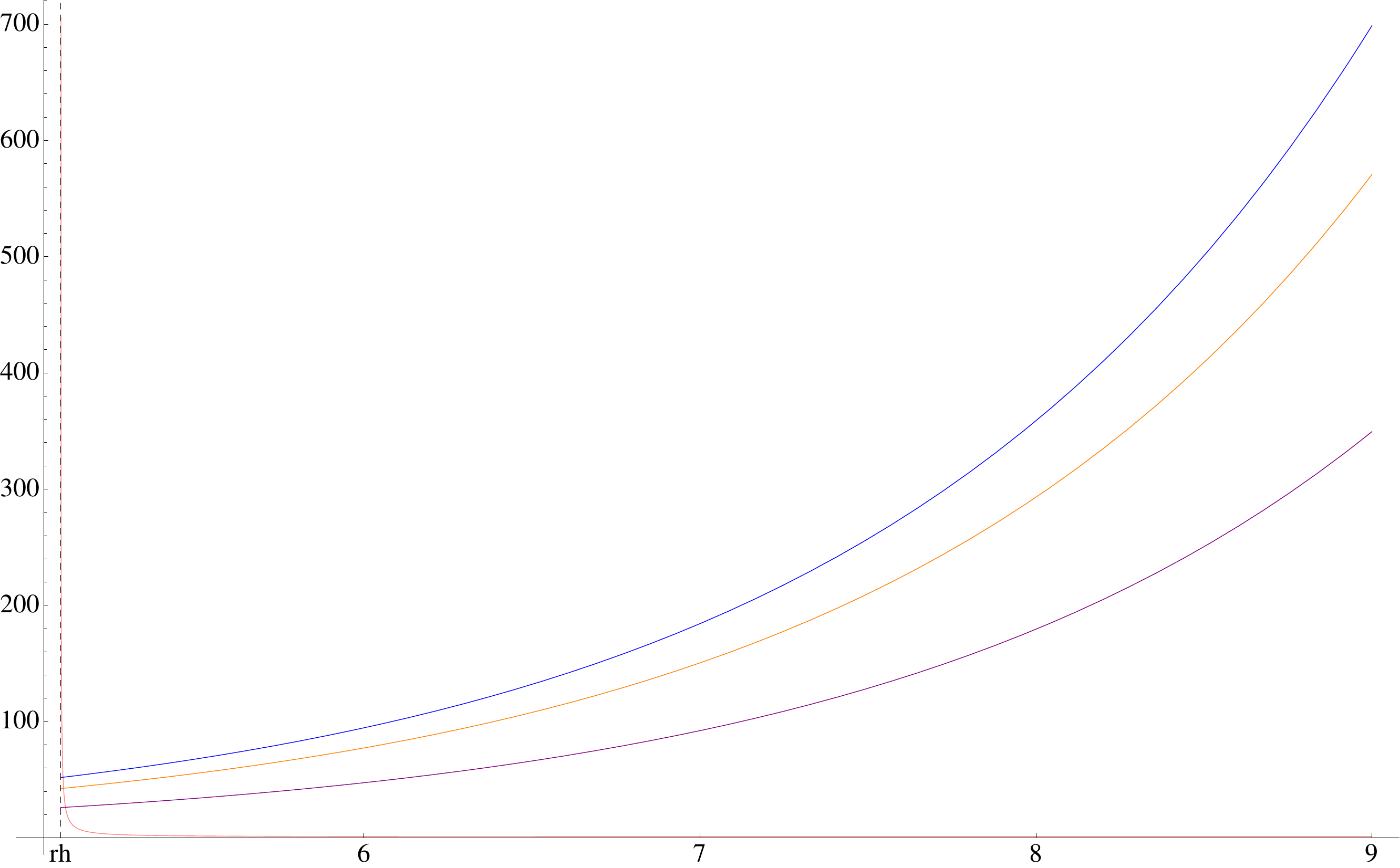}
 \caption{\small{Horizon-shot and UV-shot solutions for metric functions at $s=1$, $c_+=3$, $C_2=800000$. $e^{k}$ - orange, $e^{g}$ - blue, $e^{h}$ - purple, $e^{8x}$ - pink}}
\label{goodSolEntireInterval_c3}
\end{center}
\end{minipage}
\begin{minipage}{.495\textwidth}
%\captionsetup{width=0.8\textwidth}
\begin{center}
 \includegraphics[width=7.9cm]{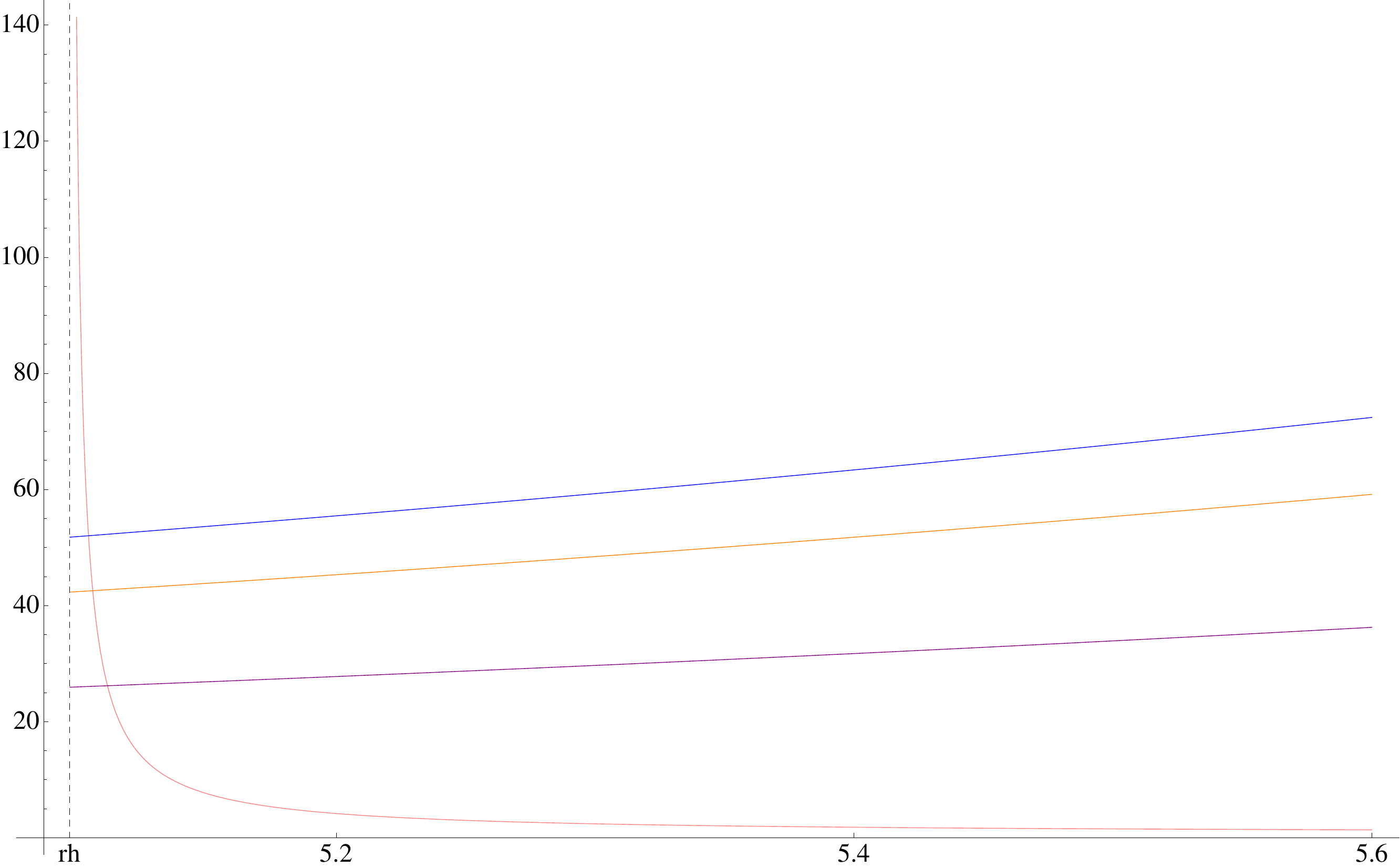}
 \caption{\small{Horizon-shot and UV-shot solutions for metric functions at $s=1$, $c_+=3$, $C_2=800000$. Zoomed into region near horizon at $\r _h \sim$ 5.097.}}
\label{goodSolZoomed_c3}
\end{center}
\end{minipage}
\end{figure}

\begin{figure}[!h]
\begin{minipage}{.495\textwidth}
%\captionsetup{width=0.8\textwidth}
 \begin{center}
 \includegraphics[width=7.9cm]{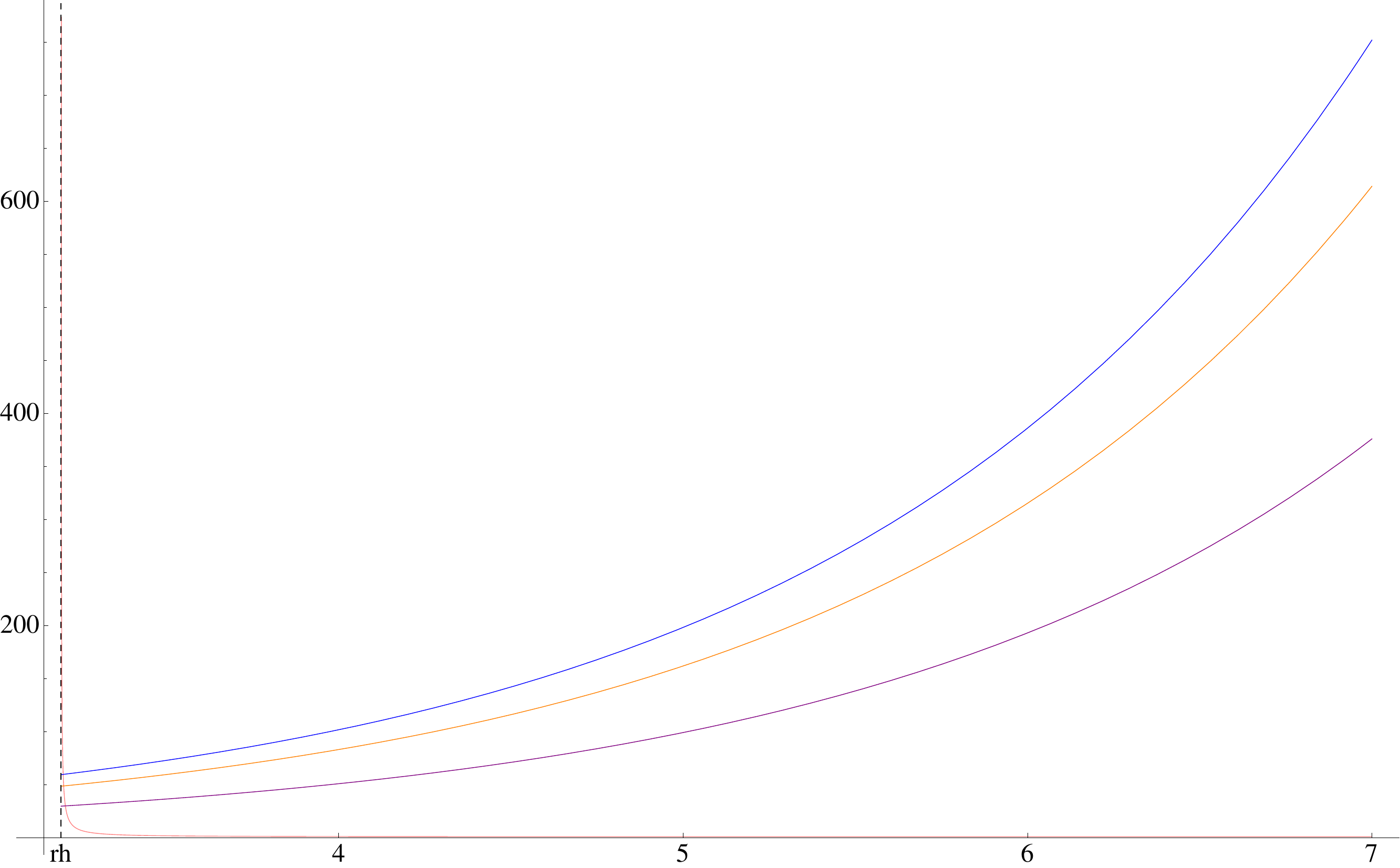}
 \caption{\small{Horizon-shot and UV-shot solutions for metric functions at $s=1$, $c_+=50$, $C_2=5000$.}}
\label{goodSolEntireInterval_c50}
\end{center}
\end{minipage}
\begin{minipage}{.495\textwidth}
%\captionsetup{width=0.8\textwidth}
\begin{center}
 \includegraphics[width=7.9cm]{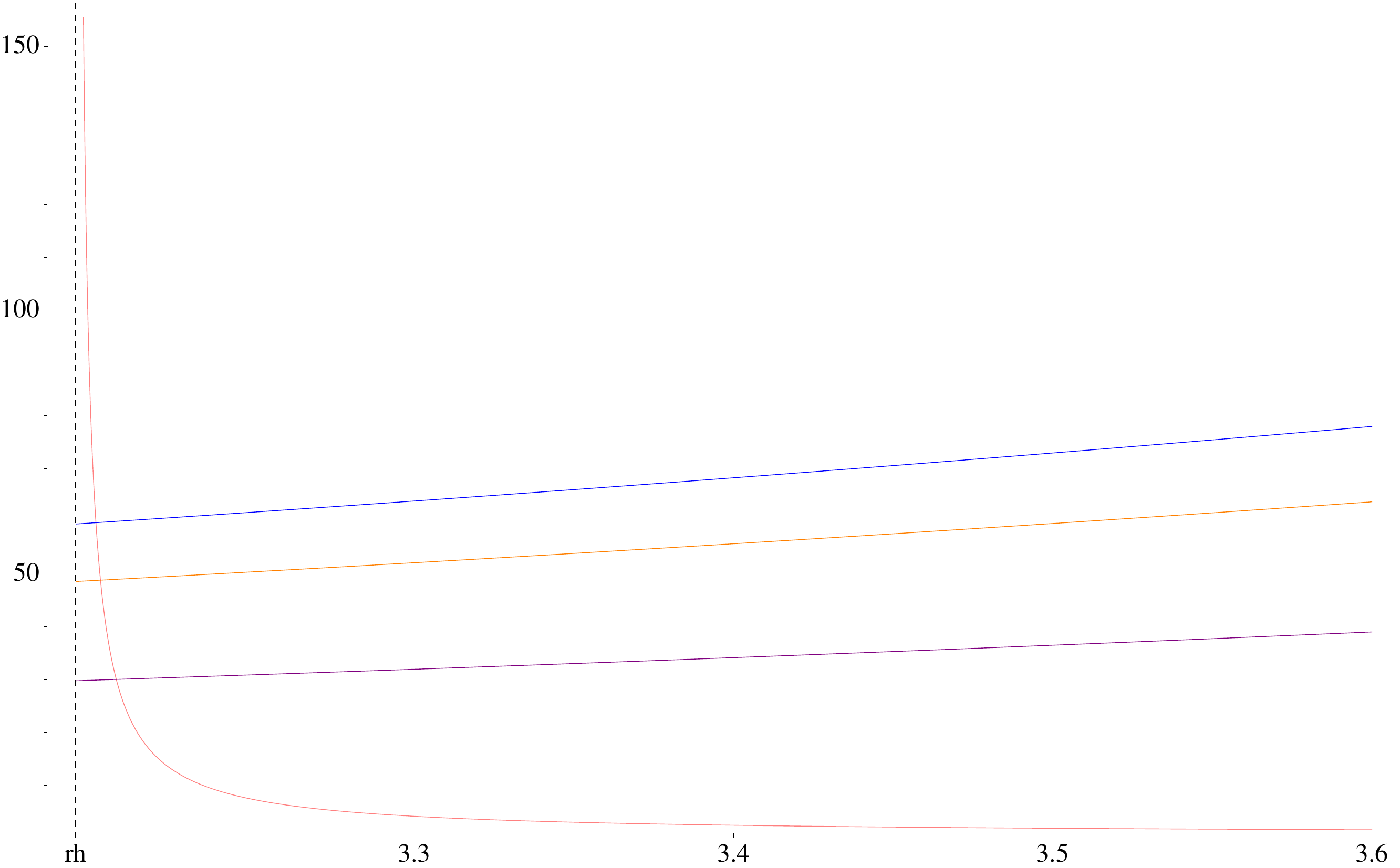}
 \caption{\small{Horizon-shot and UV-shot solutions for metric functions at $s=1$, $c_+ =50$, $C_2=5000$. Zoomed into region near horizon at $\r _h \sim$ 3.19}}
\label{goodSolZoomed_c50}
\end{center}
\end{minipage}
\end{figure}

With these  parameters we shoot from the horizon and show that our  solutions also satisfy the  constraint  coming from reparametrization invariance, $T+ U=0$. 
We use $\texttt{WorkingPrecision}=40$ and obtain solutions that satisfy $T+ U < \, \sim 10^{-6} $ throughout the interval. Figs.(\ref{goodSolEntireInterval_c3} - \ref{goodSolZoomed_c50}) show two solutions for different $(c_+,C_2)$ values.

Let us comment on the relation between the numerical solutions presented here and some known cases.  The non-extremal non-flavored with stabilized dilaton  exists in the literature only for $a(\rho)=0$ \cite{HEATING}. We can reproduce the  results of \cite{HEATING} with  improved  numerics. The constraint is much better satisfied now, $ T+U <~10^{-6}$ here as opposed to  $ T+U <~10^{-3}$ in \cite{HEATING}. In addition, the mismatch function is of order $10^{-7}$ here and was $10^{-4}$ in \cite{HEATING}.  
The extremal flavored solutions with stabilized dilaton \cite{WARPED} used a slightly different framework, they reduce the problem to an unknown master function\footnote{The master function is obtained form the BPS equations and thus, is not a formalism we can use in the non-extremal case} for which they solve numerically.  This makes difficult a direct comparison of the numerics. However, we can  compare with their  UV and IR expansions for $ e^{2g},\  e^{2h},\  e^{2k},\ $  etc. Since we use  their UV deformed by a non-extremality parameter $C_2$,  our UVs automatically reduce to the ones in \cite{WARPED} when $C_2=0$. The IR is a bit more delicate. Their solutions are singular in the IR, $ e^{2g},\  e^{2h},\  e^{2k},\ $   go like $\log r$ as $r\rightarrow 0$. 
If we set $C_2=0$ in our code we run into a singularity  at $\rho\rightarrow 0$  and reproduce the solutions of \cite{WARPED}. Also note that we find  regular horizons only for very large values of the non-extremality parameter; for small values of $C_2$ we run into singularities. 
%The reason for this has to do with the fact that the $a$ and $b$ functions that come from the UV and horizon shot solutions differ over the entire interval even in the best matched case (although they are typically small in magnitude compared to $(x,g,h,k,f)$). This is in contrast to the metric functions, whose forward and backward shot solutions tend to agree quite well away from the horizon in the best matched case.
%For some values of $c$ and $C_2$, Including $a$ and $b$ allows for $m_1$ to be minimized to only about $10^{-2}$, and the resulting horizon shot solutions for $(x,g,h,k,f)$ differ significantly from the UV shot solutions. For these same $c$ and $C_2$ values, excluding $a$ and $b$ from the mismatch function allows \texttt{NMinimize} to reduce the match to values of order $10^{-7}$, and the resulting horizon shot solutions for $(x,g,h,k)$ agree very closely with the UV shot solutions.  For either method of matching, the resulting $b$ horizon shot solution differs from the UV shot solution over the entire interval --- in particular, $b$ tends to exponentially diverging behavior at large $\rho$. However, $b_0$ can be tuned to eliminate this divergence, with negligible effect on the behavior of the other functions $(x,g,h,k,f)$ --- see Figs.\ref{bnottuned_c3} and \ref{btuned_c3}. Thus our strategy will be to minimize a mismatch function which does not include $a$ and $b$, in order to obtain horizon and UV shot solutions for  $(x,g,h,k)$ which agree as well as possible. Our new mismatch function, $m_2$, is then:

\noindent


\subsection{Temperature of the solutions}
\label{section-solutiontemperature}

The expression for the horizon temperature of our solutions follows from the standard prescription of analytically continuing the solution to Euclidean time and imposing the periodicity of this coordinate. The temperature is then given by $T = 1/\beta$, where the period $\beta$ is determined by requiring the regularity of the Euclideanized metric at the horizon. We use the numerically determined parameters described in section \ref{nearhorasymp} to describe the metric at the horizon; of these parameters, only $x_1$ and $k_0$ enter into the expression for the temperature, which is found to be
\be
\label{horizontemp}
T_{hor}=\frac{1}{4 \pi} \frac{x_1}{\sqrt{k_0}}.
\ee
In \cite{HEATING} it was shown that  the horizon temperature is the same for  pre- and post-rotated solutions. It can be checked that this is still the case here. 

\begin{figure}[h]
\begin{minipage}{.495\textwidth}
 \begin{center}
 \includegraphics[width=8cm]{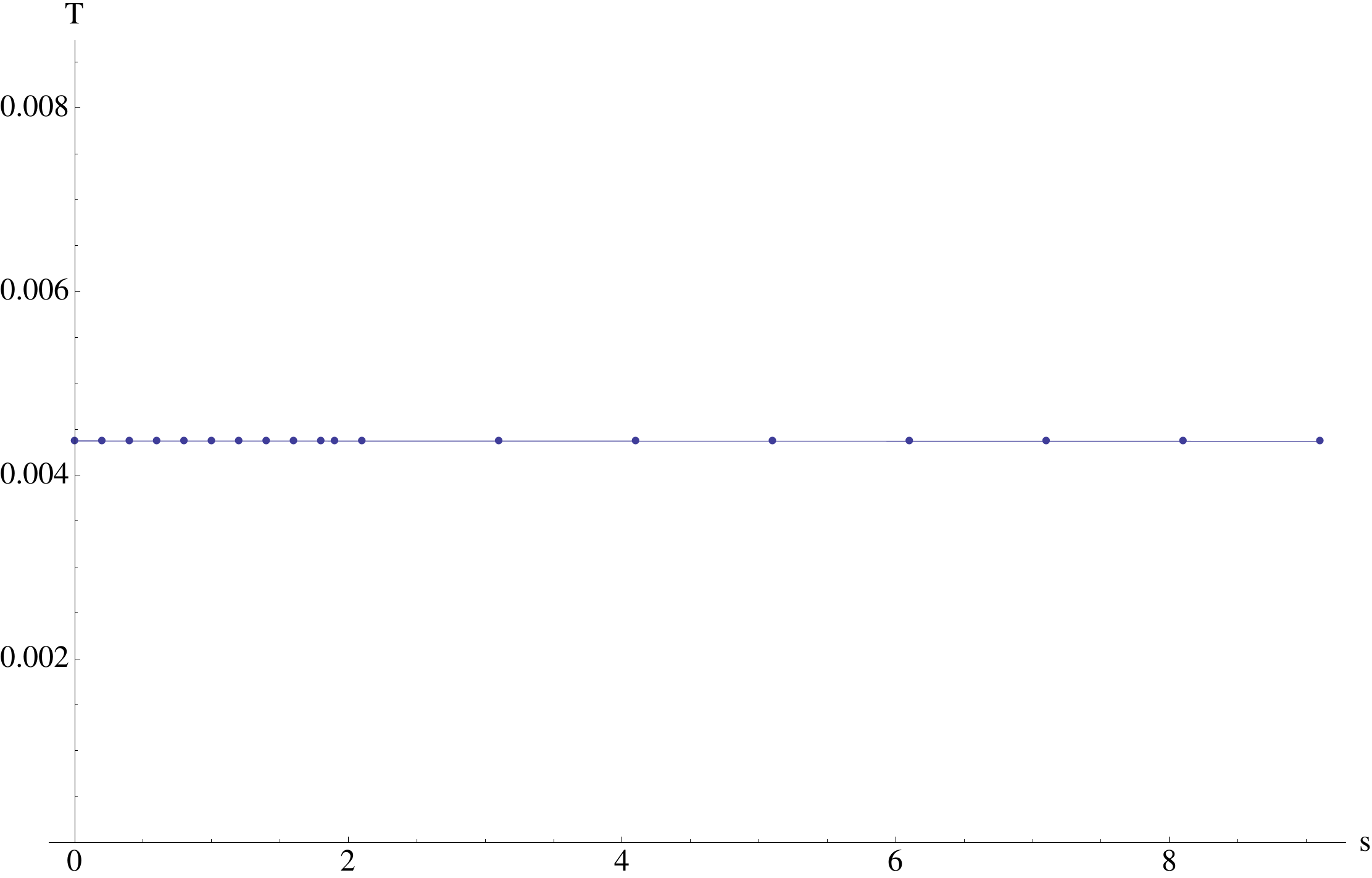}
 \caption{\small{Temperature at the horizon versus $N_f/N_c$. $c_+ = 50, C_2 = 5000$.}}
\label{tempFlavc_50_C2_5000}
\end{center}
\end{minipage}
\begin{minipage}{.495\textwidth}
\begin{center}
 \includegraphics[width=8cm]{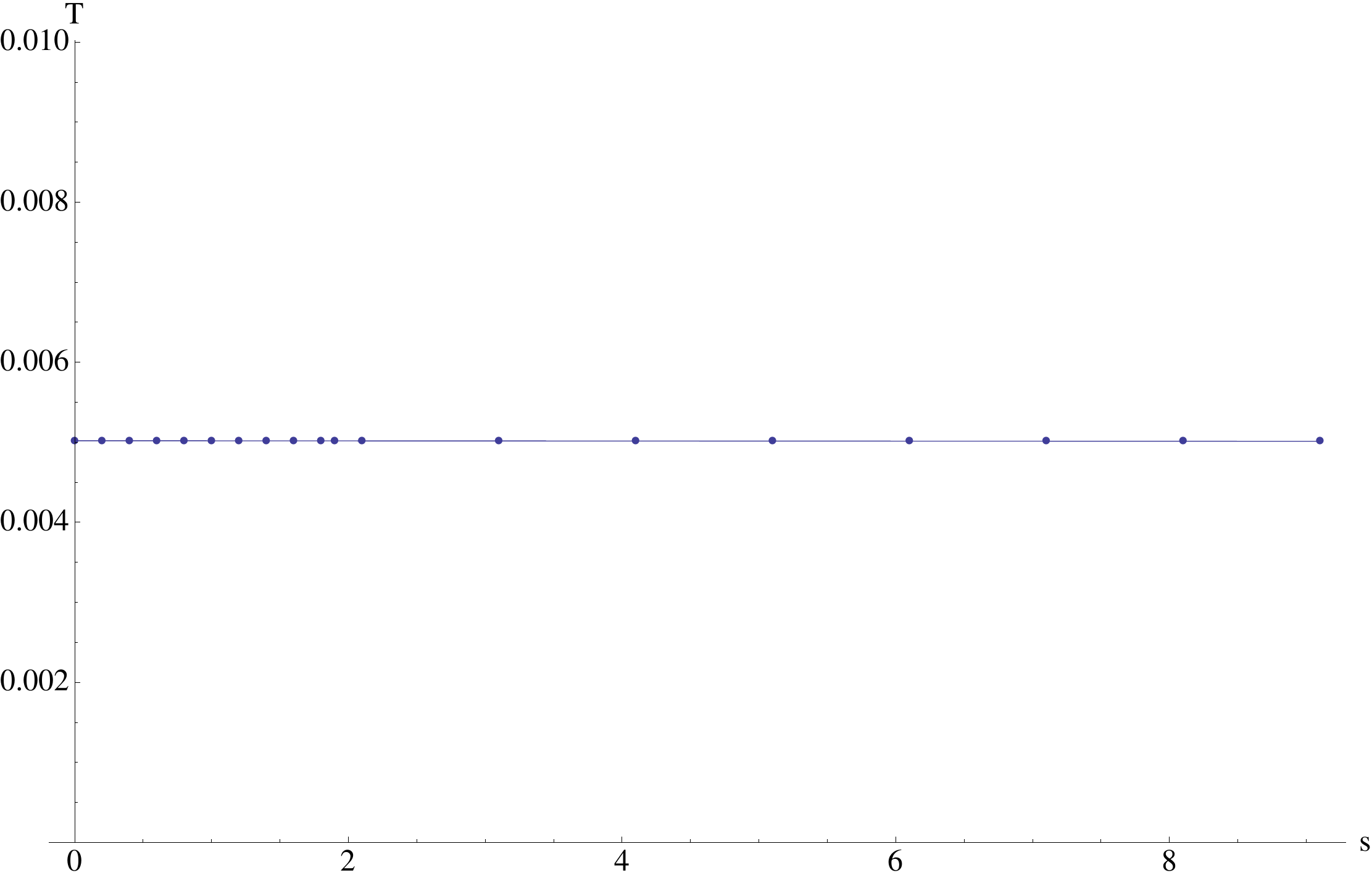}
 \caption{\small{Temperature at the horizon versus $N_f/N_c$. $c_+ = 3, C_2 = 800000$.}}
\label{tempFlavc_3_C2_800000}
\end{center}
\end{minipage}
\end{figure}

In Figs.(\ref{tempFlavc_50_C2_5000}) and (\ref{tempFlavc_3_C2_800000}) we see that, for fixed values of $c_+$ and $C_2$, the horizon temperature has negligible dependence on $s\equiv N_f/N_c$. As we vary $0 \leq s <\, \sim 10$ the temperature remains constant; we conclude that the temperature is not affected by the backreaction of the flavor branes.

\begin{figure}[!h]
\begin{minipage}{.495\textwidth}
\begin{center}
\includegraphics[width=8cm]{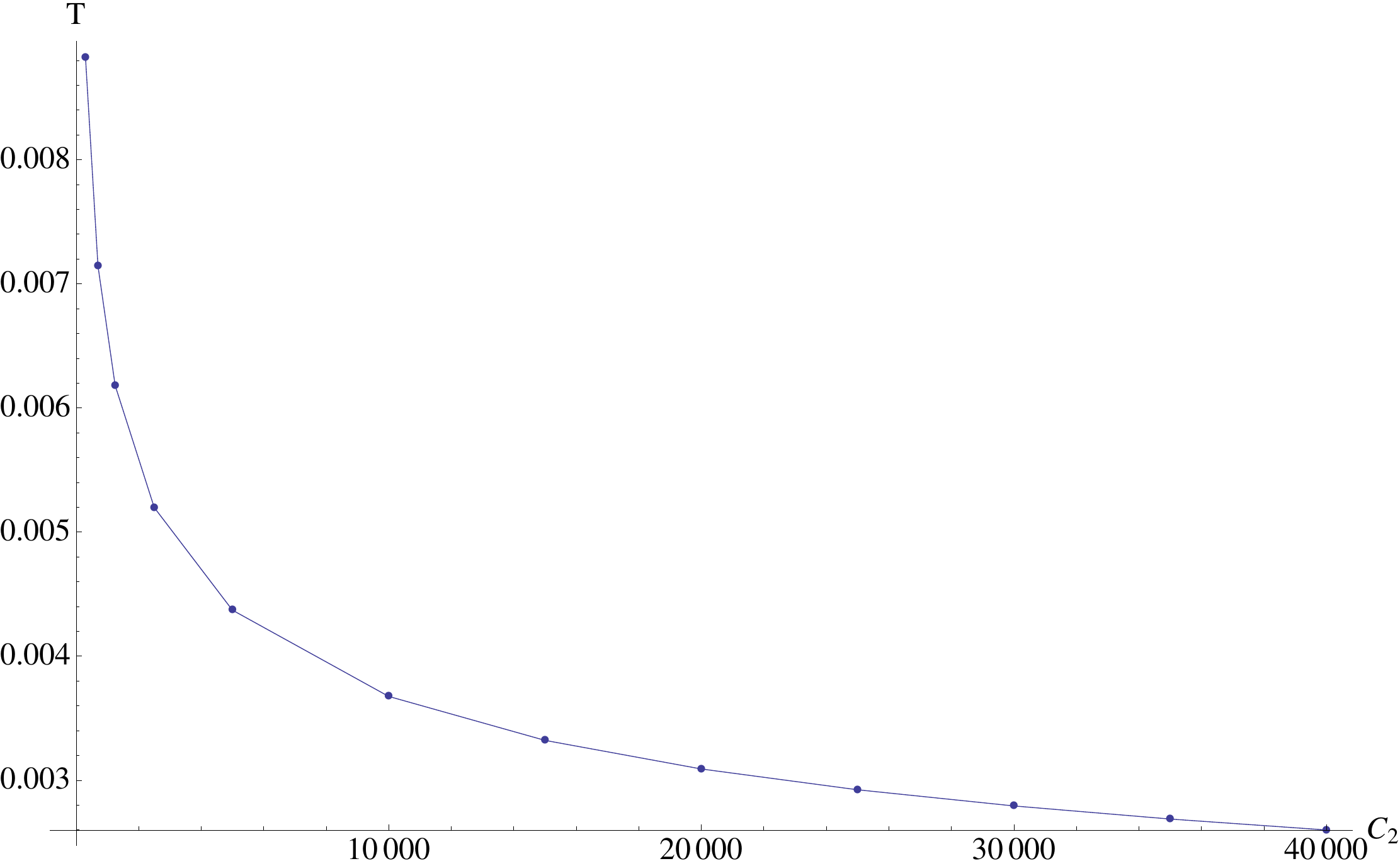}
\caption{\small{Temperature at the horizon versus $C_2$. $c_+ = 50, s = 1$.}}
\label{tempC2_50}
\end{center}
\end{minipage}
\begin{minipage}{.495\textwidth}
\begin{center}
\includegraphics[width=8cm]{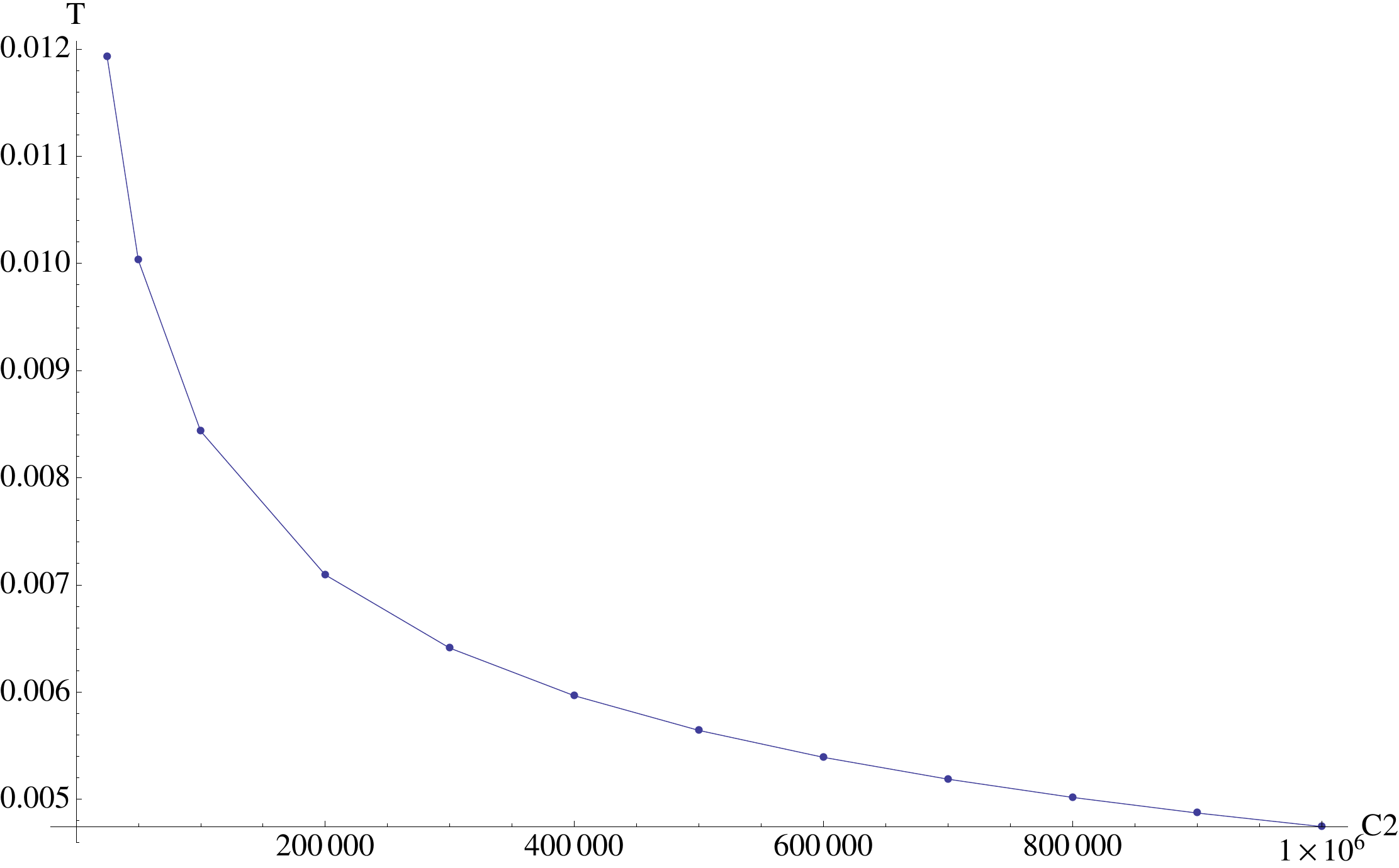}
\caption{\small{Temperature at the horizon versus  $C_2$. $c_+ = 3, s = 1$.}}
\label{tempC2_3}
\end{center}
\end{minipage}
\end{figure}

\begin{figure}[!h]
\begin{minipage}{.495\textwidth}
\begin{center}
\includegraphics[width=8cm]{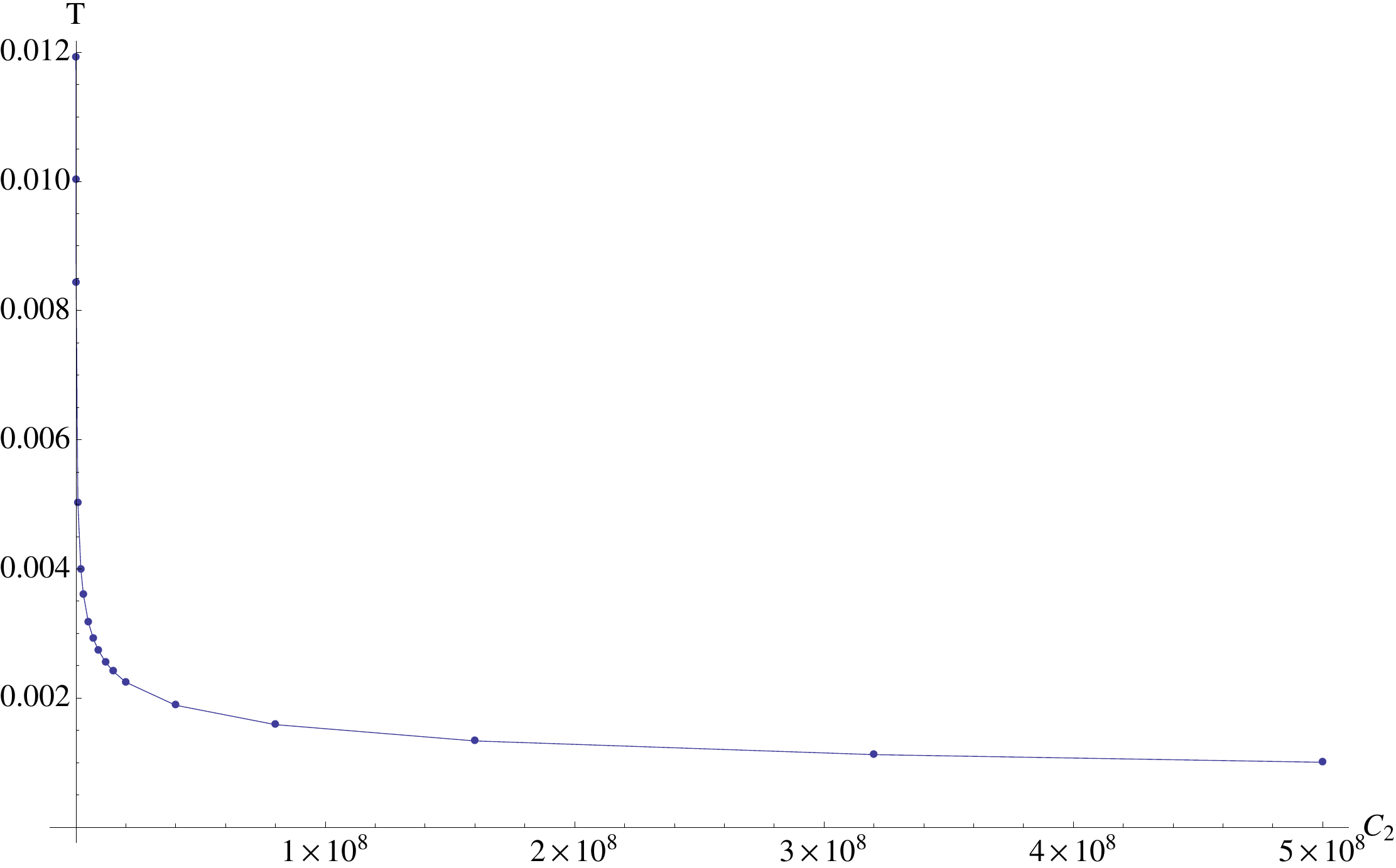}
\caption{\small{Temperature at the horizon versus for very large values of $C_2$. $c_+ = 3, s = 1$. }}
\label{tempLargeC2}
\end{center}
\end{minipage}
\begin{minipage}{.495\textwidth}
\begin{center}
\includegraphics[width=8cm]{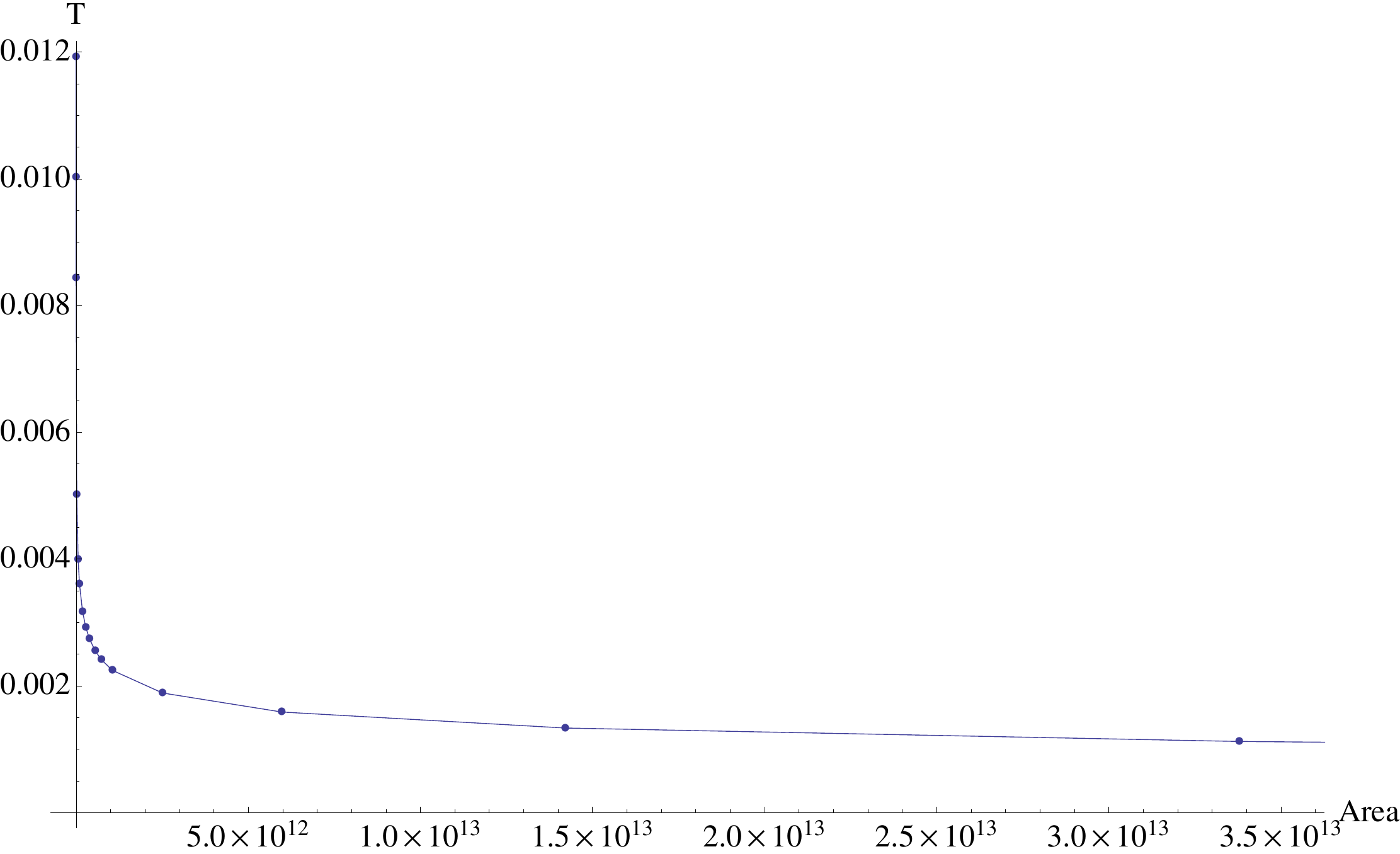}
\caption{\small{Temperature at the horizon versus $\mathcal{A}/V_3$. $c_+ = 3, s = 1$.}}
\label{tempArea}
\end{center}
\end{minipage}

\end{figure}

In Figs.(\ref{tempC2_3}), (\ref{tempC2_50}) and (\ref{tempLargeC2}) we show how the horizon temperature varies with $C_2$ for fixed values of $c_+$ and $s$. In Fig. (\ref{tempArea}) we present the temperature as a function of the  horizon area.  Note that as  $C_2 $ decreases  the temperature increases indicating that it diverges in the extremal limit.  Also,  for $C_2 \rightarrow \infty$ figure (\ref{tempLargeC2}) shows that the temperature goes to a very small but non-zero value, $T_c$. This behavior is reminiscent of the wrapped five branes black holes of  \cite{GTV}. In that framework  $T_c$ is the Hagedorn temperature of the little string theory and is also a critical temperature at which a first order phase transition  occurs (chiral symmetry restoration and deconfinement) \cite{GTV}. A similar phase transition could exist here. If  that is the case, the U-dualities described in the following  section will probably map it to a  phase transition in Klebanov-Strassler. This is an issue that deserves further study.

%, where the horizon area is
%\be
%\mathcal{A} = 8 \pi^3 f_0^{1/2} h_0 g_0 k_0^{1/2} V_3, 
%\ee
%and given in terms of the horizon expansion parameters and $V_3 = \text{Vol}(\textbf{R}^3)$. 

%%%%%%%%%%%%%%%%%%%%%%%%%%%%%%%%%%%%%%%%%%%%%%%%%%%%%%
%%%%%%%%%%%%%%%%%%%%%%%%%%%%%%%%%%%%%%%%%%%%%%%%%%%
\section{Non-extremal backgrounds with  flavored resolved deformed conifold asymptotics}
\label{section_rotating}
In \cite{Maldacena:2009mw} the authors presented a solution generating technique that takes a IIB background of  $D5$ branes wrapped on the $S^2$ of the resolved conifold (non-trivial $F_{(3)} $ and $\phi$) and produces a background wich also has   $D3$  charge (non trivial $F_{(3)}, F_{(5)}, H_{(3)}$ and $\phi$). After a scaling limit the resulting geometry represents the baryonic branch of the Klebanov-Strassler solution. The initial space has topology $R^{1,3}\times \mathcal{M}_6$ and preserves $\mathcal{N}=1$ supersymmetry . The algorithm consists of performing three T-dualities in the $R^3$ directions, lifting to M-theory, boosting with rapidity $\beta$ in the eleventh direction, reducing back to ten dimensions and finally T-dualizing back in $R^3$. This procedure is equivalent to a rotation in the space of Killing spinors. We will therefore refer  to this algorithm indistinctively  as a {\it chain of dualities} or as a {\it rotation}.

 In the previous section we found non-extremal solutions representing  a background of $N_c$ wrapped color $D5$ branes and $N_f$ smeared $D5$  flavor branes $(N_f/N_c \sim 1)$. In this section we want to apply the procedure developed in \cite{Maldacena:2009mw} to these backgrounds. 

 After applying the rotation procedure, taking $\beta \to \infty$ to decouple the gravitational modes, and performing the rescalings as described in \cite{HEATING}, we obtain the transformed solutions in Einstein frame (see Appendix \ref{appendix-UDuality} for details):
\bea
& & ds_{IIB}^2  =  N_c 
\Big[e^{-\phi/2} \mathcal{H}^{-1/2}( -e^{-8x} dt^2 + dx_i dx^i  )+
 e^{3\phi/2} \mathcal{H}^{1/2} ds_6^2 \Big],\nonumber\\
 & & F_{(3)} = \frac{ N_c }{4 } f_{(3)},\;\;\; H_{(3)} =
 -e^{-4x} \frac{  N_c }{4}\, e^{2 \phi}\,  *_6 f_{(3)},
\nonumber \\
& & F_{(5)} =   - N_c^2 ( 1+ *_{10})  \left[\volf \wedge
d\left(\frac{e^{-4x}}{\mathcal H}\right)\right],
\label{after_rotation}
\eea 
where
\be
\mathcal{H}^{1/2}= \sqrt{e^{-2\phi}-e^{-8x}}
\label{zsazsa}
\ee
and the explicit forms of $f_{(3)}$ and  $ds_6^2$ are  given in equations \eqref{eq:F3ansatz} and \eqref{eq:baseMetric} respectively. Note that, unlike more standard finite temperature  solutions where the non-extremality factor  $e^{-8x}$ enters only in the $g_{rr}$ and $g_{tt}$ elements of the metric, here  the warp factor  $\mathcal{H}$ and the fluxes $F_{(5)}$ and $H_{(3)}$ depend on $e^{-8x}$. \footnote{ Note also that in \cite{HEATING} the  factor of  $e^{-8x}$ in $H_{(3)}$ trivially cancelled with a similar one coming from the six dimensional Hodge dual. This is not the case here. The reason is that in \cite{HEATING} the authors considered a particular point in the baryonic branch ($a=0$) while in the present work we consider the general case,  $a(\r)$ (see  equation \eqref{eq:F3-app-ansatz}) }

The use of U-dualities is a well stablished solution generating technique in supergravity. In the presence of smeared flavored branes one could question the validity and meaning of the uplift to eleven dimensions. Two pieces of knowledge provide useful insight on this issue:
\begin{enumerate}
	\item In \cite{Gaillard:2009kz} the authors studied the uplift of smeared $D6$ flavor  branes to M-theory. They argued that the violation of the Bianchi identity in ten dimensions implies that the eleven dimensional-geometry will no longer be Ricci flat; it is no longer of $G_2$ holonomy but it does carry $G_2$ structure. Thus, the flavors appear in eleven dimensions as intrinsic torsion. This eleven dimensional theory is no longer maximally supersymmetric (it is $1/8$ BPS) and therefore it is no longer unique. This is an interesting possibility for the interpretation of the uplift of smeared flavor branes. For us, however, the important point is that once the reduction to ten dimensions is carried out the correct smearing form is recovered. 

	\item In \cite{WARPED} the authors studied a transformation to generate $SU(3)$ structure solutions of IIB supergravity starting from non-K\"ahler backgrounds describing wrapped $D5$ branes with additional flavor (smeared) $D5$ branes. Working with the BPS equations and the $SU(3)$ structure of the backgrounds they presented a solution generating technique that amounts to a  rotation  in the space of Killing spinors. This rotation procedure is well defined  (even in the presence of smeared flavor branes) and is equivalent to the chain of U-dualities alluded to above. 
\end{enumerate}
 
In the present work we are dealing with a non-extremal deformation of \cite{WARPED}. However, since the backgrounds we are working with are not supersymmetric, we cannot use the BPS formalism of  \cite{WARPED}. On the other hand, we are not interested in the interpretation of the background in eleven dimensions, so we will not attempt to address the questions presented in \cite{Gaillard:2009kz}. Instead, guided by the results of \cite{WARPED}, we will forge ahead and apply the chain of dualities to the flavored, non extremal backgrounds with stabilized dilaton found in section \ref{section-new_unrotated_solutions}. However in order to claim that the outcome of these dualities is still a solution of IIB plus sources, we have to show that the rotated backgrounds are indeed a solution of the EOMs.

\subsection{Equations of motion with $D3$ and $D5$ charges and smeared flavor branes}

In this section we will verify  that given the EOMs before the rotation procedure is applied, the EOMs after the rotation are also satisfied. The EOMs before rotation were presented in section \ref{subsection-non_extremal_flavored_backgrounds} and derived in Appendix \ref{appendix-EOMs}.
After the rotation procedure is carried out, the supergravity background contains $H_{(3)}$ and $F_{(5)}$ fluxes in addition to the $F_{(3)}$ that was present before rotation. The full type IIB action, supplemented by the DBI action for the smeared flavor branes, is then given by

\be
	S = S_{IIB} + S_{\text{sources}}
\ee
where
% NOTE: the \vphantom's in the following are to make the brackets line up.
	\begin{align}
		S_{IIB} = \frac{1}{2 \kappa_{10}^2}\int &\sqrt{-g}  R -   \frac{1}{4 \kappa_{10}^2} \int \left( \vphantom{\frac{1}{2}} d\phi\wedge *d\phi + 
		 e^{-\phi} H_{(3)} \wedge *H_{(3)} \right. \nn\\
		 &+\left. e^{\phi} F_{(3)} \wedge *F_{(3)} + \frac{1}{2} F_{(5)} \wedge *F_{(5)} - C_{(4)} \wedge F_{(3)} \wedge H_{(3)}\right),
	\end{align}
and
\begin{align}
	S_{\text{sources}}=-\frac{4 T_5}{(4 \pi)^2} \int&\left( \volf \wedge d\rho \wedge d\psi \; e^{\phi /2} \sqrt{|g_{ab}+e^{-\phi/2}B_{ab}|}\right.\nonumber\\
&-\left.\vphantom{\sqrt{|e({-B}|}} C_{(6)}+C_{(4)}\wedge B_{(2)}\right)\wedge \Xi_{(4)}\nonumber\\
%&=&-\frac{s\, T_5}{(4 \pi)^2}\int d^{10}x \sin \theta \sin \tilde{\theta} e^{\phi /2} |\text{Det}|^{1/2}\left[g_{ab}+e^{-\phi/2}B_{ab}\right]\nonumber\\
%&+&\int (C_{(6)}-C_{(4)}\wedge B_{(2)})\wedge \Xi_{(4)}
%\label{eq:sourceMundane}
\end{align}
Note that the metric is in Einstein frame and that $g_{ab}$ and $B_{ab}$ respectively  represent  the pullbacks of the metric and $B_{(2)}$ to the flavor brane world volume. Recall that  before rotation  $H_{(3)}=0$ and thus, the DBI action was only the pullback of the metric.  After rotation  we need to specify the form of the NS potential,  $B_{(2)}$. In the extremal case ($T=0$), this potential should reduce to the one in \cite{WARPED}. We propose the following ans\"atz:
\be\label{eq:generalB2}
B_{(2)}= b_1(\r)  \tilde{\omega_3} \wedge d\r + b_2(\r) e_1\wedge e_2 + b_3 e_1\wedge \tilde\omega_2 + b_4(\r) e_2\wedge \tilde\omega_1 + b_5(\r) \tilde\omega_1\wedge \tilde\omega_2.
\ee
To make the gauge degrees of freedom more apparent it is convenient to parameterize $B_{(2)}$ in a slightly different way. Let us introduce
$b_2(\r)=\widetilde{b_2}(\r)+ (1- a(\r)^2)b_5(\r)$.
It is then  easy to verify that 
\be
B_{(2)}= B_{(2)}|_{b_5=0} - d\left[ b_5(\r) \tilde{\omega_3}\right]
\ee
with $b_5(\r)$ an undetermined function. It is clear that  any $b_5(\r)$ results in the same $H_{(3)}$, and that $b_5$ is just  a gauge degree of freedom. Demanding  that  $H_{(3)}= d[B_{(2)}]$, we get 5 equations. This system of equations can be reduced to just one differential equation. We choose a gauge such that $b_1(\r)$ coincides with the one in \cite{WARPED} (see Appendix \ref{appendix-form_of_B2} for details).

The flavor brane world volume is $dt \wedge dx_1 \wedge dx_2 \wedge dx_3 \wedge d\rho \wedge d\psi  \equiv \volf \wedge d\rho \wedge d\psi$, and the smearing form is given by $\Xifour = \frac{s}{4} \sin \theta \sin \tilde \theta \,d \theta \wedge d \varphi \wedge d \tilde \theta \wedge d \tilde \varphi$,
where we have set $\alpha ' = g_s = 1$. As was the case before rotation, the $C_{(6)} \wedge \Xifour$ term in the source action vanishes. 
%\footnote{ {******** Internal Footnote: the form of $C_6$ isn't given in Casero:2006pt, WARPED, or HAGEDORN, and it doesn't play a part in the EOMs. I assume that it can be shown that its general form vanishes when wedged with $\Xifour$. Do we need to concern ourselves with presenting its general form here when those other papers did not mention it?}}
Note that in the EOMs that follow from this action, the contribution from the Wess-Zumino term in the source action is exactly cancelled by that from the Chern-Simons term in the bulk.

The Bianchi identities for the RR and NS-NS forms after rotation are

\be
d\Fthree =\Xifour, \qquad  d\Hthree=0, \qquad  d\Ffive + \Fthree \wedge \Hthree = \Btwo \wedge \Xifour 
\ee

\noindent
The first is unchanged from before the rotation, and the latter two can be checked to hold given the EOMs before rotation. Note that this does not give us the full expression for $\Btwo$, but only the components which are orthogonal to $\Xifour$, namely $B_{\rho \psi}$. However $\Btwo$ only appears in the source action, and there it is either wedged with $\Xifour$, or pulled back to the flavorbrane world volume (which is orthogonal to $\Xifour$). Therefore only this orthogonal component will be relevant for our equations of motion. A more detailed derivation of the general form of $\Btwo$ is given in Appendix \ref{appendix-form_of_B2}.

The EOMs for the fluxes after rotation are
\be
d \left( e^{-\phi} * \Hthree \right) = F_{(3)} \wedge F_{(5)} + \frac{e^{-8x}}{\mathcal{H}}\; \volf \wedge \Xifour 
\ee
\be
d \left( e^{\phi} * F_{(3)} \right) = - \Hthree \wedge F_{(5)}
\ee

\noindent
where the term with $\mathcal{H}$ in the first equation comes from varying the source term. These can be checked to hold given the EOMs before rotation.

\noindent
The dilaton EOM is
\bea
d\left(* d\phi\right) &=& \frac{e^{\phi}}{2}\Fthree \wedge *\Fthree - \frac{e^{-\phi}}{2}\Hthree \wedge *\Hthree\nonumber +  \left(2 \kappa_{10}^2 \right)\frac{\delta \mathcal{L}_{\text{sources}}}{\delta \phi }
\eea
where
\begin{align}
\frac{\delta \mathcal{L}_{\text{sources}}}{\delta \phi }&=
\frac{1}{2}e^{\phi /2}\left(\sqrt{|g_{ab}+e^{-\phi/2}B_{ab}|}-\frac{|g_{(4)}| e^{-\phi} B_{\rho\psi}^2}{\sqrt{|g_{ab}+e^{-\phi/2}B_{ab}|}}\right)\  \volf \wedge d\rho \wedge d\psi \wedge \Xifour
\end{align}
and can be checked to hold given the EOMs before rotation.

\noindent
Lastly, we have the Einstein equations
\begin{align}
R_{\mu\nu} - \frac{1}{2}g_{\mu\nu}R =\ & \frac{1}{2}\left(\partial_{\mu} \phi \partial_{\nu} \phi - \frac{1}{2} g_{\mu\nu}\partial_{\lambda}\phi \partial^{\lambda}\phi \right) + \frac{1}{12}e^{\phi}\left(3 F_{\mu \alpha \beta} F_{\nu}^{ \alpha \beta} - \frac{1}{2} g_{\mu\nu}F_{(3)}^2 \right)\nonumber\\
& + \frac{1}{12}e^{-\phi}\left(3 H_{\mu \alpha \beta} H_{\nu}^{ \alpha \beta} - \frac{1}{2} g_{\mu\nu}H_{(3)}^2 \right) + \frac{1}{96}\left(F_{\mu \alpha \beta \gamma \delta}  F_{\nu}^{\alpha \beta \gamma \delta}\right) + T^{\,\text{sources}}_{\mu \nu}
\end{align}
%\begin{align}
%R_{\mu\nu}&=\frac{1}{2}\partial_{\mu} \phi \partial_{\nu} \phi + \frac{1}{96}F_{\mu\alpha\beta\gamma \delta} F_{\nu}^{\alpha\beta\gamma\delta}
% + \frac{1}{4} ( e^{-\phi} H_{\mu\alpha\beta} H_{\nu}^{\alpha\beta} + e^\phi  F_{\mu\alpha\beta} F_{\nu}^{\alpha\beta} ) \nn\\
% & - \frac{1}{48}g_{\mu\nu} ( e^{-\phi} H_{\alpha\beta\gamma} H^{\alpha\beta\gamma} + e^\phi  F_{\alpha\beta\gamma} F^{\alpha\beta\gamma} )+ T^{flavor}_{\mu\nu}\end{align} 

\noindent
where the variation of the Chern-Simons term in the bulk action is cancelled by the variation of the Wess-Zumino term in the source action. Here, $T^{\,\text{sources}}_{\mu \nu}$ is obtained from (after rotation)
\bea
T^{\mu\nu}_{\text{sources}}&=&\frac{2 \kappa_{10}^2}{\sqrt{-g_{(10)}}} \frac{\delta \mathcal{L}_{\text{sources}}}{\delta g_{\mu \nu}},
\eea

\noindent
and  has  components:
\begin{align}
&T^{\,\text{sources}}_{t t}=\frac{s\,e^{-2 g-2 h-8 x-4 \phi }}{2 \mathcal{H}^2}\,\,, \qquad\qquad T^{\,\text{sources}}_{x_i x_j}=-\frac{s\,\eta_{ij} e^{-2 g-2 h-4 \phi } }{2 \mathcal{H}^2}\,, \nn \\
&T^{\,\text{sources}}_{\rho\rho}=-\frac{s}{2} e^{-2 g-2 h+2 k+8 x},\qquad\qquad T^{\,\text{sources}}_{\psi\psi}=-\frac{s}{8} e^{-2 g-2 h+2 k}\,\,,\nn\\
&T^{\,\text{sources}}_{\varphi\psi}=-\frac{s}{8} e^{-2 g-2 h+2 k} \cos\theta\,\,,\qquad \qquad T^{\,\text{sources}}_{\varphi\varphi}=-\frac{s}{8} e^{-2 g-2 h+2 k}\cos^2\theta \,\,,\nn\\
&T^{\,\text{sources}}_{\tilde\varphi\psi}=-\frac{s}{8}e^{-2h-2g+2k} \cos\tilde\theta\,\,,\qquad \qquad
T^{\,\text{sources}}_{\tilde\varphi\tilde\varphi}=-\frac{s}{8}e^{-2h-2g+2k}
\cos^2 \tilde \theta \,\,,\nn\\
&T^{\,\text{sources}}_{\varphi\tilde\varphi}=-\frac{s}{8}e^{-2h-2g+2k}\cos\theta
\cos\tilde\theta  \,.\nonumber\\
\label{Tflavoraft}
\end{align}

\noindent
Again, the Einstein equations can be checked to hold given the EOMs before rotation.

\subsection{Rotating  the solutions}
 
It is now a simple procedure to use equations (\ref{after_rotation}) along with our numerical solutions to produce numerical solutions for the background after rotation. Before presenting and commenting on the numerical results  let look at the large $\rho$  behavior. %These are shown in Figs.(\ref{figrhorhoflavc50} -- \ref{newfig2})\footnote{The UV behavior of the rotated solutions is extremely sensitive to the precise UV behavior of the seed solutions, due to the $\mathcal{H}$ factor. For this reason, we have generated these plots from the UV-shot seed solutions.}; here the rotated metric has been written in the form

\subsubsection{Asymptotics after the rotation}

 In order to write down the form of the metric  the UV  after the rotation, we identify from \eqref{after_rotation},
 \begin{align}
	 &g_{xx}= e^{-\phi/2}\mathcal{H}^{-1/2}\qquad g_{t t} = e^{-\phi/2 -8 x}\mathcal{H}^{-1/2} \qquad g_{\rho \rho}= e^{3\phi/2}\mathcal{H}^{1/2}  e^{2 k + 8 x}\nonumber\\
&g_{\theta\theta}=e^{3\phi/2}\mathcal{H}^{1/2}  e^{2 h}  \qquad g_{\tilde\theta \tilde\theta}=e^{3\phi/2}\mathcal{H}^{1/2} \frac{ e^{2 g}}{4} \qquad g_{\psi\psi}=e^{3\phi/2}\mathcal{H}^{1/2} \frac{ e^{2 k}}{4}  \nonumber\\
& \label{eq:identify-metric-after}
\end{align}

Using  \eqref{eq:UVexpansion} and \eqref{eq:identify-metric-after} we  obtain, 
\be
g_{xx} \sim \frac{\sqrt{\frac{2c_+}{3}}  e^{2 \rho /3}}{\sqrt{s}}-\frac{e^{-2 \rho /3}}{24 \sqrt{6c_+} s^{3/2}} \bigg(16 c_+^2 C_2+3 \left(-2+s (2-16 \rho )+16 \rho +s^2 (1+4 \rho)\right)\bigg)+\cdots\NO
\ee
\be
-g_{tt} \sim \frac{\sqrt{\frac{2c_+}{3}} e^{2 \rho /3}}{\sqrt{s}}-\frac{e^{-2 \rho /3}}{24 \sqrt{6c_+} s^{3/2}} \bigg(16 c_+^2 C_2+3 \left(-2+s (2-16 \rho )+16 \rho +s^2 (1+4 \rho)\right)\bigg)+\cdots\NO
\ee
\be
g_{\rho\rho} \sim \sqrt{\frac{2c_+}{3}} e^{2 \rho /3} \sqrt{s}+\frac{e^{-2 \rho /3}}{24 \sqrt{6c_+} \sqrt{s}} \bigg(-6+16 c_+^2 C_2+s (6-48 \rho )+48 \rho +3 s^2 (-11+4 \rho)\bigg) + \cdots\NO
\ee
\be
g_{\theta\theta} \sim \sqrt{\frac{3c_+}{2}} e^{2 \rho /3} \sqrt{s}+\frac{e^{-2 \rho /3}}{64 \sqrt{6c_+} \sqrt{s}} \bigg(-6+16 c_+^2 C_2 + s^2 (9-36 \rho )+48 \rho +6 s (-7+8 \rho )\bigg)+ \cdots\NO
\ee
\be
g_{\tilde\theta\tilde\theta} \sim \sqrt{\frac{3c_+}{2}} e^{2 \rho /3} \sqrt{s}+\frac{e^{-2 \rho /3}}{64 \sqrt{6c_+} \sqrt{s}} \bigg(-6+16 c_+^2 C_2+48 \rho -18 s (-3+8 \rho )+s^2 (-39+60 \rho )\bigg)+\cdots\NO
\ee
\be\label{eq:asympt_after}
g_{\psi\psi} \sim \frac{\sqrt{c_+} e^{2 \rho /3} \sqrt{s}}{2 \sqrt{6}}+\frac{e^{-2 \rho /3}}{96 \sqrt{6c_+}\sqrt{s}} \bigg(-6+16 c_+^2 C_2+s (6-48 \rho )+48 \rho +3 s^2 (-11+4 \rho )\bigg)+\cdots\\
\ee

It is clear that this expansion is not valid as $s\rightarrow 0$. However, this does not mean that we cannot obtain the expected KS asymptotics in that limit. It just indicates that we should take the  $s\rightarrow 0$ limit before performing the rotation. Indeed, setting  $s=0$ before the rotation in \eqref{eq:UVexpansion} and defining $c_+ \sqrt{8} A(\rho) = \sqrt{24 \rho + 8 C_2 c_+^2 -3}$ to avoid clutter we obtain the following metric after rotation\footnote{The effect of flavor in the UV can be heuristically understood by looking at the warp factor $
\mathcal{H}^2=e^{-2\phi} -e^{-8x} $ 
and at the expansions \eqref{eq:UVexpansion}. In the flavored case $\mathcal{H}^2\sim e^{-4 \rho/3} \frac{s }{c_+} + \mathcal{O}(e^{-8\rho/3}) $ while in the unflavored case $\mathcal{H}^2\sim p(C2,c_+)  e^{-8 \rho/3}  + \mathcal{O}(e^{-4\rho }) $, where $p(\rho;C2,c_+)$ denotes a polynomial in $\rho$.
}
, 

%The asymptotics of the rotated unflavored case were discussed in \cite{HEATING}, but ours have a slightly different form due to differing signs for the $C_2$ parameter. We define a quantity
%\be
%c_+ \sqrt{8} A(\rho) = \sqrt{24 \rho + 8 C_2 c_+^2 -3}
%\ee
%\noindent
%in terms of which the unflavored UV asymptotics of the rotated metric functions become\footnote{Note that $g_{tt} = -\mathcal{\tilde H}^{1/2} e^{-8x}$ after rotation. $-g_{tt}$ below then gives the expansion of the positive expression $\mathcal{\tilde H}^{1/2} e^{-8x}$.}
\begin{alignat}{1}
\label{gxx_unflavored_rotated}
g_{xx} \sim \frac{e^{4 \rho/3}}{A(\rho)} +\frac{e^{-4 \rho /3}}{2048 c_+^4A(\rho)^3} \bigg(&16 c_+^2 C_2 \left(185-272 \rho +128 \rho ^2\right)\NO\\
&+3 \left(-847+3504 \rho -3456 \rho ^2+2048 \rho ^3\right)\bigg)+\mathcal{O}(e^{-8 \rho/3})\NO
\end{alignat}
\begin{alignat}{1}
-g_{tt} \sim \frac{e^{4 \rho/3}}{A(\rho)} +\frac{e^{-4 \rho /3}}{2048 c_+^4 A(\rho)^3} \bigg(&-2048 c_+^4 C_2^2+16 c_+^2 C_2 \left(233-656 \rho +128 \rho ^2\right)\NO\\
&+3 \left(-847+3504 \rho -3456 \rho ^2+2048 \rho ^3\right)\bigg)+\mathcal{O}(e^{-8 \rho/3})\NO
\end{alignat}
\begin{alignat}{2}
&g_{\rho\rho} \sim \frac{2c_+}{3} A(\rho)+\mathcal{O}(e^{-8 \rho/3})\quad\quad& &g_{\theta\theta} \sim \frac{c_+}{4}A(\rho) +  \frac{e^{-4 \rho /3}}{4} A(\rho)(2 \rho -1)+\mathcal{O}(e^{-8 \rho/3})\NO
\end{alignat}
\begin{alignat}{2}
&g_{\psi\psi} \sim \frac{1}{6} c_+ A(\rho)+\mathcal{O}(e^{-8 \rho/3})\quad\quad& &g_{\tilde\theta\tilde\theta} \sim \frac{c_+}{4}A(\rho) -  \frac{e^{-4 \rho /3}}{4} A(\rho)(2 \rho -1)+\mathcal{O}(e^{-8 \rho/3}).\NO\\
\end{alignat}
\noindent 
\noindent
Using a variable $u = e^{2\rho/3}$, the metric reads at leading order
\beq
ds^2 = \frac{ u^2}{A(u)} (dx_idx^i ) +\frac{ 3  c A(u) }{2 } (\frac{d u^2}{u^2} +  ds^2_{T^{1 1}}) +   \cdots \mathcal{O}(u^{-2}),
\eeq
\noindent
where $A(u)=\frac{3}{c_+\sqrt{2}} \sqrt{  \log u + \frac{2}{9}C_2 c_+^2 - \frac{1}{12}}$ contains the typical $\log$ behavior of  KS asymptotics.
. Note that,  since numerically the  work of finding non-extremal solutions is done  before the rotation  it is trivial for us to set $s$ to any value, however small, and then rotate. 
%The addition of the  flavor branes $s=N_f/N_c \sim causes the rotated asymptotics to depart drastically from the Klebanov-Strassler form in the UV. To first sub-leading order, we find
%\be
%g_{xx} \sim \frac{\sqrt{\frac{2c}{3}}  e^{2 \rho /3}}{\sqrt{s}}-\frac{e^{-2 \rho /3}}{24 \sqrt{6c} s^{3/2}} \bigg(16 c^2 C_2+3 \left(-2+s (2-16 \rho )+16 \rho +s^2 (1+4 \rho)\right)\bigg)+\cdots\NO
%\ee
%\be
%-g_{tt} \sim \frac{\sqrt{\frac{2c}{3}} e^{2 \rho /3}}{\sqrt{s}}-\frac{e^{-2 \rho /3}}{24 \sqrt{6c} s^{3/2}} \bigg(16 c^2 C_2+3 \left(-2+s (2-16 \rho )+16 \rho +s^2 (1+4 \rho)\right)\bigg)+\cdots\NO
%\ee
%\be
%g_{\rho\rho} \sim \sqrt{\frac{2c}{3}} e^{2 \rho /3} \sqrt{s}+\frac{e^{-2 \rho /3}}{24 \sqrt{6c} \sqrt{s}} \bigg(-6+16 c^2 C_2+s (6-48 \rho )+48 \rho +3 s^2 (-11+4 \rho)\bigg) + \cdots\NO
%\ee
%\be
%g_{\theta\theta} \sim \sqrt{\frac{3c}{32}} e^{2 \rho /3} \sqrt{s}+\frac{e^{-2 \rho /3}}{64 \sqrt{6c} \sqrt{s}} \bigg(-6+16 c^2 C_2 + s^2 (9-36 \rho )+48 \rho +6 s (-7+8 \rho )\bigg)+ \cdots\NO
%\ee
%\be
%g_{\tilde\theta\tilde\theta} \sim \sqrt{\frac{3c}{32}} e^{2 \rho /3} \sqrt{s}+\frac{e^{-2 \rho /3}}{64 \sqrt{6c} \sqrt{s}} \bigg(-6+16 c^2 C_2+48 \rho -18 s (-3+8 \rho )+s^2 (-39+60 \rho )\bigg)+\cdots\NO
%\ee
\subsubsection{Rotated flavored solutions}

\begin{figure}
\begin{minipage}[t]{.495\textwidth}
%\captionsetup{width=0.8\textwidth}
 \begin{center}
 \includegraphics[width=7.3cm]{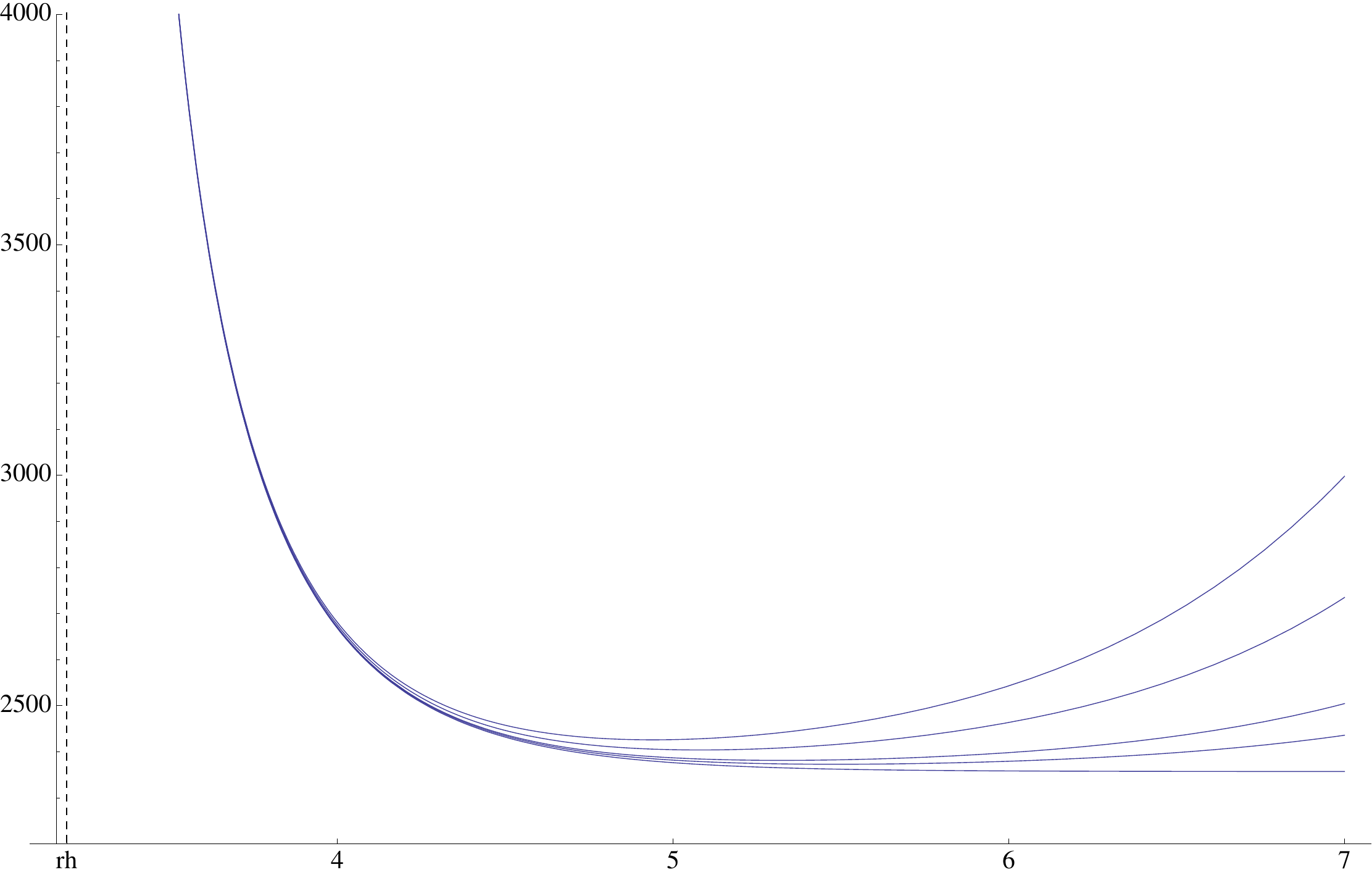}
 % graph-c1-97-new.pdf: 240x167 pixel, 72dpi, 8.47x5.89 cm, bb=0 0 240 167
 \caption{\small{$g_{\rho\rho}$ metric element for solutions after rotation, $c_+=50,\  C_2= 5000$. From bottom to top: $s=0,1,19/10,51/10,91/10$.}}
\label{figrhorhoflavc50}
\end{center}
\end{minipage}
\begin{minipage}[t]{.495\textwidth}
%\captionsetup{width=0.8\textwidth}
\begin{center}
 \includegraphics[width=7.3cm]{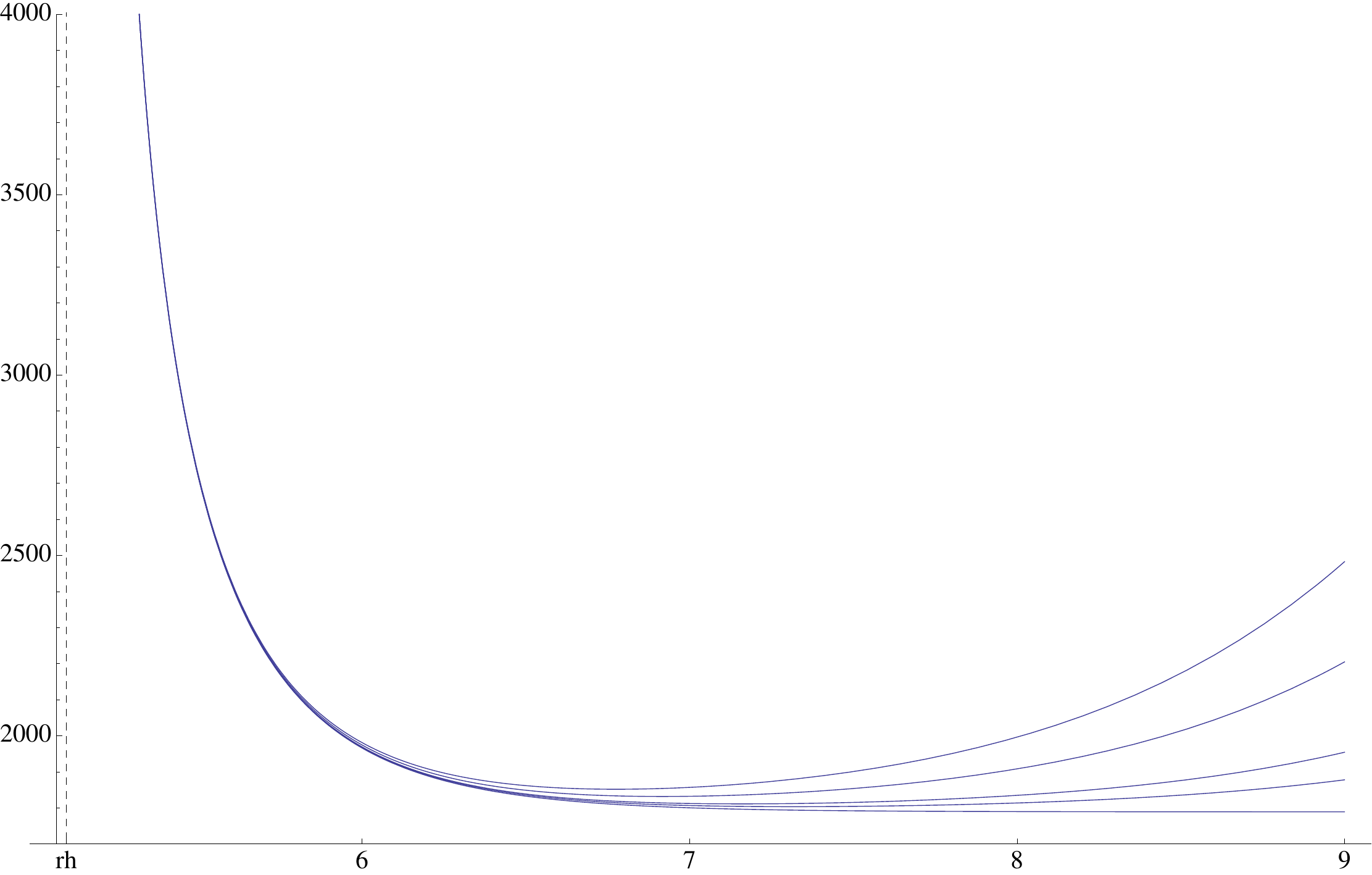}
 % graph-c125-new.pdf: 240x161 pixel, 72dpi, 8.47x5.68 cm, bb=0 0 240 161
 \caption{\small{$g_{\rho\rho}$ metric element for solutions after rotation, $c_+=3,\  C_2= 800000$. From bottom to top: $s=0,1,19/10,51/10,91/10$.}}
\label{figrhorhoflavc2}
\end{center}
\end{minipage}
\end{figure}

\begin{figure}
\begin{minipage}[t]{.495\textwidth}
%\captionsetup{width=0.8\textwidth}
 \begin{center}
 \includegraphics[width=7.3cm]{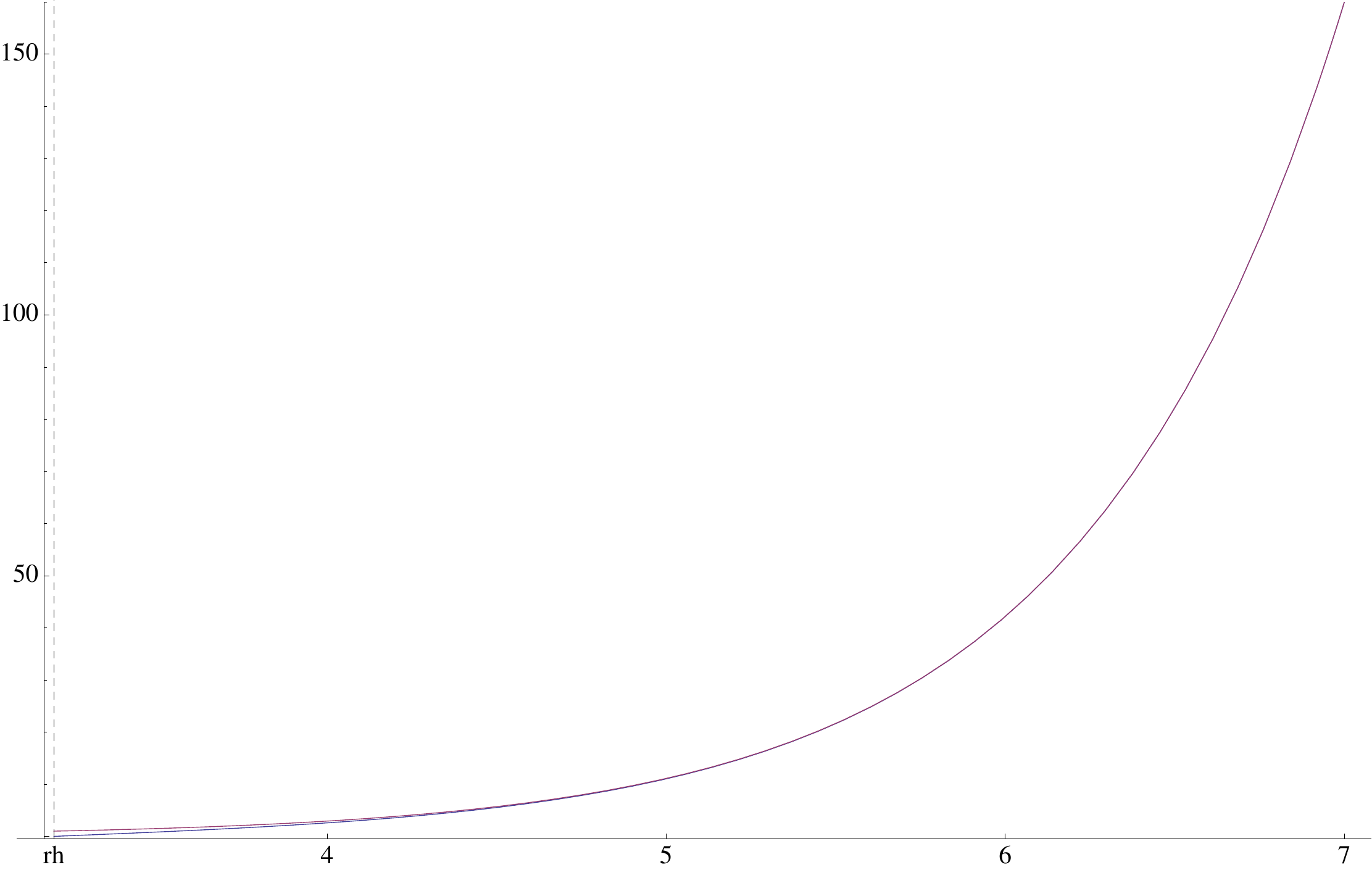}
 % graph-c1-97-new.pdf: 240x167 pixel, 72dpi, 8.47x5.89 cm, bb=0 0 240 167
 \caption{\small{$g_{xx}$ (red) and $g_{tt}$ (blue) metric elements for the unflavored solution after rotation: $c_+=50,\  C_2= 5000, s=0$.}
}
\label{figxxttflavc50}
\end{center}
\end{minipage}
\begin{minipage}[t]{.495\textwidth}
%\captionsetup{width=0.8\textwidth}
\begin{center}
 \includegraphics[width=7.3cm]{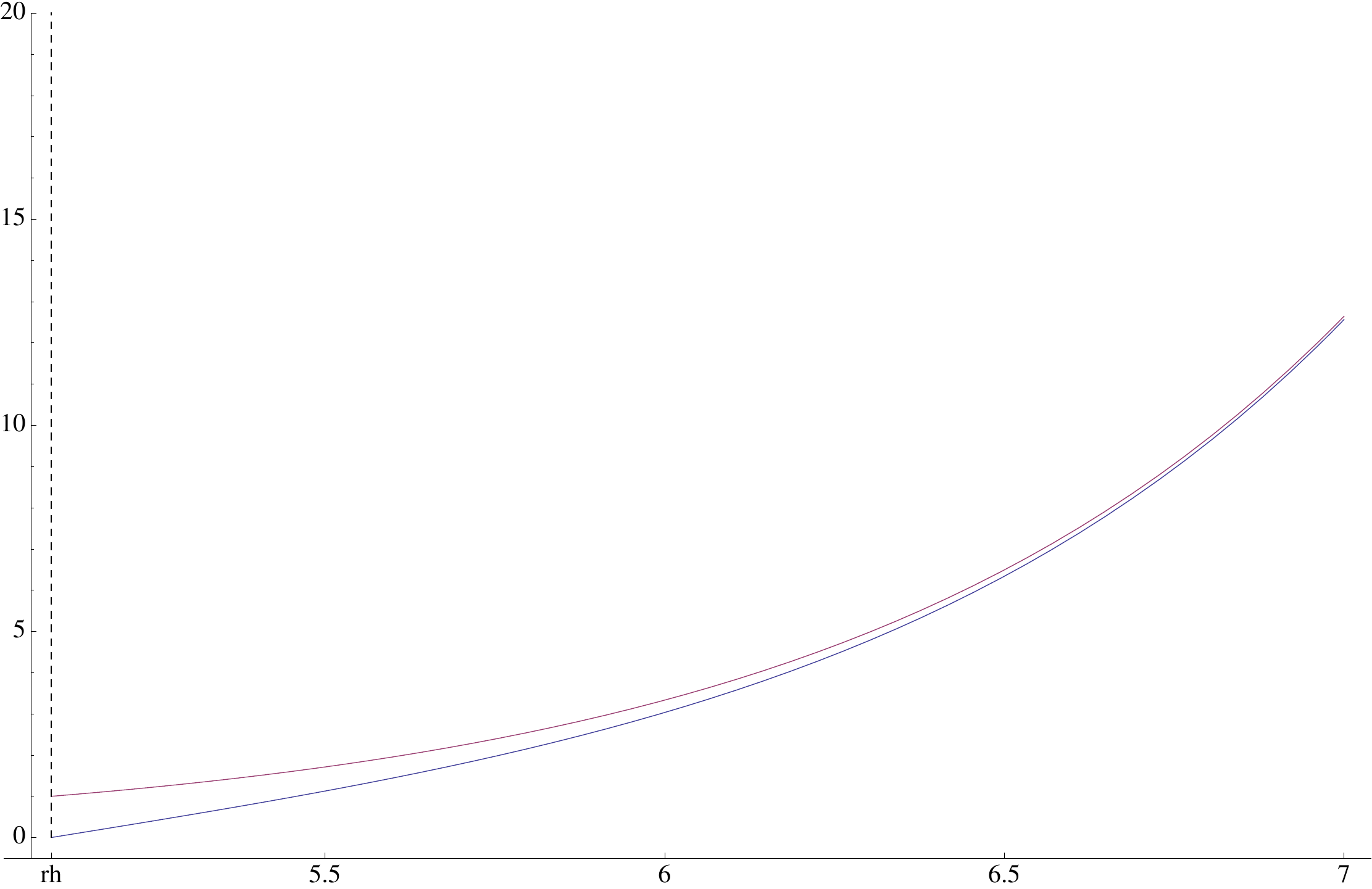}
 % graph-c125-new.pdf: 240x161 pixel, 72dpi, 8.47x5.68 cm, bb=0 0 240 161
 \caption{\small{$g_{xx}$ (red) and $g_{tt}$ (blue) metric elements for the unflavored solution after rotation: $c_+=3,\  C_2= 800000, s=0$.}}
\label{fig:grr2after}
\end{center}
\end{minipage}
\end{figure}

\begin{figure}
\begin{minipage}[t]{.495\textwidth}
%\captionsetup{width=0.8\textwidth}
 \begin{center}
 \includegraphics[width=7.3cm]{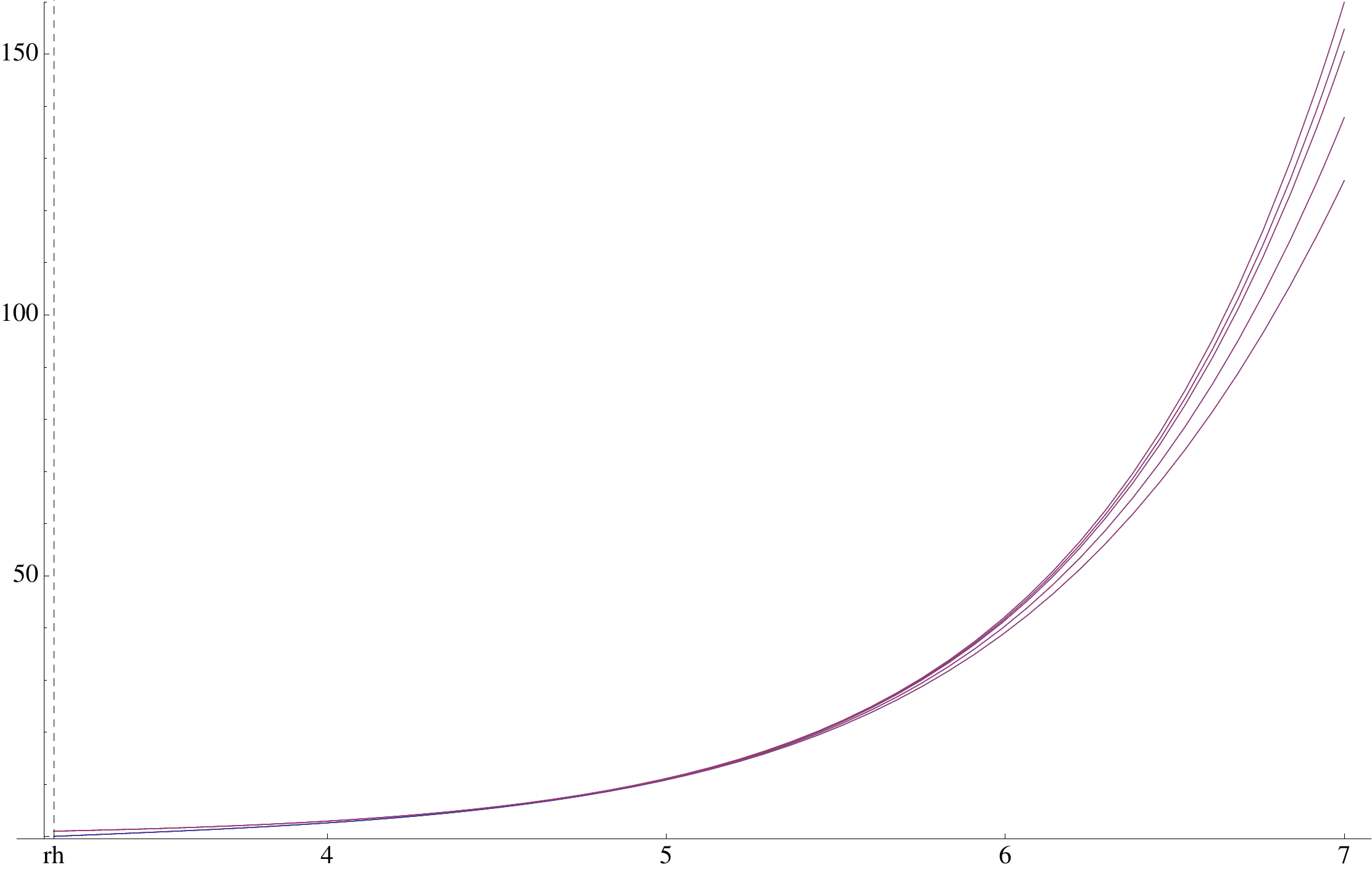}
 % graph-c1-97-new.pdf: 240x167 pixel, 72dpi, 8.47x5.89 cm, bb=0 0 240 167
 \caption{\small{$g_{xx}$ (red) and $g_{tt}$ (blue) metric elements for flavored solutions after rotation: $c_+=50,\  C_2= 5000$. From top to bottom: $s=0,1,19/10,51/10,91/10$.}
}
\label{figxxttflavc2}
\end{center}
\end{minipage}
\begin{minipage}[t]{.495\textwidth}
%\captionsetup{width=0.8\textwidth}
\begin{center}
 \includegraphics[width=7.3cm]{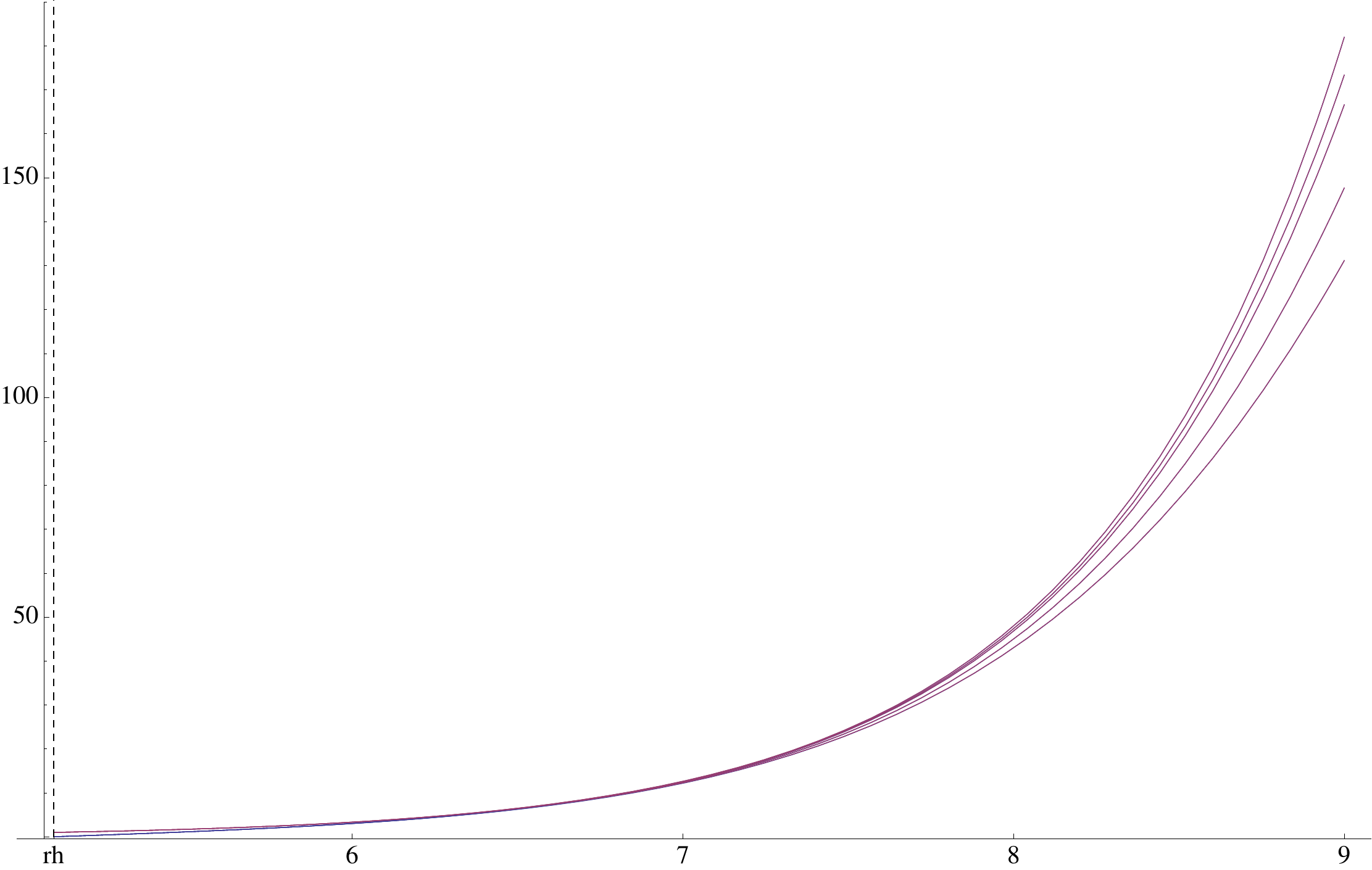}
 % graph-c125-new.pdf: 240x161 pixel, 72dpi, 8.47x5.68 cm, bb=0 0 240 161
 \caption{\small{$g_{xx}$ (red) and $g_{tt}$ (blue) metric elements for flavored solutions after rotation: $c_+=3,\  C_2= 800000$. From top to bottom: $s=0,1,19/10,51/10,91/10$.}}
\label{fig:grr2after}
\end{center}
\end{minipage}
\end{figure}

\begin{figure}
\begin{minipage}[t]{.8\textwidth}
%\captionsetup{width=0.8\textwidth}
 \begin{center}
 \includegraphics[width=12cm]{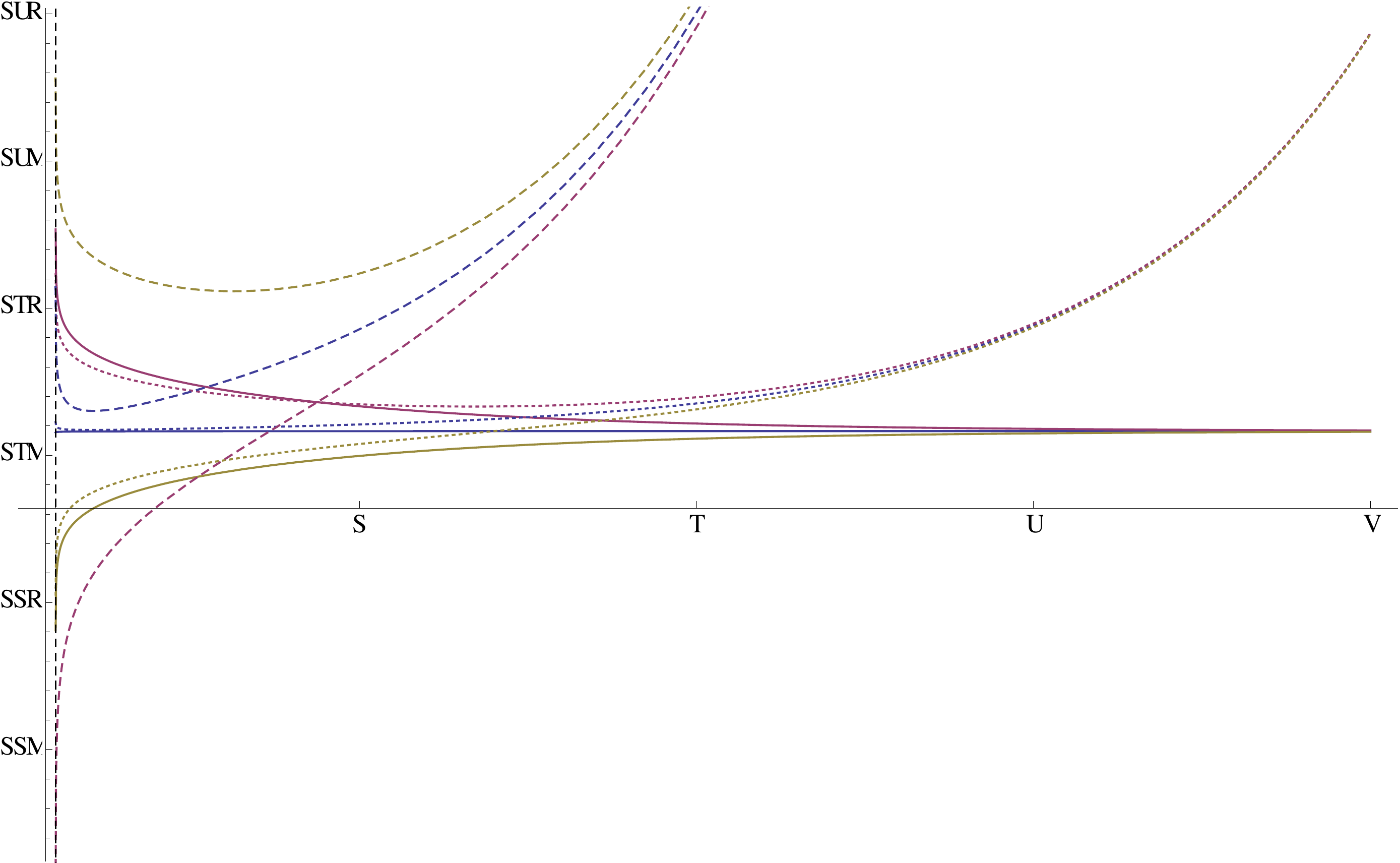}
 % graph-c1-97-new.pdf: 240x167 pixel, 72dpi, 8.47x5.89 cm, bb=0 0 240 167
\caption{\small{$\frac{3}{2} g_{\psi\psi}$ (blue), $g_{\theta\theta}$ (red), and  $g_{\tilde\theta\tilde\theta}$ (yellow) metric elements for flavored solution after rotation. The different groups correspond toi $s=0$ (solid), $s=2/5$ (dotted) and  $s=61/10$ (dashed).  $c_+=3$,\  $C_2= 800000$.}}
% \caption{\small{$g_{\rho\rho}$ metric element for solutions after rotation, $c_+=50,\  C_2= 5000$. From bottom to top: $s=0,1,19/10,51/10,91/10$.}}
\label{newfig1}
\end{center}
\end{minipage}
\end{figure}

\begin{figure}
\begin{minipage}[t]{.8\textwidth}
%\captionsetup{width=0.8\textwidth}
 \begin{center}
 \includegraphics[width=12cm]{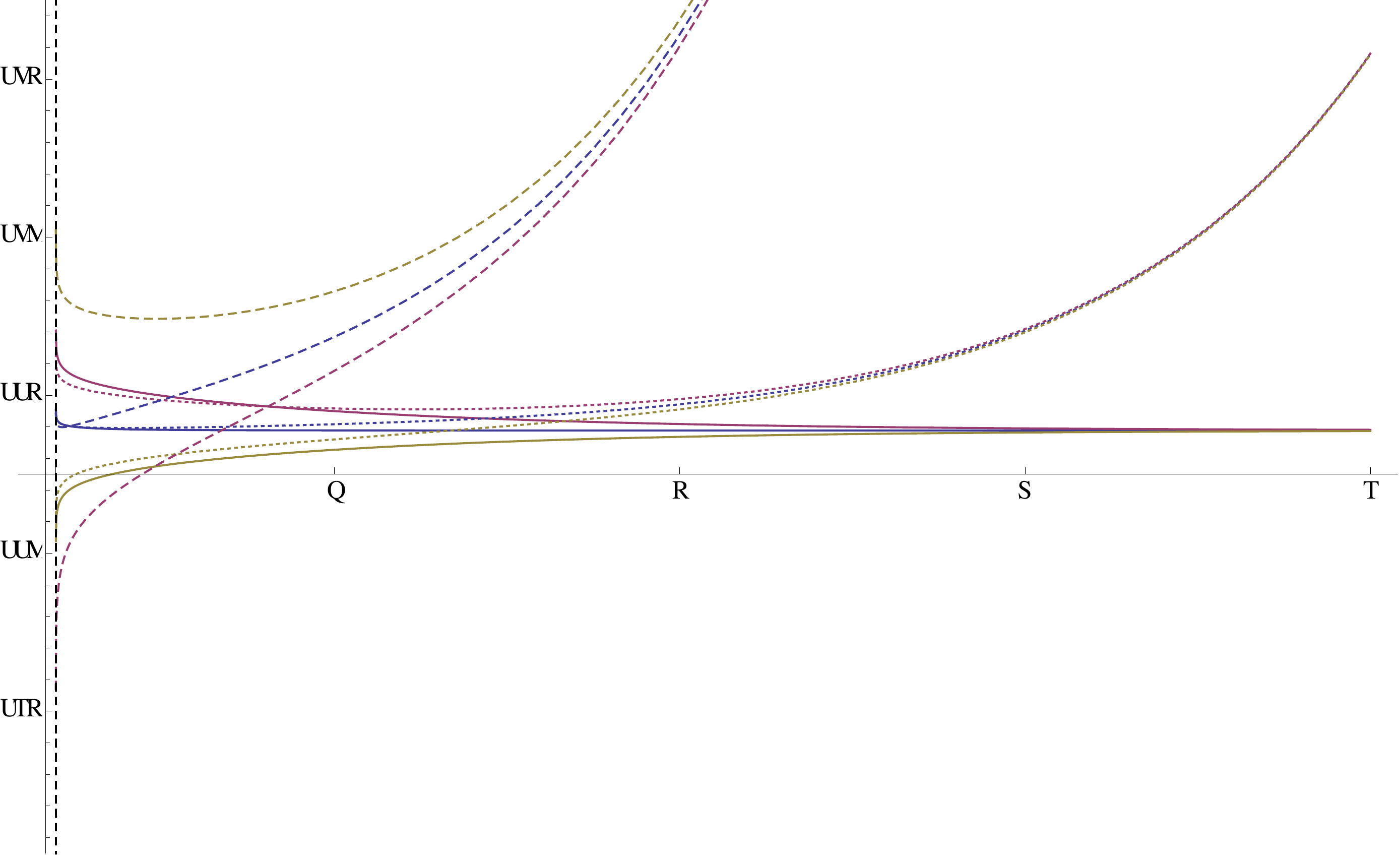}
 % graph-c1-97-new.pdf: 240x167 pixel, 72dpi, 8.47x5.89 cm, bb=0 0 240 167
 \caption{\small{$\frac{3}{2} g_{\psi\psi}$ (blue), $g_{\theta\theta}$ (red), and  $g_{\tilde\theta\tilde\theta}$ (yellow) metric elements for flavored solution after rotation. The different groups correspond to $s=0$ (solid), $s=2/5$ (dotted) and  $s=61/10$ (dashed). $ c_+=50$,\  $C_2= 50000$.}}
\label{newfig2}
\end{center}
\end{minipage}
\end{figure}

After rotating the numerical solutions obtained in Section \ref{section-new_unrotated_solutions} we obtain Figs.(\ref{figrhorhoflavc50} -- \ref{newfig2})\footnote{The UV behavior of the rotated solutions is extremely sensitive to the precise UV behavior of the seed solutions, due to the $\mathcal{H}$ factor. For this reason, we have generated these plots from the UV-shot seed solutions.}.
Note that,
\begin{itemize}
	\item The addition of flavor alters the behavior of the dilaton and the $g_{xx}, g_{tt}, g_{\rho\rho}$ parts of the metric in the UV, but not in the IR. On the other hand, flavor affects the metric of the compact manifold at all energy scales, as shown in Figs.(\ref{newfig1}, \ref{newfig2}).
	\item  Comparing $g_{\theta \theta}$ and $g_{\tilde\theta \tilde\theta}$ for $s=61/10$ and $s=2/5$ in  Figs.(\ref{newfig1}, \ref{newfig2}) we can see that the behavior of the curves has been swaped. For $s=61/10$, $g_{\tilde\theta\tilde\theta} $ grows as we approach the horizon while $g_{\theta\theta}$ decreases. For  $s=2/5$  it is the other way around. This swapping occurs as we cross $s=2$.	\item From \eqref{after_rotation} we can see that after the rotation the compact part of the metric is still an expanding $T^{1,1}$. This is also shown in(\ref{newfig1}, \ref{newfig2}) where we can see that for large $\rho$ we have $g_{\theta\theta}\sim g_{\tilde\theta\tilde\theta}\sim \frac{3}{2} g_{\psi\psi}$.
\end{itemize}


%%%%%%%%%%%%%%%%%%%%%%%%%%%%%%%%%%%%%%%5
\section{Energy and specific heat}
\label{section-energy}
In order to assess the thermodynamic stability  of our solutions, we need to determine the specific heat, $C_v$. This is obtained from the expression for the ADM energy of our solutions and their temperature via the standard thermodynamic relation $C_v = \partial E / \partial T$. $T$ is the numerically determined temperature given in equation \eqref{horizontemp}, and $E$ is given by the expression for the conserved ADM energy \cite{HAWKINGHOROWITZ} (where $16 \pi G_{10}= 2 \kappa_{10}^2 = (2 \pi)^7$)

\be
E_{ADM} = -\frac{1}{8 \pi G_{10}} \int_{S_t^{\infty}}\sqrt{|g_{00}|}(^8K - ^8K_0) d S_t^\infty.
\ee 

\noindent
Here, $^8K$ and $^8K_0$ are the extrinsic curvatures of  two 8-dimensional submanifolds $S_t$ of $\Sigma_t$ at constant $\rho$. $\Sigma_t$ is a constant-time 9d spatial slice of the entire geometry, and we take for $S_t$ the boundary manifold  $S_t^\infty$ as $\rho \to \infty$. $^8K$  corresponds to our finite temperature backgrounds, whereas  $^8K_0$ is the extrinsic curvature of a reference background \cite{HAWKINGHOROWITZ, Cotrone:2007qa}. For the reference background, we take one of the family of zero-temperature BPS solutions discussed in \cite{WARPED,HoyosBadajoz:2008fw}. The UV asymptotics of these backgrounds are summarized in Appendix \ref{appendix-UV asymptotics SUSY}.

Our Euclideanized 10d metric reads
\be
ds^2 = g_{00}d\tau^2 + g_{xx}dx_i dx^i + g_{\rho\rho}d\rho^2 + g_{\theta\theta}(e_1^2 + e_2^2) + g_{\tilde\theta\tilde\theta}(\tilde\om_1^2 +\tilde\om_2^2)  + g_{\psi\psi}\,\tilde\omega_3^2.
\ee
\noindent
The metric on a constant time slice $\Sigma_t$, and on its 8-dimensional boundary $S_t^\infty$ are
\be
ds^2_{\Sigma_t} = g_{xx}dx_i dx^i + g_{\rho\rho}d\rho^2 + g_{\theta\theta}(e_1^2 + e_2^2) + g_{\tilde\theta\tilde\theta}(\tilde\om_1^2 +\tilde\om_2^2)  + g_{\psi\psi}\,\tilde\omega_3^2
\ee
\be
ds^2_{S_t^\infty} = g_{xx}dx_i dx^i + g_{\theta\theta}(e_1^2 + e_2^2) + g_{\tilde\theta\tilde\theta}(\tilde\om_1^2 +\tilde\om_2^2)  + g_{\psi\psi}\,\tilde\omega_3^2,
\ee
\noindent
where the metric functions in Einstein frame are
\begin{alignat}{2}
g_{00}&= e^{\phi/2}e^{-8x}\quad\quad\quad\quad&g_{00_{B}}&= e^{\phi_{B}/2}\nu_{B}\NO\\
g_{xx}&= e^{\phi/2}&g_{xx_{B}}&= e^{\phi_{B}/2}\NO\\
g_{\rho\rho}&= e^{\phi/2}e^{8x}e^{2k}&g_{\rho\rho_{B}}&= e^{\phi_{B}/2}e^{2k_{B}}\NO\\
g_{\theta\theta}&= e^{\phi/2}e^{2h}&g_{\theta\theta_{B}}&= e^{\phi_{B}/2}e^{2h_{B}}\NO\\
g_{\tilde\theta\tilde\theta}&= \frac{1}{4}e^{\phi/2}e^{2g}&g_{\tilde\theta\tilde\theta_{B}}&= \frac{1}{4}e^{\phi_{B}/2}e^{2g_{B}}\NO\\
g_{\psi\psi}&=  \frac{1}{4}e^{\phi/2}e^{2k}&g_{\psi\psi_{B}}&=  \frac{1}{4}e^{\phi_{B}/2}e^{2k_{B}}\NO\\
\end{alignat}
\noindent
before rotation, and
\begin{alignat}{2}
g_{00}&= e^{-\phi/2}\mathcal{H}^{-1/2} e^{-8x}\quad\quad\quad\quad&g_{00_{B}}&= e^{-\phi_{B}/2}\mathcal{H}_{B}^{-1/2}\nu_{B}\NO\\
g_{xx}&= e^{-\phi/2}\mathcal{H}^{-1/2}&g_{xx_{B}}&= e^{-\phi_{B}/2}\mathcal{H}_{B}^{-1/2}\NO\\
g_{\rho\rho}&= e^{3\phi/2}\mathcal{H}^{1/2} e^{8x}e^{2k}&g_{\rho\rho_{B}}&= e^{3\phi_{B}/2}\mathcal{H}_{B}^{1/2}\,e^{2k_{B}}\NO\\
g_{\theta\theta}&= e^{3\phi/2}\mathcal{H}^{1/2}e^{2h}&g_{\theta\theta_{B}}&= e^{3\phi_{B}/2}\mathcal{H}_{B}^{1/2}e^{2h_{B}}\NO\\
g_{\tilde\theta\tilde\theta}&= \frac{1}{4}e^{3\phi/2}\mathcal{H}^{1/2}e^{2g}&g_{\tilde\theta\tilde\theta_{B}}&= \frac{1}{4}e^{3\phi_{B}/2}\mathcal{H}_{B}^{1/2}e^{2g_{B}}\NO\\
g_{\psi\psi}&=  \frac{1}{4}e^{3\phi/2}\mathcal{H}^{1/2}e^{2k}&g_{\psi\psi_{B}}&=  \frac{1}{4}e^{3\phi_{B}/2}\mathcal{H}_{B}^{1/2}e^{2k_{B}}\NO\\
\end{alignat}

\noindent
after rotation; `$B$' subscripts indicate the BPS backgrounds. Here $\mathcal{H} = (e^{-2\phi}- e^{-8x})$ and $\mathcal{H}_{B} = (e^{-2\phi_{B}} -\kappa_2^2)$, where $\kappa_2$ is a free parameter with $0 < \kappa_2 < e^{-\phi_\infty}$ that parametrizes the rotation, see \cite{WARPED}. In terms of the 11d boost parameter $\beta$, we have $\kappa_2 = e^{-\phi_\infty}\tanh{\beta}$. Note that for the BPS solutions $e^{-8x}=1$, and we have also allowed for an arbitrary rescaling of the $\tau$ coordinate via a constant parameter $\nu_{B}$, reflecting the freedom in choosing the period $\beta$ of the Euclidean time coordinate for these backgrounds.

It is now straightforward to calculate the extrinsic curvature of an 8-manifold $S_t$ at  constant $\rho$. We obtain
\be
\label{extcurvbefore}
^8K_{bef} = e^{-\frac{\phi(\rho)}{4}-k(\rho)-4 x(\rho)} \left(2 \phi'(\rho)+2 g'(\rho)+2 h'(\rho)+k'(\rho)\right)
\ee
for the background before rotation, and
\begin{align}
\label{extcurvafter}
^8K_{aft} &=\frac{1}{2}\frac{\mathcal{H}(\rho)}{\mathcal{H}(\rho)^{5/4}} e^{-\frac{3 \phi(\rho)}{4}-k(\rho)-4 x(\rho)}\NO\\
&+ \frac{1}{\mathcal{H}(\rho)^{1/4}} e^{-\frac{3 \phi(\rho)}{4}-k(\rho)-4 x(\rho)} \bigg(3 \phi'(\rho)+2 g'(\rho)+2 h'(\rho)+k'(\rho)\bigg)
\end{align}

\noindent
for the rotated background.\footnote{The BPS backgrounds will have $e^{-4x}=1$.} %The expressions for the BPS backgrounds are obtained from these by replacing $\mathcal{H},\phi,k,g, \text{and }h$ with their BPS versions, and setting $x=0$. 

The expressions for $dS_t$ are
\begin{align}
dS_{t_{bef}}= \frac{1}{8}e^{2\phi}e^{2g+2h+k} d V_3 \wedge dS_{\MM_6}\quad\quad\quad dS_{t_{aft}}=\frac{1}{8}\mathcal{H}^{1/2}e^{3\phi}e^{2g+2h+k} d V_3 \wedge dS_{\MM_6}
\end{align}

\noindent
%where  $\mathcal{H},\phi,k,g, \text{and }h$ are either the finite temperature or BPS versions.
where $d V_3 = dx_1 \wedge dx_2 \wedge dx_3$ is the volume form on the Minkowski spatial directions, and $dS_{\MM_6} = (\sin\theta \sin \tilde\theta) d\theta \wedge d\varphi \wedge d\tilde\theta \wedge d\tilde\varphi\wedge d\psi$ is the volume form on the compact cycles. Putting all of these together and integrating over the compact $\theta, \varphi, \tilde\theta, \tilde \varphi, \psi$ directions we obtain expressions for the ADM energy before and after the rotation:
%\begin{align}
%e_{bef}=\frac{E_{bef}}{V_3} = -\pi^2 \int_{S_t^{\infty}}\bigg[&e^{-8x}e^{2\phi+2g+2h}\bigg(2(\phi' + g'+h')+k'\bigg)\NO\\
%&-\sqrt{\nu_{BPS}}e^{2\phi_{BPS}+2g_{BPS}+2h_{BPS}}\bigg(2 (\phi_{BPS}'+g_{BPS}'+h_{BPS}')+k_{BPS}'\bigg)\bigg]
%\end{align}
%\begin{align}
%e_{aft} = -\pi^2  \int_{S_t^{\infty}}\bigg[&e^{-8x}e^{2\phi+2g+2h}\bigg(\frac{H'}{2H}+\big(3 \phi' + 2(g'+h') + k' \big)\bigg)\NO\\
%& -\sqrt{\nu_{BPS}}e^{2\phi_{BPS}+2g_{BPS}+2h_{BPS}}\bigg(\frac{H'_{BPS}}{2 H_{BPS}}+\big(3 \phi_{BPS}' + 2(g_{BPS}'+h_{BPS}') + k_{BPS}' \big)\bigg)\bigg]
%\end{align}
\begin{align}
e_{bef}=\frac{E_{bef}}{V_3} = -\frac{1}{8 \pi^4} \bigg[&e^{-8x}e^{2\phi+2g+2h}\bigg(2(\phi' + g'+h')+k'\bigg)\NO\\
&-\sqrt{\nu_{B}}\,e^{2\phi_{B}+2g_{B}+2h_{B}}\bigg(2 (\phi_{B}'+g_{B}'+h_{B}')+k_{B}'\bigg)\bigg]\Bigg\vert_{\rho \to \infty}
\end{align}
\begin{align}
e_{aft} = -\frac{1}{8 \pi^4}  \bigg[&e^{-8x}e^{2\phi+2g+2h}\bigg(\frac{\mathcal{H}'}{2\mathcal{H}}+\big(3 \phi' + 2(g'+h') + k' \big)\bigg)\NO\\
& -\sqrt{\nu_{B}}\,e^{2\phi_{\small{B}}+2g_{B}+2h_{B}}\bigg(\frac{\mathcal{H}'_{B}}{2 \mathcal{H}_{B}}+\big(3 \phi_{B}' + 2(g_{B}'+h_{B}') + k_{B}' \big)\bigg)\bigg]\Bigg\vert_{\rho \to \infty}
\end{align}

\noindent
Here $V_3$ = Vol($R^3$), and we have expressed the results as energy densities per unit flat 3-volume.

We evaluate these expressions at a large but finite value of $\rho = \rho_*$ by inserting the UV asymptotics of the %pre-rotated
BPS and finite temperature solutions. The BPS asymptotics are given in terms of the free variables $Q_o, c_+, c_-, \rho_0$ and $f_{1,0}$, and are listed in Appendix \ref{appendix-UV asymptotics SUSY}. The finite temperature asymptotics were given in section \ref{FTUVexpansions}, and are a deformation of a BPS background with $Q_o= -N_c + N_f/2$ and $\rho_0=0$. 

In order to calculate the ADM energy, we first match the geometries and matter fields of the BPS and finite temperature backgrounds at $\rho_*$, this amounts to matching the metric functions $g_{00},g_{xx},g_{\theta\theta},g_{\tilde\theta\tilde\theta},g_{\psi\psi}$, as well as the dilaton. Only after this matching is performed we  let  $\rho_* \to \infty$.

We begin with the case before rotation.  To avoid confusion with the parameters $c_+$ and $c_-$ in the finite temperature solutions, we will write the BPS $c_+$ and $c_-$ parameters in bold as $\bpscp$ and $\bpscm$. First we pick some fixed values of $c_+, C_2$ and $s$ to specify the finite temperature background. We begin the matching by using the freedom in $\nu_B$ to set $\nu_B = e^{-8x(\rho_*)}$. Next we match $e^{\phi/2}=e^{\phi_B/2}$ to $\mathcal{O}(e^{-12 \rho_*/3})$ by setting
\begin{equation}
f_{1,0} = 1 - e^{-12 \rho_*/3}\frac{C_2 s}{c_+}.
\end{equation}
As a result, $g_{00}$ and $g_{xx}$ are now matched to $\mathcal{O}(e^{-12 \rho_*/3})$. Now, we match $e^{2k}=e^{2k_B}, e^{2h}=e^{2h_B},$ and $e^{2g}=e^{2g_B},$ respectively, to $\mathcal{O}(e^{-8 \rho_*/3})$, by setting
\begin{align}
\bpscp &= c_+ + e^{-12 \rho_*/3}\frac{11 C_2 s}{48}\nn\\
%\bpscp &= c_+ + e^{-12 \rho_*/3}\left(\frac{-\bpscm}{32 c_+^2} + \frac{C_2 s \rho_*}{12}\right)\nn\\
Q_0 &= -1+\frac{s}{2}-\frac{1}{16}e^{-8 \rho_*/3} C_2 (-5+4 \rho_*)(-2+s)\nn\\ 
%Q_0 &= -1+\frac{s}{2} + e^{-8 \rho_*/3}\left(\frac{-40 c_+^2 C_2+3\bpscm +42 c_+^2 C_2 s}{64 c_+^2} + -\frac{1}{8} C_2 (-4+3 s) \rho_*\right)\nn\\
\bpscm &= \frac{2}{3} c_+^2 (-11 C_2 s+4 C_2 s \rho_*),
\end{align}
\noindent
which results in the matching of $g_{\theta\theta},g_{\tilde\theta\tilde\theta},g_{\psi\psi}$ to $\mathcal{O}(e^{-8 \rho_*/3})$.

With these definitions, the dilaton and all of the metric functions $g_{00},g_{xx},g_{\theta\theta},g_{\tilde\theta\tilde\theta},g_{\psi\psi}$ are matched up to (at least) $\mathcal{O}(e^{-8 \rho_*/3})$. The freedom in the $\rho_0$ parameter in the BPS solutions is not needed, and we set it equal to zero as was done for the finite temperature solutions. Taking $\rho_* \to \infty$, the resulting ADM energy density is finite, independent of $s$, and equal to
\be\label{eq:ebefore}
%e_{bef} =  \frac{5}{12} c_+^2 C_2 \pi ^2
e_{bef} =  \frac{5 c_+^2 C_2}{96 \pi^4}
\ee

%The freedom in the BPS parameters $Q_o, \bpscp$ and $\bpscm$ can be used to match the $g_{\theta\theta},g_{\tilde\theta\tilde\theta},g_{\phi\phi}$ metric functions at the boundary, whereas $f_{1,0}$ can be used to match the dilaton. Since $g_{xx_{B}}=e^{-\phi_{B}/2}$, it will also matched by this choice of $f_{10}$. It remains to match the $g_{00}$ component of the metric, which we accomplish by choosing $\nu_{B}$ to equal the finite temperature expression for $e^{-8x}$.  The freedom in the $\rho_0$ parameter in the BPS solutions is not needed, and we set it equal to 0 as was done for the finite temperature solutions. After these matchings are performed, the resulting ADM energy density is finite, independent of $s$, and equal to

After the rotation, the presence of $\mathcal{H}$ causes the mixing of the UV asymptotics in the metric functions to become slightly more involved. The rotation also causes the metric function expansions to become unwieldy at higher order, so we only perform the matching to $\mathcal{O}(e^{-2 \rho_*})$ which  is enough to render the ADM energy finite. 
%As before, we begin by setting $\nu_B = e^{-8x(\rho_*)}$, but we now match $\mathcal{H}=\mathcal{H_B}$ by setting $\kappa_2 = e^{-4x(\rho_*)}$. The remainder of the process proceeds as above by expanding $f_{1,0}, \bpscp,$ and $Q_0$\footnote{Matching the metrics to this order does not require us to fix $\bpscm$.} in powers of $e^{-2 \rho_*/3}$, and
Despite these technical complications the process proceeds  as above. Namely, before taking the $\rho_* \rightarrow \infty$ limit we fix coefficients so that the BPS and finite temperature metrics match at the boundary, we then take the $\rho_* \rightarrow \infty$ limit and obtain a finite energy given by
%\begin{align}
%f_{1,0} &= 1 - e^{-12 \rho_*/3}\frac{C_2 s}{c_+}\nn\\
%\bpscp &= c_+\nn\\
%Q_0 &= -1 + \frac{s}{2}
%\end{align}
%Again, $\rho_0$ is not needed and we set it to zero. The ADM energy remains independent of $s$, and we find %the matching the $g_{xx}$ metric function and matching the dilaton require two independent conditions. We achieve this by first using $f_{10}$ to match $ \phi_{B}$ to its finite temperature value, and then setting $\kappa_2$ equal to $e^{-8x}$ of the finite temperature solution so that $\mathcal{H} = \mathcal{H}_{B}$, whence $g_{xx}$ is also matched. We next match $g_{\theta\theta},g_{\tilde\theta\tilde\theta}$ and $g_{\phi\phi}$ by adjusting $Q_o, \bpscp$ and $\bpscm$ as in the non-rotated case, and finally $g_{00}$ is matched by again setting  $\nu_{B}= e^{-8x}$. The resulting ADM energy for the rotated solutions is
\be\label{eq:eafter}
%e_{aft}=\frac{1}{2} c_+^2 C_2 \pi ^2
e_{aft}=\frac{5 c_+^2 C_2 }{96 \pi^4}.
\ee
Thus, the  energy remains the same as before the rotation. 

The above analysis was done for flavored finite-temperature backgrounds with $s \neq 0$. However, since the results do not depend on $s$ they should also hold for backgrounds without flavor {\it i.e} for  non-extremal deformations that, after the rotation, do have  Klebanov-Strassler asymptotics. We have explicitly checked that this is indeed the case. 
%a similar calculation yields the same value of the ADM energy before and after rotation:

%\be
%e_{bef,\,\, s=0} = e_{aft,\,\, s=0} = \frac{5 c_+^2 C_2}{96 \pi^4}
%\ee

\begin{figure}[t]
\centering{\includegraphics[height=7.6cm]{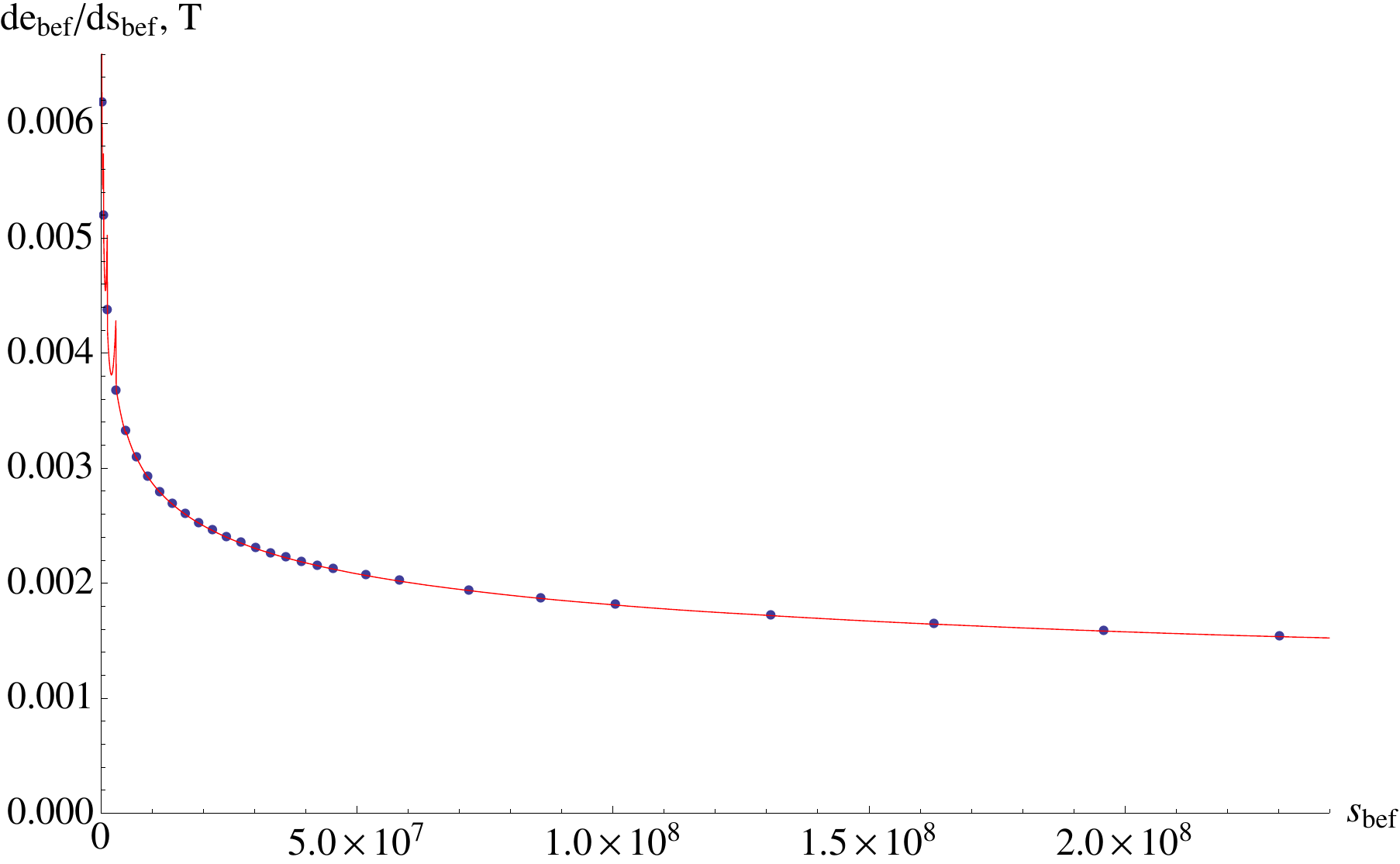}}
\caption{$de_{bef}/d\mathfrak{s}_{bef}$ (\color{red} red line\color{black}) and $T_h$ (\color{blue} blue dots\color{black}) plotted vs. $\mathfrak{s}_{bef}$, for the flavored case with $c_+=50$, $s=1$.}
\label{dedsandt_before}
\end{figure}

 In Fig. \eqref{dedsandt_before} we plot  $d e/d \mathfrak{s}$ \footnote{ $\mathfrak{s}$ denotes the entropy density $S/V_3$ and should not be confused with the flavor parameter $s$.}
and $T$ as functions of the entropy density (or equivalently the ``radius" of the horizon) for solutions before the rotation. The overlap of the two curves show that the numerical solutions satisfy the first law of thermodynamics \be
 de = T d\mathfrak{s}.
 \ee
To verify that this is also the case  after the rotation recall that we have already shown that  the temperature and energy density remain the same
 \eqref{eq:ebefore},\eqref{eq:eafter},\eqref{horizontemp}. It is easy to see that the entropy density is also unaffected by the rotation,

\be
\mathfrak{s}_{bef}=\frac{S_{\text{bef}}}{V_3} =  \left.\frac{e^{2 \phi}e^{2h+2g+k}}{4 \pi^3}\right|_{\rho=\rho_h}
\ee
\be
\label{Saft}
\mathfrak{s}_{aft}=\frac{S_{\text{aft}}}{V_3} = \left.\frac{e^{3 \phi} \mathcal{H}^{1/2}e^{2h+2g+k} }{4 \pi^3}\right|_{\rho=\rho_h} = \left.\frac{e^{3 \phi} (e^{-2\phi} - e^{-8x})^{1/2} e^{2h+2g+k}}{4 \pi^3}\right|_{\rho=\rho_h} = \mathfrak{s}_{\text{bef}} 
\ee
\\
where the last equality in equation (\ref{Saft}) follows from the fact that $e^{-8x}$ vanishes at the horizon.

Thus, the thermodynamical quantities $e, \mathfrak{s}$ and $T$ are not modified by the rotation and the first law will still hold. 

\begin{figure}[h]
\begin{minipage}{.495\textwidth}
%\captionsetup{width=0.8\textwidth}
 \begin{center}
 \includegraphics[width=7.9cm]{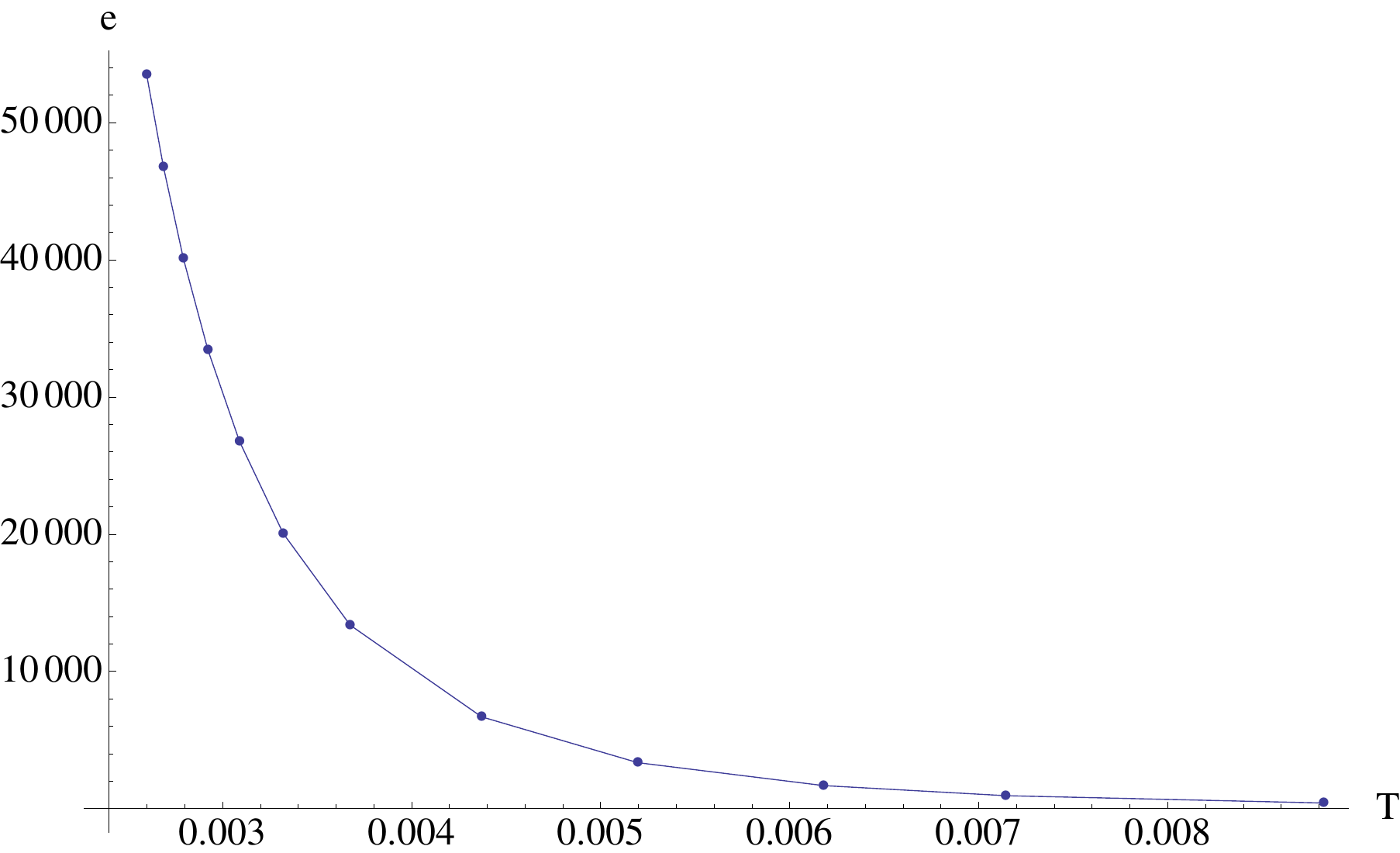}
 \caption{\small{ADM energy density versus horizon temperature ($s=1,\, c_+ = 50$)}}
\label{CV_after_c_50}
\end{center}
\end{minipage}
\begin{minipage}{.495\textwidth}
%\captionsetup{width=0.8\textwidth}
\begin{center}
 \includegraphics[width=7.9cm]{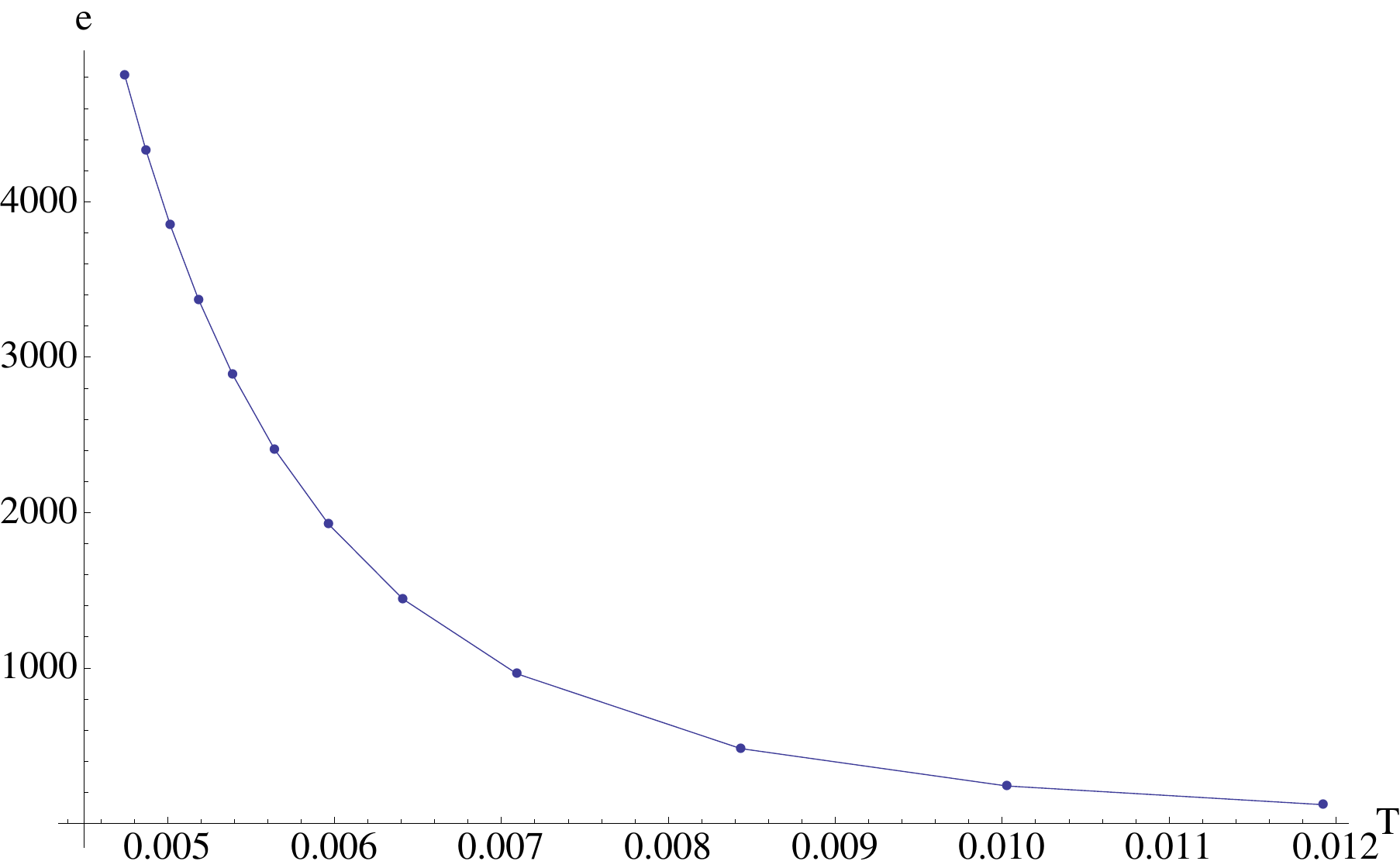}
 \caption{\small{ADM energy density versus horizon temperature ($s=1,\, c_+ = 3$)}}
\label{CV_after_c_3}
\end{center}
\end{minipage}
\end{figure}

As discussed in section \ref{section-solutiontemperature}, our numerics associate a horizon temperature with a choice of $c_+$ and $C_2$. In Figs.\eqref{CV_after_c_50} and \eqref{CV_after_c_3} we plot the energy density versus the associated temperature for a solution with $s=1$. The slope of these plots is the specific heat $C_v$ and is seen to be negative both before and after the rotation.
Note that the fact  that the first law is satisfied implies that we would have gotten the same answer had we chosen  to use  holographic renormalization methods. 

\begin{figure}[h]
\begin{minipage}{.495\textwidth}
%\captionsetup{width=0.8\textwidth}
 \begin{center}
 \includegraphics[width=7.9cm]{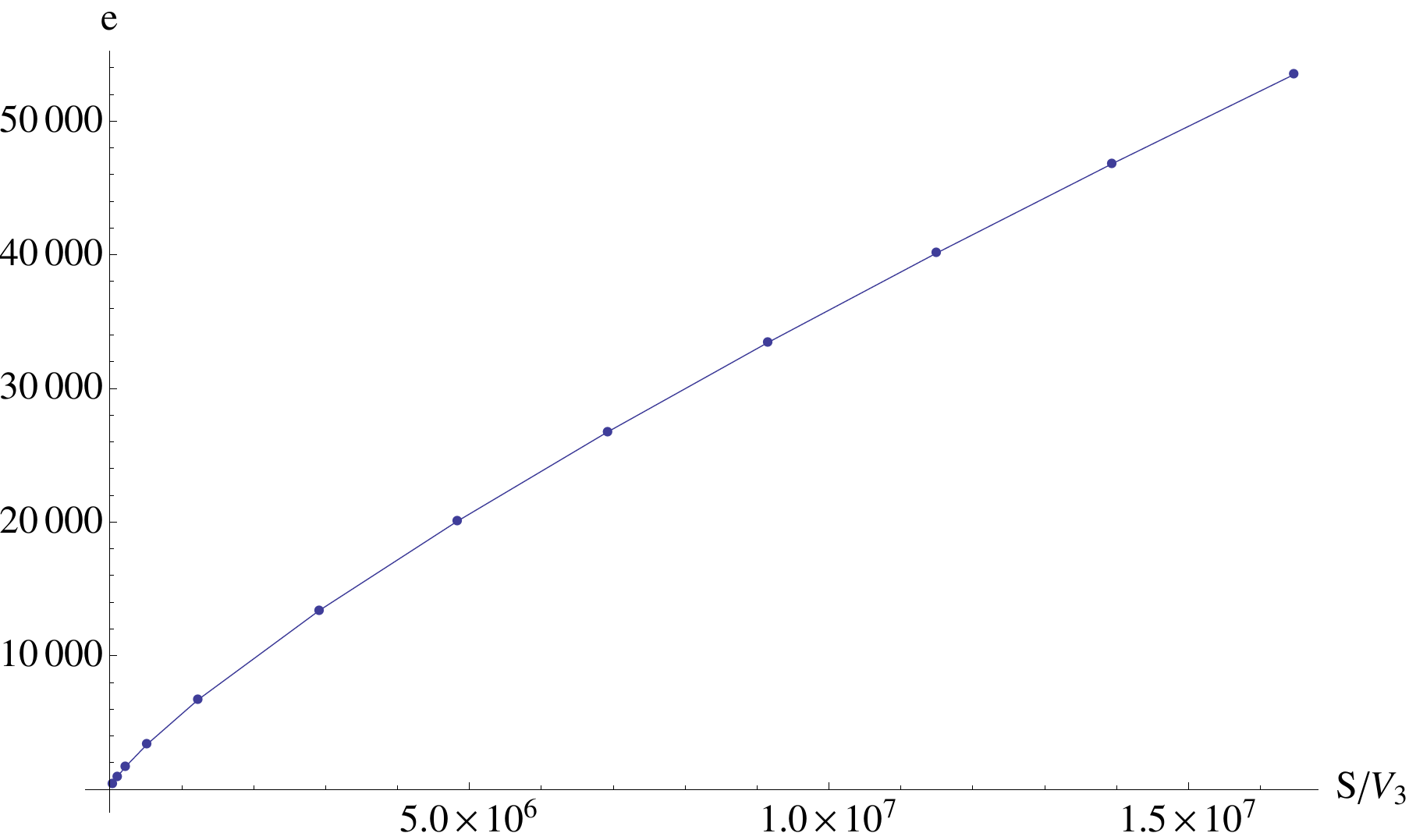}
 \caption{\small{ADM energy density versus entropy density ($s=1,\, c_+ = 50$)}}
\label{EvsS_after_c_50}
\end{center}
\end{minipage}
\begin{minipage}{.495\textwidth}
%\captionsetup{width=0.8\textwidth}
\begin{center}
 \includegraphics[width=7.9cm]{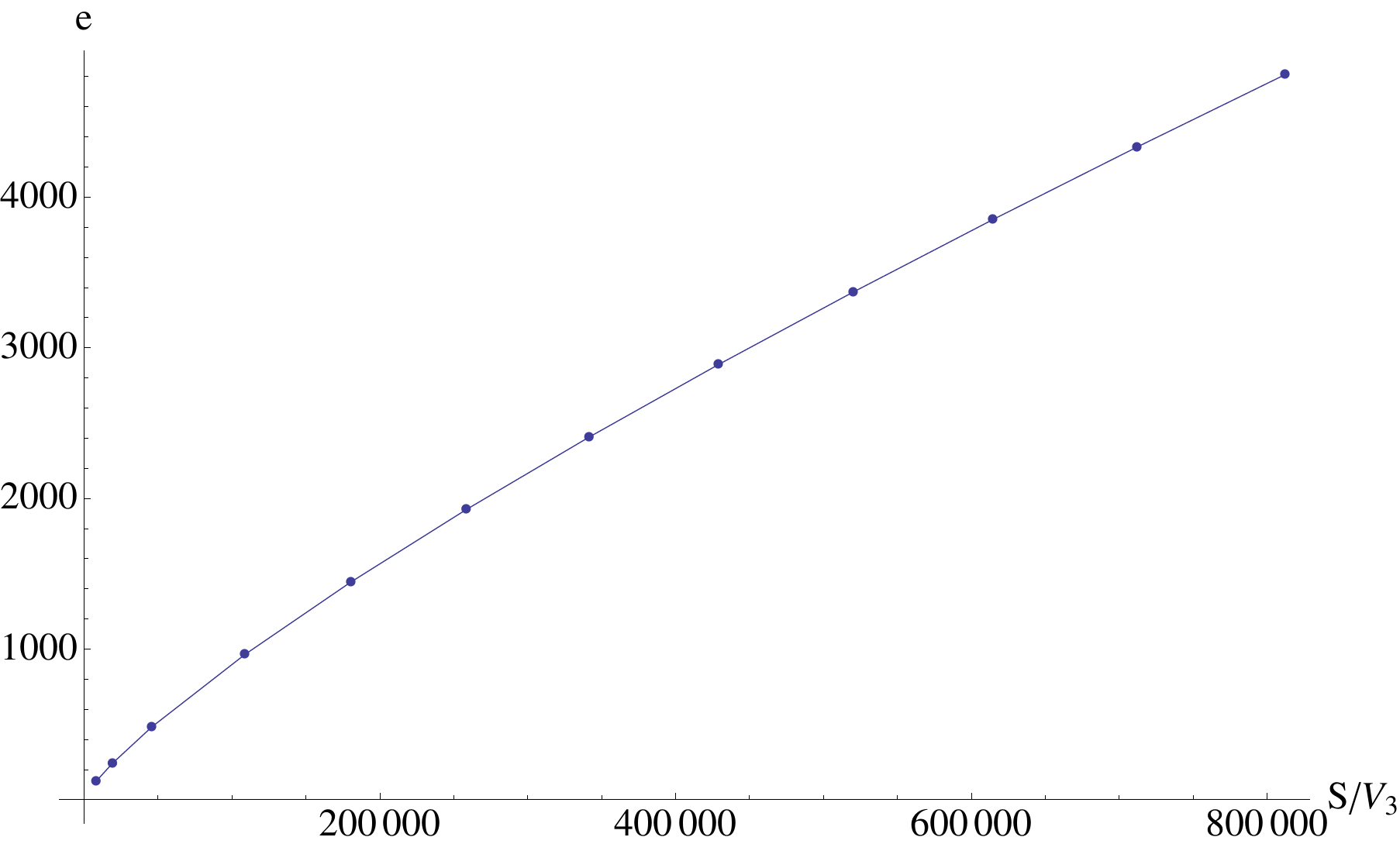}
 \caption{\small{ADM energy density versus entropy density ($s=1,\, c_+ = 3$)}}
\label{EvsS_after_c_3}
\end{center}
\end{minipage}
\end{figure}

The backreaction of the flavor branes does not play any role in the stability analysis  as the temperature's dependence on $s$ is essentially negligible compared to its dependence on $C_2$ and neither the energy nor the entropy depend on $s$.
In Figs.\eqref{EvsS_after_c_50} and \eqref{EvsS_after_c_3} we show the energy density versus the entropy density $S/V_3$. Note that in Figs.\eqref{EvsS_after_c_50} and \eqref{EvsS_after_c_3} the behavior of the energy vs. entropy is linear for large values of $S$ (or equivalently for large values of the horizon ``radius"). From the first law of thermodynamics, $dE= T dS$, we expect this to happen when the temperature is constant.

Thus, we have shown that, despite of having very different asymptotics, the backgrounds after the rotation have thermodynamical properties similar to the $D5$ wrapped branes system.

%\be
%h_{1,1}=\frac{1}{2} - \frac{s}{4}, \quad\quad x_{5,0}=0,\quad\quad h_{1,0}=\frac{-3(-4+10 s + 7 s^2 -4 f_{3,0}k_{0,0}^2)}{32(-2+s)}
%\ee
%for the flavored case.

\section{Conclusions}
Let us summarize what we have done in  this paper and point out possible future directions. 
\begin{itemize}
	\item We found  finite temperature solutions describing $N_c$ D5 branes wrapped on the $S^2$ of the resolved conifold. Unlike the generic wrapped D5 branes backgrounds the ones presented here have a dilaton that does not blow up at infinity but stabilizes to a finite value. 
	\item We identify a set of 11 UV and 6 IR parameters that  determine the  asymptotics of the solutions to {\it any} order.
	\item It was  not our goal to explore all the possible families of solutions in the 17 parameter space  but to restrict ourselves to those that have a UV  behavior similar to the ones in \cite{WARPED}. Therefore,  we fixed several of the UV parameters to appropriate values. We then solved the EOM's numerically using these UV expansions as boundary conditions and demanding regularity at the horizon. We also imposed that the constraint coming from reparametrization invariance is satisfied, to order $10^{-6}$ or better,  throughout all the interval. 

	\item Using U-dualities and these backgrounds as ``seeds" we generate solutions with $D3$ and $D5$ charge. To avoid issues related to the uplift of the smeared flavor branes, we  show explicitly that after the U-dualities are applied  the background obtained is  a solution of the EOM's of $D3$ and $D5$ brane sources. 
	\item In the absence of  temperature the backgrounds obtained are dual to  interesting field theories that exhibit Seiberg dualities, Higgsing and confinement \cite{WARPED}. It is tempting to think of the non-extremal solutions found here as dual to   finite temperature versions of the field theories  in  \cite{WARPED}. However,  for non-zero temperature we have to first study the thermodynamical stability. We proceed to do that and  find that the specific heat is negative and thus, they are unstable.

	\item The solutions in \cite{HEATING}  are a particular case (no flavor, $a(\r)$=0) of ours, thus the thermodynamical instability found here applies also to \cite{HEATING}. 
	\item We show that our numerical solutions satisfy the First Law of Thermodynamics.
	\item We find that the temperature grows as the non-extremality parameter decreases. Also, as we increase the area of the horizon the temperature goes  to a small non-zero value. A similar behavior was found non-extremal MN  \cite{GTV} and indicates the possibility of a phase transition. It is  interesting to pursue this issue further and explore if the dualities can be used to map a phase transition in a wrapped fivebrane background to one in non-extremal KS  \cite{Buchel:2010wp}. Note that the temperature dependence of $F_{(5)}$ and the warp factor  necessary to describe the chirally symmetric phase, $T>>T_c$, \cite{Gubser:2001ri} emerges naturally in this framework due to the dualities. 

	\item It would be interesting to study wether there is a connection between our results and the ones existing in the literature regarding non-extremal fractional branes at an orbifold point \cite{Bertolini:2002de}.

	\item In \cite{Buchel:2005nt} it was found that the lowest quasinormal mode in non-extremal  Maldacena-Nu\~nez is tachyonic. In this paper we have studied more general backgrounds than the one in\cite{Buchel:2005nt}. The negative specific heat seems to indicate that a  tachyonic mode  is also present in these non-extremal  flavored backgrounds, it would be interesting to confirm this. 

\end{itemize} 

In \cite{WARPED},\cite{Conde:2011aa},\cite{Conde:2011ab},\cite{Elander:2011mh}  it was shown that  solutions of type IIB supergravity with sources  generalize the Klebanov-Strassler baryonic branch;  the dual field theory  exhibits Seiberg dualities, Higgsing and confinement at different scales and was conjectured to  describe a mesonic branch. One of the motivations of the present work was to find gravity duals of these interesting field theories. We find that starting from non-extremal wrapped $D5$ branes and applying the same dualities used to generate the extremal backgrounds produces thermodynamically unstable backgrounds.  What is then the finite temperature gravity dual of the field theories found in \cite{WARPED},\cite{Conde:2011aa},\cite{Conde:2011ab},\cite{Elander:2011mh}? This fascinating question will undoubtedly lead to interesting  physics and remains open. 

\section{Acknowledgments}
We  thank Carlos Nu\~nez   for  many  useful discussions, correspondence  and comments. We also thank Alex Buchel and  Leopoldo Pando Zayas for comments on this  manuscript.  The work of E.C. and S.Y. is
partially supported by the National Science Foundation 
under Grant No. PHY-0969020. E.C. also acknowledges support 
of CONACyT grant CB-2008-01-104649 and CONACyT's 
High Energy Physics Network.

\appendix
\newpage

\section{Appendix: The equations of motion}
\label{appendix-EOMs}
In this appendix we briefly review the metric \ansatz and solution method of \cite{GTV},  which we use to derive the Einstein equations and constraint given in section \ref{subsection-non_extremal_flavored_backgrounds} for the flavored non-extremal backgrounds before rotation.\\

\noindent
The metric \ansatz in \cite{GTV}, in Einstein frame,  reads
\ba
 ds^2_{IIB}&=& -Y_1 dt^2 +  Y_2 (dx_1^2 + dx_2^2 + dx_3^2) +Y_3 d\rho^2 + Y_4 (d\theta^2+\sin^2\theta d\varphi^2)\nonumber\\
 &+& Y_5 \left(\omega_1+a\, d\theta)^2 + (\omega_2-a \sin\theta d\varphi)^2\right) + Y_6 (\om_3 +
\cos\theta d\varphi)^2.\nonumber
\ea
\noindent
The $F_{(3)}$ gauge field \ansatz is given by
\begin{align}
F_{(3)}& = \frac{N_c}{4} \Bigg[ -(\om_1 + b\, d\theta)\wedge (\om_2 - b \sin \theta d\varphi) \wedge (\om_3 +\cos \theta d \varphi) \nonumber\\
& + b' dr \wedge (-d \theta \wedge \om_1 + \sin \theta d\varphi \wedge \om_2) + \left( 1-b^2-\frac{N_f}{N_c} \right) \sin \theta d \theta \wedge d\varphi \wedge \om_3\Bigg].
\end{align}
\noindent
where the  $Y_i$, $a$, and $b$ are functions of $\rho$ only. %Note that $F_3$ is not closed due to the presence of the smeared flavor branes. %%Furthermore, we assume that the dilaton is also a function of $\rho$ and define $\phi = Y_9(\rho)$.
\\
\noindent
The complete action for type IIB supergravity with smeared flavor branes is then
\begin{equation}
S=S_{\text{grav}}+ S_{\text{sources}},
\end{equation}
with  $S_{\text{grav}}$ and $ S_{\text{sources}}$ given by equations \eqref{eq:Sgrav} and \eqref{eq:SflavorSmeared}.

% {********{comment on how the C6 term vanishes}}.

We plug the metric and $F_{(3)}$ \ansatz into this action and integrate over all coordinates except $\rho$, drop the overall volume factor, and end up with a one-dimensional action with Lagrangian
\begin{equation}\label{eq:lagrangian}
L= \sum_{i,j} G_{ij}(Y) Y_i'Y_j'-U(Y)\equiv T-U.
\end{equation}
We can express the $Y_i$ in terms of nine other functions to make $G_{ij}$ diagonal:
\begin{equation}\label{eq:Ys}
\begin{array}{ccc}
  Y_1=e^{2z-6x}, & Y_2=e^{2z+2x}, & Y_3=e^{10y-2z+2l}, \\
  Y_4=e^{2y-2z+2p+2q}, & Y_5=e^{2y-2z+2p-2q}, & Y_6=e^{2y-2z-8p}, \\
  Y_7=a, & Y_8= b, & Y_9=\phi.
\end{array}
\end{equation}

\noindent
We end up with the following expressions:
\begin{eqnarray}
T&=&e^{-l}\left(5y'^2-3x'^2-2z'^2-5p'^2-q'^2 -\frac14\,e^{-4q} a'^2-\frac{N_c^2}{64} \, e^{\phi+4z-4y-4p} b'^2-\frac18 \phi'^2 \right)\, , \nonumber \\
U&=&\frac18\,e^{l} \bigg[e^{8y}\left\{e^{-12p}\, [e^{4q}+e^{-4q}(a^2-1)^2+2a^2(1- e^{10p-2q})^2] -8e^{-2p}\cosh\ 2q \right\}  \nonumber \\
&+& \! \! \frac{N_c^2}{16} e^{\phi+4z+4y+4p} \! \left\{e^{4q} \! + e^{-4q} \! \left(\!a^2 \! \! - \! 2a b + \! 1 \! -\frac{N_f}{N_c} \! \right)^2 \! \!
\! + 2(a \! - \! b)^2 \! \right\} \! + \! N_c e^{\phi/2 + 2z + 6y - 4p} \bigg].\nonumber\\
\label{TminusU}
\end{eqnarray}
Note that the $N_c$ dependence can be absorbed by shifting the dilaton, which we will do from now on. We will also write the ratio of the number of flavor branes to color branes as $s \equiv N_f/N_c$. Since $l$ has no kinetic term, it is a pure gauge degree of freedom reflecting the remaining reparametrization invariance. Varying with respect to $l$ one can set it to any value.\\

\noindent
We want to make connection with the metric \ansatz given in equation \eqref{eq:baseMetric} in Einstein frame, \emph{i.e.} one in which
\begin{equation}\label{eq:YsTransformed}
\begin{array}{ccc}
  Y_1=e^{\phi/2}e^{-8x}, & Y_2=e^{\phi/2}, & Y_3=e^{\phi/2}e^{2k}e^{8x}, \\
  Y_4=e^{\phi/2}e^{2h}, & Y_5=(e^{\phi/2}e^{2g})/4, & Y_6=(e^{\phi/2}e^{2k})/4, \\
  Y_7=a, & Y_8= b, & Y_9=\phi.
\end{array}
\end{equation}

\noindent
In order to do this, we must pick the gauge $l= -4p - 4y + 4x + \log{2}$, and also make the transformations
\bea
&&y = \frac{1}{10} (4 g + 4 h + 2 k - 10 x + 5 \phi - 3 \log{4}), \;\; p=\frac{1}{10} (g + h - 2 k + \log{2}),\nonumber\\
&& q = \frac{1}{2}(-g + h + \log{2}),\;\;z =\frac{1}{4}(-4 x + \phi)
\eea

\noindent
After these transformations, the metric \ansatz takes the desired form. In practice, we first compute the equations of motion from the Lagrangian in the $p,q,y,z,l,x, \phi$ variables, and then apply the above transformations to get equations of motion in the $k,g,h,x,\phi$ variables. This results in equations (\ref{eq:eomfirst} - \ref{constrainteqn}) for the EOMs and constraint, given in section \ref{subsection-non_extremal_flavored_backgrounds}.

We note that the equations of motion and constraint are invariant under constant shifts of the dilaton
\be
\phi \to \phi + C,
\ee
as well as shifts in the radial coordinate
\be
r \to r + r_0.
\ee

\section{Appendix: UV asymptotics}
\label{appendix-UV asymptotics}

\subsection{Finite temperature UV asymptotics}
\label{appendix-UV asymptotics FT}
In the UV, we express the metric and gauge functions as a series in powers of $e^{-2 \rho/3}$:

\begin{alignat}{2}
\label{UV expansion form}
e^{2h} =& \sum_{i=0}^{\infty} \sum_{j=0}^{i} h_{i,j} \; \rho^j \; e^{4(1-i) \rho /3}\quad\quad\quad\quad& e^{4\phi} =& \sum_{i=1}^{\infty} \sum_{j=0}^{i} f_{i,j} \; \rho^j \; e^{4(1-i) \rho /3}\nonumber\\
e^{2g} =& \sum_{i=0}^{\infty} \sum_{j=0}^{i} g_{i,j} \; \rho^j \; e^{4(1-i) \rho /3}& e^{8x}=&\sum_{i=1}^{\infty} \sum_{j=0}^{i} x_{i,j} \; \rho^j \; e^{2(1-i) \rho /3}\nonumber\\\
e^{2k} =& \sum_{i=0}^{\infty} \sum_{j=0}^{i} k_{i,j} \; \rho^j \; e^{4(1-i) \rho /3}& a =& \sum_{i=1}^{\infty} \sum_{j=0}^{i} a_{i,j} \; \rho^j \; e^{2(1-i) \rho /3}\nonumber\\
&\quad& b =& \sum_{i=1}^{\infty} \sum_{j=0}^{i} b_{i,j} \; \rho^j \; e^{2(1-i) \rho /3}.
\end{alignat}

The coefficients $h_{i,j}, \cdots, b_{i,j}$ are not all free, and inserting these expansions into the EOMs and constraint equation (\ref{eq:eomfirst} - \ref{constrainteqn}) allows us to express them in terms of 11 independent coefficients, which we list here along with their order in the expansion:

\begin{alignat}{4}
  &x_{1,0}\; (\text{const}),\quad& &x_{5,0}\; (e^{-8 \rho/3}),\quad& &f_{1,0}\;  (\text{const}),\quad& &f_{3,0}\; (e^{-8 \rho/3}),\NO\\
  &a_{2,0}\; (e^{-2 \rho/3}), & &a_{4,0}\; (e^{-2 \rho}), & &b_{4,0}\; (e^{-2 \rho}),& & \NO\\
  &k_{0,0}\; (e^{4 \rho/3}), &  &k_{3,0}\; (e^{-8 \rho/3}), & & & & \NO\\\
  &h_{1,0}\; (\text{const}), & &h_{1,1}\; (\rho).&   & &  & 
\end{alignat}

Here we present expressions for the remaining non-zero dependent coefficients in the UV expansions, up to $\co\left(\text{poly}(\r) \times e^{-10 \rho/3}\right)$. In order to simplify some of the higher order coefficients in what follows, we will assume that $x_{1,0}=1$ and $a_{2,0}=0$. \footnote{The latter condition is required for a supersymmetric solution, and we set it equal to zero also in our finite temperature solutions, such that $C_2$ is the only parameter deforming the BPS backgrounds.} 
\\
\\
\noindent
$\bullet \mathcal{O} (e^{4 \r /3})$:
\be
h_{0,0} = \frac{3}{8}k_{0,0}\quad g_{0,0}=\frac{3}{2}k_{0,0}
\ee

\noindent
$\bullet \mathcal{O} (\text{const})$:
\be
k_{1,0}=\frac{s}{2},\quad g_{1,0}=\frac{9 s}{4}-4 h_{1,0},\quad g_{1,1}=-4 h_{1,1},
\ee
\noindent
$\bullet \mathcal{O} (e^{-4 \r /3})$:
\bea
k_{2,0}&=&\frac{1}{36 k_{0,0}}\bigg(9+9 s^2-256 h_{1,0}^2+9 s (-1+16 h_{1,0}-9 h_{1,1})\NO\\
&+&288 h_{1,0} h_{1,1}-216 h_{1,1}^2-\frac{9 f_{3,0} k_{0,0}^2}{f_{1,0}}\bigg)\NO\\
k_{2,1}&=&\frac{3 (-2+s)^2+8 (9 s-32 h_{1,0}) h_{1,1}+144 h_{1,1}^2}{18 k_{0,0}}\NO
\eea
\bea
k_{2,2}&=&-\frac{64 h_{1,1}^2}{9 k_{0,0}}\NO\\
g_{2,0}&=&\frac{1}{24 k_{0,0}}\bigg(-3+60 s^2+256 h_{1,0}^2-3 s (11+48 h_{1,0}-33 h_{1,1})\NO\\
&-&96 h_{1,0} h_{1,1}+120 h_{1,1}^2-\frac{9 f_{3,0} k_{0,0}^2}{f_{1,0}}\bigg)\NO\\ 
g_{2,1}&=&\frac{3 (-2+s)^2+8 (-9 s+32 h_{1,0}) h_{1,1}-48 h_{1,1}^2}{12 k_{0,0}}\quad g_{2,2}=\frac{32 h_{1,1}^2}{3 k_{0,0}}\NO\\
h_{2,0}&=&\frac{1}{96 k_{0,0}}\bigg(-3+24 s^2+256 h_{1,0}^2-96 h_{1,0} h_{1,1}+120 h_{1,1}^2\NO\\
&-&3 s (-13+48 h_{1,0}+15 h_{1,1})-\frac{9 f_{3,0} k_{0,0}^2}{f_{1,0}}\bigg)\NO\\
h_{2,1}&=&\frac{3 (-2+s)^2+8 (-9 s+32 h_{1,0}) h_{1,1}-48 h_{1,1}^2}{48 k_{0,0}}\NO\\
 h_{2,2}&=&\frac{8 h_{1,1}^2}{3 k_{0,0}}\quad f_{2,0}=-\frac{2 s f_{1,0}}{k_{0,0}}
\eea

\noindent
$\bullet \mathcal{O} (e^{-2 \r})$:
\bea
b_{4, 1}= (2 - s) a_{4, 0}
\eea

\noindent
$\bullet \mathcal{O} (e^{-8 \r/3})$:
\bea
k_{3,1}&=&\frac{1}{432 f_{1,0} k_{0,0}^2}\Bigg[s f_{1,0} \bigg[103 s^2-4 s (-125+352 h_{1,0}+304 h_{1,1})\NO\\
&+&4 \Big(1+1024 h_{1,0}^2+64 h_{1,0} (-7+25 h_{1,1})+4 h_{1,1} (-73+91 h_{1,1})\Big)\bigg]\NO\\
&+&24 s (x_{5,0} f_{1,0}-f_{3,0}) k_{0,0}^2+576 f_{1,0} k_{0,0}^3 a_{4,0}^2\Bigg]\NO
\eea
\bea
k_{3,2}&=&\frac{s \bigg[(-2+s)^2-8 (14+11 s-64 h_{1,0}) h_{1,1}+400 h_{1,1}^2\bigg]}{54 k_{0,0}^2}
\NO\\
k_{3,3}&=&\frac{256 s h_{1,1}^2}{81 k_{0,0}^2}\NO\\
g_{3,0}&=&-x_{5,0} h_{1,0}+\frac{f_{3,0} h_{1,0}}{f_{1,0}}+\frac{3}{4} x_{5,0} h_{1,1}+s^2 \bigg(-\frac{313}{192 k_{0,0}^2}-\frac{5 h_{1,0}}{3 k_{0,0}^2}+\frac{373 h_{1,1}}{96 k_{0,0}^2}\bigg)\NO\\
&+&s \bigg(\frac{5}{8} x_{5,0}-\frac{f_{3,0}}{2 f_{1,0}}+\frac{313}{192 k_{0,0}^2}+\frac{2 h_{1,0}}{k_{0,0}^2}-\frac{8 h_{1,0}^2}{3 k_{0,0}^2}-\frac{79 h_{1,1}}{48 k_{0,0}^2}-\frac{35 h_{1,0} h_{1,1}}{3 k_{0,0}^2}-\frac{131 h_{1,1}^2}{48 k_{0,0}^2}\bigg)\NO\\
&+&\frac{1225 s^3}{768 k_{0,0}^2}-\frac{7 h_{1,0}}{3 k_{0,0}^2}-\frac{7 h_{1,1}}{4 k_{0,0}^2}+\frac{32 h_{1,0}^2 h_{1,1}}{3 k_{0,0}^2}+\frac{40 h_{1,0} h_{1,1}^2}{3 k_{0,0}^2}+\frac{7 h_{1,1}^3}{k_{0,0}^2}-\frac{3}{4} k_{3,0}+\frac{3}{8} k_{0,0} a_{4,0}^2\NO\\
g_{3,1}&=&-x_{5,0} h_{1,1}+\frac{f_{3,0} h_{1,1}}{f_{1,0}}+s^2 \bigg(-\frac{317}{144 k_{0,0}^2}+\frac{16 h_{1,0}}{9 k_{0,0}^2}+\frac{4 h_{1,1}}{9 k_{0,0}^2}\bigg)\NO\\
&+&s \bigg(-\frac{1}{24} x_{5,0}+\frac{f_{3,0}}{24 f_{1,0}}+\frac{191}{144 k_{0,0}^2}+\frac{52 h_{1,0}}{9 k_{0,0}^2}-\frac{64 h_{1,0}^2}{9 k_{0,0}^2}+\frac{145 h_{1,1}}{36 k_{0,0}^2}\NO\\
&-&\frac{148 h_{1,0} h_{1,1}}{9 k_{0,0}^2}-\frac{511 h_{1,1}^2}{36 k_{0,0}^2}\bigg)\NO\\
&+&\frac{89 s^3}{576 k_{0,0}^2}-\frac{8 h_{1,0}}{3 k_{0,0}^2}-\frac{7 h_{1,1}}{3 k_{0,0}^2}+\frac{64 h_{1,0} h_{1,1}^2}{3 k_{0,0}^2} +\frac{40 h_{1,1}^3}{3 k_{0,0}^2}-k_{0,0} a_{4,0}^2\NO\\
g_{3,2}&=&s^2 \bigg(\frac{1}{18 k_{0,0}^2}+\frac{5 h_{1,1}}{9 k_{0,0}^2}\bigg)+s \bigg(-\frac{1}{18 k_{0,0}^2}+\frac{38 h_{1,1}}{9 k_{0,0}^2}-\frac{64 h_{1,0} h_{1,1}}{9 k_{0,0}^2}-\frac{74 h_{1,1}^2}{9 k_{0,0}^2}\bigg)\NO\\
&-&\frac{s^3}{72 k_{0,0}^2}-\frac{8 h_{1,1}}{3 k_{0,0}^2}+\frac{32 h_{1,1}^3}{3 k_{0,0}^2}\NO\\
g_{3,3}&=&-\frac{64 s h_{1,1}^2}{27 k_{0,0}^2}\NO\\
h_{3,0}&=&\frac{1}{4} x_{5,0} h_{1,0}-\frac{f_{3,0} h_{1,0}}{4 f_{1,0}}-\frac{3}{16} x_{5,0} h_{1,1}+s^2 \bigg(-\frac{25}{768 k_{0,0}^2}+\frac{5 h_{1,0}}{4 k_{0,0}^2}+\frac{259 h_{1,1}}{384 k_{0,0}^2}\bigg)\NO\\
&+&s \bigg(\frac{1}{64} x_{5,0}+\frac{f_{3,0}}{64 f_{1,0}}+\frac{61}{768 k_{0,0}^2}-\frac{2 h_{1,0}}{3 k_{0,0}^2}-\frac{2 h_{1,0}^2}{3 k_{0,0}^2}-\frac{247 h_{1,1}}{192 k_{0,0}^2}+\frac{h_{1,0} h_{1,1}}{12 k_{0,0}^2}+\frac{229 h_{1,1}^2}{192 k_{0,0}^2}\bigg)\NO\\
&-&\frac{287 s^3}{3072 k_{0,0}^2}+\frac{7 h_{1,0}}{12 k_{0,0}^2}+\frac{7 h_{1,1}}{16 k_{0,0}^2}-\frac{8 h_{1,0}^2 h_{1,1}}{3 k_{0,0}^2}-\frac{10 h_{1,0} h_{1,1}^2}{3 k_{0,0}^2}-\frac{7 h_{1,1}^3}{4 k_{0,0}^2}\NO\\
&-&\frac{3}{16} k_{3,0}-\frac{9}{32} k_{0,0} a_{4,0}^2
\NO
\eea

\bea 
h_{3,1}&=&\frac{1}{4} x_{5,0} h_{1,1}-\frac{f_{3,0} h_{1,1}}{4 f_{1,0}}+s^2 \bigg(-\frac{101}{576 k_{0,0}^2}+\frac{7 h_{1,0}}{9 k_{0,0}^2}+\frac{16 h_{1,1}}{9 k_{0,0}^2}\bigg)\NO\\
&+&s \bigg(-\frac{1}{96} x_{5,0}+\frac{f_{3,0}}{96 f_{1,0}}-\frac{25}{576 k_{0,0}^2}+\frac{h_{1,0}}{9 k_{0,0}^2}-\frac{16 h_{1,0}^2}{9 k_{0,0}^2}\NO\\
&-&\frac{23 h_{1,1}}{144 k_{0,0}^2}-\frac{37 h_{1,0} h_{1,1}}{9 k_{0,0}^2}-\frac{79 h_{1,1}^2}{144 k_{0,0}^2}\bigg)\NO\\
&-&\frac{127 s^3}{2304 k_{0,0}^2}+\frac{2 h_{1,0}}{3 k_{0,0}^2}+\frac{7 h_{1,1}}{12 k_{0,0}^2}-\frac{16 h_{1,0} h_{1,1}^2}{3 k_{0,0}^2}-\frac{10 h_{1,1}^3}{3 k_{0,0}^2}-\frac{1}{4} k_{0,0} a_{4,0}^2\NO\\
h_{3,2}&=&s^2 \bigg(\frac{1}{72 k_{0,0}^2}+\frac{17 h_{1,1}}{36 k_{0,0}^2}\bigg)+s \bigg(-\frac{1}{72 k_{0,0}^2}-\frac{5 h_{1,1}}{18 k_{0,0}^2}-\frac{16 h_{1,0} h_{1,1}}{9 k_{0,0}^2}-\frac{37 h_{1,1}^2}{18 k_{0,0}^2}\bigg)\NO\\
&-&\frac{s^3}{288 k_{0,0}^2}+\frac{2 h_{1,1}}{3 k_{0,0}^2}-\frac{8 h_{1,1}^3}{3 k_{0,0}^2}\NO\\
h_{3,3}&=&-\frac{16 s h_{1,1}^2}{27 k_{0,0}^2}\quad f_{3,1}=-\frac{8 f_{1,0}}{3 k_{0,0}^2}+\frac{8 s f_{1,0}}{3 k_{0,0}^2}-\frac{2 s^2 f_{1,0}}{3 k_{0,0}^2}
\eea

\noindent
$\bullet \mathcal{O} (e^{-10 \r/3})$:
\bea
a_{6,0}&=&-\frac{3 s a_{4,0}}{4 k_{0,0}}+\frac{8 h_{1,0} a_{4,0}}{3 k_{0,0}}\quad a_{6,1}=\frac{8 h_{1,1} a_{4,0}}{3 k_{0,0}}\NO\\
b_{6,0}&=&\frac{33 s^2 a_{4,0}}{128 k_{0,0}}+s \bigg(\frac{3 b_{4,0}}{8 k_{0,0}}+\frac{21 a_{4,0}}{64 k_{0,0}}-\frac{3 h_{1,0} a_{4,0}}{2 k_{0,0}}-\frac{45 h_{1,1} a_{4,0}}{32 k_{0,0}}\bigg)\NO\\
b_{6,1}&=&-\frac{3 s^2 a_{4,0}}{8 k_{0,0}}+s \bigg(\frac{3 a_{4,0}}{4 k_{0,0}}-\frac{3 h_{1,1} a_{4,0}}{2 k_{0,0}}\bigg)
\eea

\subsection{Supersymmetric UV asymptotics}

Here we summarize the BPS, zero-temperature UV asymptotics (up to $\mathcal{O}(e^{-4 \rho /3})$) of the backgrounds corresponding to D5s wrapped on the $S^2$ of the resolved conifold, with the addition of smeared D5 flavor sources, as described in \cite{HoyosBadajoz:2008fw, WARPED}. These serve as matching conditions for our finite temperature UV asymptotics. In addition, they furnish us with a family of reference backgrounds that are used in the energy calculations of section \ref{section-energy}.

The BPS asymptotics are given in terms of the free variables $Q_0, c_+,c_-, \rho_{0}$ and $f_{1,0}$ after performing a UV expansion of \emph{e.g.} equations (3.5) of \cite{WARPED}.\footnote{In \cite{WARPED}, the parameter $\phi_0$ is used instead of $f_{1,0}$. We remove the $c_+$ dependence of the asymptotically constant part of the dilaton by setting $e^{4\phi_0} = 2 f_{1,0} c_+^3/3$. See equation (B.46) of \cite{WARPED}.} The finite temperature asymptotics of section \ref{FTUVexpansions} are a deformation of these BPS asymptotics with $\rho_0 = 0,\ Q_0=-N_c+N_f/2$ and $f_{1,0}=1$. In the energy calculations of section \ref{section-energy}, we also set $\rho_0=0$, but $Q_0,\ c_+,\ c_-$ and $f_{1,0}$ are left free to allow for the matching.

What follows, then, is the form of the UV BPS asymptotics with $\rho_0=0$, but with $Q_0,\ c_+,\ c_-$ and $f_{1,0}$ left free.\footnote{$c_-$ appears only at $\mathcal{O}(e^{-8 \rho /3})$ and higher in the metric functions, which we omit here for brevity.}

\bea
e^{2k}&=&\frac{2}{3} c_+ e^{4 \rho /3}+\frac{s}{2}+\frac{1}{96c_+}e^{-4 \rho /3} \bigg[-64 Q_0^2-16 (3-4 \rho )^2+16 s (3-4 \rho )^2\nn\\
&&+32 Q_0 (-2+s) (-3+4 \rho )+s^2 \left(-27+96 \rho -64 \rho ^2\right)\bigg]\nn\\
e^{2h}&=&\frac{1}{4} c_+ e^{4 \rho /3}+\frac{1}{32} \left(8 Q_0+9 s-8 (-2+s) \rho \right)\nn\\
&&+\frac{1}{256 c_+}e^{-4 \rho /3} \bigg[64 Q_0^2+9 \left(16-16 s+3 s^2\right)-128 Q_0 (-2+s) \rho +64 (-2+s)^2 \rho ^2\bigg]\nn\\
e^{2g}&=&c_+ e^{4 \rho /3}-Q_0+\frac{9 s}{8}+(-2+s) \rho\nn\\
&&+\frac{1}{64 c_+}e^{-4 \rho /3} \bigg[64 Q_0^2+9 \left(16-16 s+3 s^2\right)-128 Q_0 (-2+s) \rho +64 (-2+s)^2 \rho ^2\bigg]\nn\\
e^{4\phi}&=&f_{10}-\frac{3 e^{-4 \rho /3} s f_{10}}{c_+}+\frac{3 f_{1,0}}{8 c_+^2} e^{-8 \rho /3} \bigg[4 Q_0 (-2+s)+s^2 (15-4 \rho )-2 (3+8 \rho )+2 s (3+8 \rho )\bigg]\nn\\
\nn\\
a&=&2 e^{-2\rho}\;\;\;\;\;\;\;\;\;\;b=e^{-2 \rho } (2+2 Q_0-s-2 (-2+s) \rho )
\eea

\label{appendix-UV asymptotics SUSY}

\section{Appendix: Comments on numerical procedure}

\label{appendix-numerics}

\begin{figure}[t]
\begin{minipage}{.495\textwidth}
\begin{center}
\includegraphics[width=8cm]{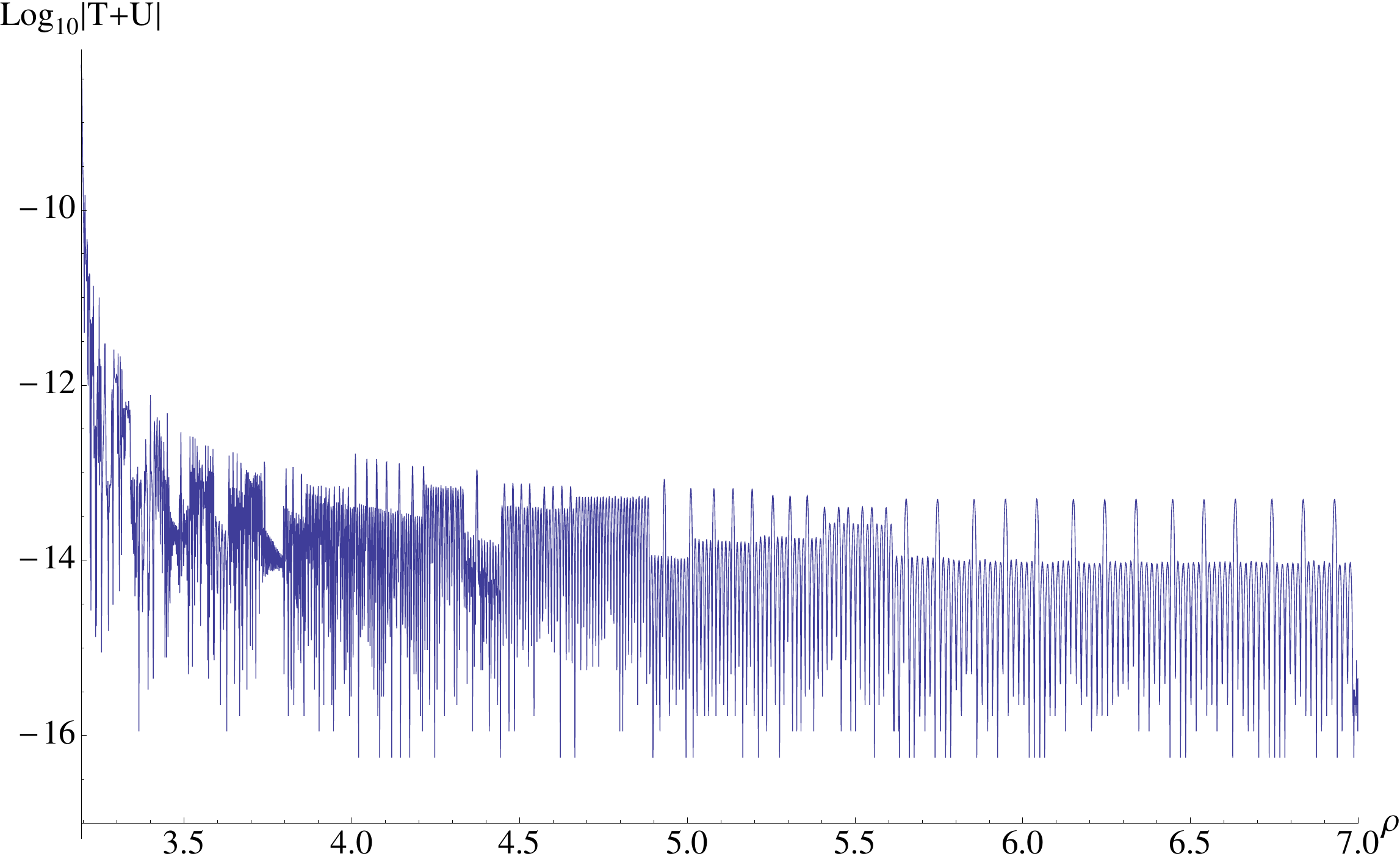}
\caption{\small{Constraint evaluated on a horizon shot solution with $c_+=50, C_2=5000,s=1$. }}
\label{goodConstraintHorizonShotEntire}
\end{center}
\end{minipage}
\begin{minipage}{.495\textwidth}
\begin{center}
\includegraphics[width=8cm]{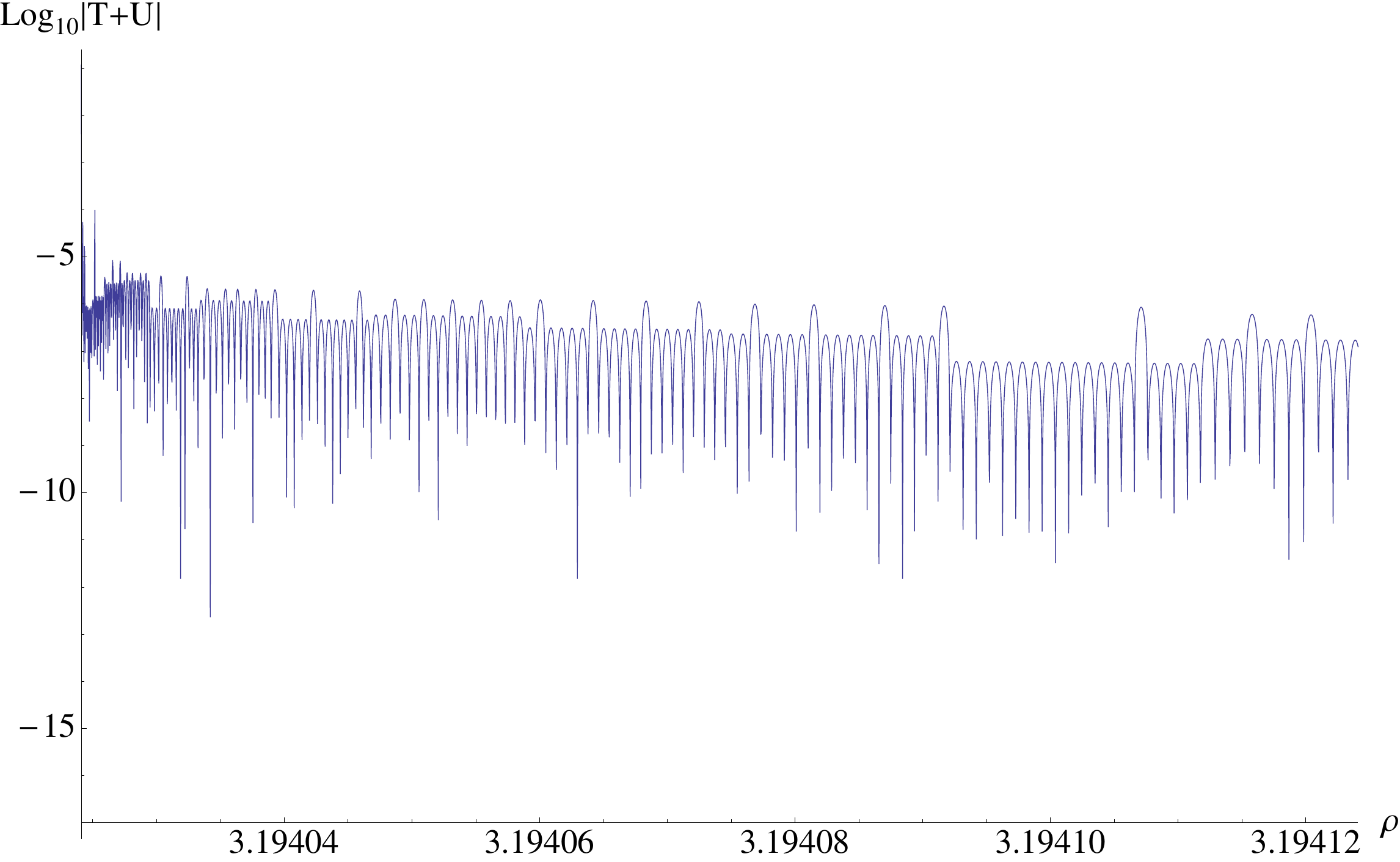}
\caption{\small{Constraint evaluated on same solution, near horizon at $\rho_h \simeq 3.194024.$ }}
\label{goodConstraintHorizonShotatHorizon}
\end{center}
\end{minipage}
%\begin{minipage}{.495\textwidth}
\begin{center}
\includegraphics[width=8cm]{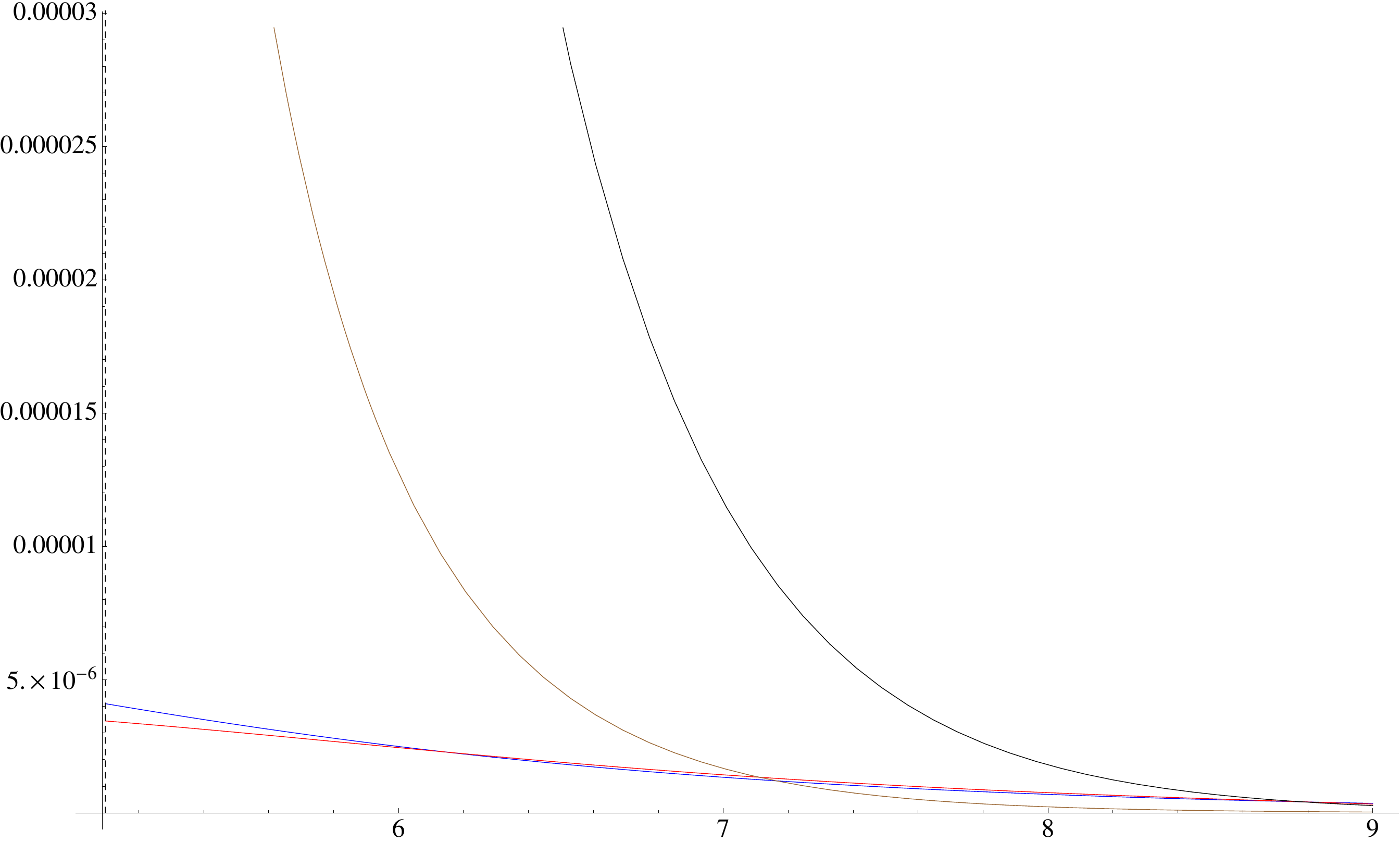}
\caption{\small{Horizon-shot and UV-shot $a$ and $b$ after tuning $b_0$. brown - UV-shot $a$, black - UV-shot $b$, blue - horizon shot $a$, red - horizon-shot $b$. $s=1$, $c=3$, $C_2=800000$.}}
\label{btuned_c3}
\end{center}
%\end{minipage}
\end{figure}

The UV-shot solutions of section \ref{section-numerics} were obtained by using Mathematica's \texttt{NDSolve}, with UV boundary conditions determined by the expansions given in equation \eqref{UV expansion form} up to $\mathcal{O}(e^{-8\rho})$ for $e^{8x}$, $\mathcal{O}(e^{-26\rho/3})$ for $a$ and $b$, $\mathcal{O}(e^{-20\rho/3})$ for $e^{2k},e^{2g}$ and $e^{2h}$, and $\mathcal{O}(e^{-8\rho})$ for $e^{4\phi}$. Using an expansion taken out to this order, NDSolve  at  \texttt{WorkingPrecision} = 70 finds a horizon radius $\r_h$ which is fairly independent of the value of  $\r_{\infty}$ chosen ($\sim$ 1 part in $10^{-5}$ as  $\r_{\infty}$ is varied from 6 to 9). The constraint equation \eqref{constrainteqn} is of order $10^{-8}$ when evaluated on a typical UV-shot solution, but climbs near the horizon; on the horizon shot solution the constraint is of order $10^{-14}$ with a maximum of $\sim 10^{-6}$ at the horizon --- see Figs. (\ref{goodConstraintHorizonShotEntire}) and (\ref{goodConstraintHorizonShotatHorizon}). The degree to which the constraint equation is violated near the horizon also depends on the size of $C_2$. This stems from the fact that many higher order terms in the UV expansion are proportional to $C_2$, so any finite order truncation of the UV asymptotics becomes less accurate for larger $C_2$ values.

When matching our UV-shot solutions to a series expansion near the horizon, we choose to exclude the $a$ and $b$ functions, the dilaton, and their derivatives from the mismatch function. The reason for excluding the $a$ and $b$ functions is that unlike the metric and dilaton functions --- whose UV and horizon shot solutions agree quite well away from the horizon in what \texttt{NMinimize} considers the best matched case --- $a$ and $b$'s UV and horizon shot solutions differ over the entire interval in the best matched case, as shown in Fig. (\ref{btuned_c3}) (although $a$ and $b$ are typically small in magnitude compared to $x,g,h,k \text{\ and\ } \phi$). The UV-shot dilaton also displays divergent behavior near the horizon, and drastically decreases the match's accuracy if included. Including it in the match also produces a horizon-shot dilaton which stabilizes to a UV value different than $e^{4 \phi}|_{\r_\infty}\sim 1$. We remedy this by using the invariance of the EOMs under a constant shift in the dilaton. After the matching is performed, we pick an $f_0$ that  gives a horizon-shot dilaton that agrees with the UV value of $e^{4 \phi}|_{\r_\infty}\sim 1$. For the solutions we have found, including $a$,  $b$ and the dilaton allows for the mismatch to be minimized to only $\sim 10^{-2}$, and the resulting horizon shot solutions for $(x,g,h,k,f)$ differ significantly from the UV shot solutions. Excluding $a$, $b$ and $\phi$ from the mismatch function, \texttt{NMinimize} is able to reduce the mismatch to $\sim 10^{-7}$, and the resulting horizon shot solutions for $(x,g,h,k,\phi,a)$ agree very closely with the UV shot solutions.  This will still leave us with a horizon-shot $b$ that tends to exponentially diverging behavior at large $\rho$. However, $b_0$ can be tuned to eliminate this divergence, with negligible effect on the behavior of the other functions $(x,g,h,k,f)$ --- see Fig. (\ref{btuned_c3}).

Performing our matching close to the horizon (e.g. $\epsilon=.15$) necessarily involves greater errors in the UV-shot solutions. This shows up as horizon-shot solutions that disagree with the UV-shot solutions near $\r_\infty$. To remedy this, we instead begin by matching at a higher $\epsilon = .7$ --- this eliminates the disagreement near $\r_\infty$.  We then use the resulting values of the matched horizon parameters as seeds for a new match performed at a lower $\eps=.15$, where in addition we constrain the allowed variance of $(x_1,g_0,h_0,k_0,f_0)$ around the seed values until the resulting $\eps=.15$ match gives horizon shot solutions which agree with the UV shot solutions at  $\r_\infty$. After doing this, we find that the high-$\eps$-seeded $\eps=.15$ match produces a mismatch of the same order ($10^{-7}$) as a non-seeded $\eps=.15$ match, with the added benefit that the high-$\eps$-seeded horizon-shot solutions are in good agreement with the UV-shot solutions at $\r_\infty$ 

The numerical method we use could undoubtedly be improved. However, as presented here it is enough to identify the behavior of the horizon temperature as a function of the non-extremality parameter $C_2$, which is necessary to study the stability of the background. Other methods have been used, such as \cite{Mahato:2007zm}.

\section{Appendix: U-Duality for the flavored, wrapped D5 branes black hole}
\label{appendix-UDuality}
Here we describe in more detail the steps of the U-duality procedure as applied to backgrounds of the type mentioned in equations (\ref{eq:baseMetric}) and (\ref{eq:F3ansatz}). We will consider the rotation in string frame.
%, and for the purpose of clarity in this appendix define $e^{-8x(\rho)}=\frh(\rho)$.
We begin with
\bea
ds^2_{IIB}&=&e^{\phi(\rho)}\Big[-e^{-8x(\r)} dt^2 + dx_1^2 + dx_2^2 + dx_3^2
\Big] +
ds_{6}^2,\nonumber\\
ds_6^2&=&
e^{\phi(\r)}\Big[e^{8x(\r)}e^{2k(\rho)}d\rho^2
+ e^{2 h(\rho)}
(d\theta^2 + \sin^2\theta d\varphi^2)\nonumber\\
&+&\frac{e^{2 g(\rho)}}{4}
\left((\omega_1+a(\rho)d\theta)^2
+ (\omega_2-a(\rho)\sin\theta d\varphi)^2\right)\nonumber\\
&+& \frac{e^{2 k(\rho)}}{4}
(\omega_3 + \cos\theta d\varphi)^2\Big],
\eea
and
\bea
F_{(3)} &=&\frac{N_c}{4}\Bigg[-(\omega_1+b(\rho) d\theta)\wedge
(\omega_2-b(\rho) \sin\theta d\varphi)\wedge
(\omega_3 + \cos\theta d\varphi)+\nonumber\\
& & b'd\rho \wedge (-d\theta \wedge \omega_1  +
\sin\theta d\varphi
\wedge
\omega_2) + (1-b(\rho)^2 - \frac{N_f}{N_c}) \sin\theta d\theta\wedge d\varphi \wedge
\omega_3\Bigg].\nonumber\\
\eea
%\bea
%F_{(3)} &=&\frac{N_c}{4}\Bigg[-(\omega_1+b(\rho) d\theta)\wedge
%(\omega_2-b(\rho) \sin\theta d\varphi)\wedge
%(\omega_3 + \cos\theta d\varphi)+\nonumber\\
%& & b'd\rho \wedge (-d\theta \wedge \omega_1  +
%\sin\theta d\varphi
%\wedge
%\omega_2) + (1-b(\rho)^2 - \frac{N_f}{N_c}) \sin\theta d\theta\wedge d\varphi \wedge
%\omega_3\Bigg]\nonumber\\
%\eea
We will % write $\phi = f$ to avoid confusion with the transformed dilatonand will 
supress  obvious $\r$ dependence  to avoid cluttering. We begin by T-dualizing in the $x_1,x_2,x_3$ directions, which results in the type IIA background
\bea
& & ds^2_{IIA}=e^{\phi} \Big[ - e^{-8x} dt^2   \Big] +
e^{-\phi}(dx_1^2 + dx_2^2 +dx_3^2)+ ds_6^2
,\nonumber\\
& & e^{2\phi_{A}} =  e^{-\phi}, \nonumber\\
& & F_{(6)}=F_{(3)} \wedge dx_1\wedge dx_2 \wedge dx_3\to F_{(4)}=
e^{2\phi}e^{-4x} *_{6}F_{(3)} \wedge dt.
\label{mamamaka}
\eea
where
\bea\label{eq:F3-app-ansatz}
*_{6}F_{(3)} = \frac{N_c e^{4x}}{8}\Bigg\{&&-\Big[e^{2g-2h}a^2\left(1 +a^2-2 a b - s \right)+ 16 e^{2h-2g} + 8 a\left(a - b\right)\Big]e_1\wedge e_2 \wedge d \rho \nonumber\\
&& + \Big[e^{2g - 2h}a\left(1+a^2 -2 a b-s \right) +4(a-b)\Big]\left(e_1 \wedge \omega_2  + e_2 \wedge \omega_1 \right)\wedge d \rho\nonumber\\
&& + e^{2g-2h}\left(1+a^2-2ab -s \right)\left(\omega_1 \wedge \omega_2 \wedge d \rho \right)\nonumber\\
&&+ e^{-8x}\, b' \left(e_1 \wedge \omega_1 + e_2 \wedge \omega_2 \right)\wedge \tilde\omega_3 \Bigg\}.
\eea
Now, we lift this to M-theory:
\bea
& & ds^2_{11}=  e^{-2\phi/3}dx_{11}^2 + e^{\phi/3}\Big[   - e^{-8x}
e^{\phi}dt^2
+
e^{-\phi}(dx_1^2 + dx_2^2 +dx_3^2)+ ds_6^2 \Big] ,\nonumber\\
& & G_{(4)}=e^{-4x}e^{2\phi}*_{6}F_{(3)} \wedge dt.
\eea
Boosting in the $(t,x_{11})$ directions according to
\beq
dt\to \cosh\beta dt-\sinh\beta dx_{11},\;\;\;\; dx_{11}\to -\sinh\beta dt
+ \cosh\beta dx_{11},
\eeq
we rewrite the boosted metric as
\bea
& & ds^2_{11}=  e^{\phi/3}\Big[ e^{-\phi}(dx_1^2 + dx_2^2 +dx_3^2)+ds_6^2
\Big] + A dt^2 + B dx_{11}^2 + C dtdx_{11},\nonumber\\
& & G_{(4)}=e^{-4x}e^{2\phi}*_{6}F_{(3)} \Big[\cosh\beta dt-\sinh\beta dx_{11}
\Big]
\label{mtheoryboosted}
\eea
where
\bea
& & A=  e^{4\phi/3}[\sinh^2\beta e^{-2\phi}- e^{-8x}\cosh^2\beta]
,\;\;B=  e^{4\phi/3}[\cosh^2\beta e^{-2\phi}- e^{-8x}\sinh^2\beta]
,\nonumber\\
& & C=  -2\cosh\beta \sinh\beta  e^{4\phi/3}[ e^{-2\phi}- e^{-8x}
].
\label{xxxx}\eea
Before reducing to IIA, it is useful to write equation \eqref{mtheoryboosted} as
\bea
& & ds^2_{11}=  B^{-1/2} \Big[  g_{tt}dt^2 + B^{1/2} e^{\phi/3}(
e^{-\phi}(dx_1^2 + dx_2^2 +dx_3^2)+ds_6^2) \Big]
+
B(dx_{11}+a_t
dt)^2,\nonumber\\
& & G_{(4)}=e^{-4x}e^{2\phi} *_{6}F_{(3)} \Big[(\cosh\beta +a_t \sinh\beta)
dt-\sinh\beta (dx_{11}+a_t dt)
\Big]
\eea
where we have defined
\beq
a_t=\frac{C}{2B},\;\;\;\; g_{tt}=\frac{4AB-C^2}{4\sqrt{B}},\;\;
e^{4\phi_{A}/3}= B.
\eeq
Now we reduce to IIA, obtaining in string frame,
\bea
& & ds^2_{IIA}=  g_{tt}dt^2 +
\sqrt{B} e^{-2\phi/3}( dx_1^2 + dx_2^2 +dx_3^2)+
\sqrt{B}e^{\phi/3}ds_6^2,\nonumber\\
& & e^{2\phi_{A}} =  B^{3/2}, \nonumber\\
& & F_{(4)}=  e^{-4x}e^{2\phi}*_{6}F_{(3)} \wedge\Big[(\cosh\beta +a_t
\sinh\beta)dt\Big]
,\nonumber\\
& & H_{(3)}=  -\sinh\beta\,e^{-4x}e^{2\phi} *_6 F_{(3)},\nonumber\\
& & F_{(2)}=a_t' d\r \wedge dt.
\eea
Next, we T-dualize back along the $x_1,x_2,x_3$ directions, and obtain
\bea\label{eq:finalconf}
& & ds^2_{IIB}=  g_{tt}dt^2 +
\frac{e^{2\phi/3}}{\sqrt{B}}(dx_3^2+ dx_1^2 + dx_2^2 )
+\sqrt{B}e^{\phi/3}ds_6^2,\nonumber\\
& & e^{2\phi_{B}} =  e^{2\phi}, \nonumber\\
& & F_{(7)}=  e^{-4x}e^{2\phi} *_{6}F_{(3)} \wedge \Big[(\cosh\beta +a_t
\sinh\beta)dt \Big]
\wedge
dx_3 \wedge dx_2  \wedge dx_1 \quad \quad F_{(3)} = *_{10} F_7
,\nonumber\\
& & H_{(3)}=  -\sinh\beta\,e^{-4x}e^{2\phi} *_6 F_{(3)},\nonumber\\
& & F_{(5)}=a_t' d\r \wedge dt\wedge dx_3\wedge dx_2\wedge dx_1 (1+*_{10}).
\eea
Finally we %set $f=\phi$,
use the definitions for $A,B,C$ and $a_t$, and take the limit $\beta \to \infty$. This is the field theory limit, where the warp factors vanish at infinity. We then rescale
\begin{equation}
\tilde N = N \cosh \beta, \qquad x_i \to \sqrt{\cosh \beta} \sqrt{\tilde N \alpha'} x_i.
\end{equation}
With all of the above, the $\beta \to \infty$ limits are finite and the final solution is given by equations (\ref{after_rotation}) and (\ref{zsazsa}) after transforming to Einstein frame \footnote{To avoid cluttering the notation,  in equations (\ref{after_rotation}) and (\ref{zsazsa}) the   rescaled quantity is called $N_c$}.
\section{Appendix: General form of $B_2$}
\label{appendix-form_of_B2}
In the extremal case \cite{WARPED}, the $SU(3)$ structure fixes the form of $B_{(2)}$. In our solutions $B_{(2)}$ should reduce to the one in \cite{WARPED} when $T=0$. The following \ansatz is compatible with that requirement:
\begin{equation}
\label{eq:generalB2}
B_{(2)}= b_1(\r)  \tilde{\omega_3} \wedge d\r + b_2(\r) e_1\wedge e_2 + b_3(\r) e_1\wedge \tilde\om_2 + b_4(\r) e_2\wedge \tilde\om_1 + b_5(\r)  \tilde\om_1\wedge  \tilde\om_2.
\end{equation}

We find it convenient to introduce a new function $\widetilde{b_2}(\r)$ and parametrize $b_2$ as,
\be
b_2(\r)=\widetilde{b_2}(\r)+ (1- a(\r)^2)b_5(\r)
\ee
Therefore, 
\be\
B_{(2)}= b_1(\r)  \tilde{\omega_3} \wedge d\r + (\widetilde{b_2}(\r)+ (1- a(\r)^2)b_5(\r))
e_1\wedge e_2 + b_3(\r) e_1\wedge \tilde\om_2 + b_4(\r) e_2\wedge \tilde\om_1 + b_5(\r)  \tilde\om_1\wedge  \tilde\om_2.
\ee
With this notation it is easy to show that 
\be
B_{(2)}= B_{(2)}|_{b_5=0} - d\left[ b_5(\r) \tilde{\omega_3}\right]
\ee
Thus, $b_5$ is really a gauge choice that does not affect the value of  $H_{(3)}$. 

From (\ref{after_rotation}) we have,
\begin{align}\label{eq:explicit_H3}
	H_{(3)}&  =
	-e^{-4x} \frac{1 }{4}\, e^{2 \phi}\,  *_6 \, f_{(3)}\nn\\ 
\end{align}
Demanding that  $H_{(3)}=d[B_{(2)}]$  results in five equations,
\begin{align}
	&\cot(\theta) (b_3 - b_4)=0\nn\\
& b_3 +a b_5- \frac{1}{8} e^{ 2\phi-8 x} b'=0 \nn \\
& 2 e^{2\phi -2g+ 2h} -(-1 + a^2) b_1+(b_3+ b_4 + 2 a b_5) a' - \widetilde{b_2}'-b_5'(1-a^2)=0\nn\\
& e^{2 \phi} ( -a + b) -2 (a b_1 + b_5 a' + b_3')=0\nn\\
&\frac{1}{8} e^{2\phi + 2g - 2h} \left( -1 + s -a^2 + 2 a b\right) + b_1-b_5'=0
\end{align}
Three of them are easily solved by setting,
\begin{align}\label{eq:bs}
&b_3 = b_4\nn\\
&b_4 = -\frac{1}{8}\left(8  a b_5 -e^{2 \phi- 8 x} b'\right)\nn \\
&b_1= \frac{1}{8} e^{-2 h+ 2g +2 \phi}\left( 1-s+ a^2 -2 a b\right) + b_5' =0 
\end{align}
we are then left with two equations, one of them is just the equation of motion  for $b(\r)$. The other one is a differential equation for $\widetilde{b_2}(\r)$,
\begin{align}\label{eq:blast}
e^{2\phi + 4g +8x}\left( s a^2 -a^4 -2 a b+ 2a^3b)\right)&+ e^{2\phi + 2g + 2h}a' b' + e^{8x + 2\phi} \left(16 e^{4 h} -e^{4 g}(-1 +s)\right)\nn\\
&\qquad -e^{8x +2 g +2 h}\widetilde{b_2}'(\r) =0
\end{align}
Note that equations (\ref{eq:bs}-\ref{eq:blast}) determine $B_2$ in terms of an arbitrary function $b_5$ that we have shown is just a gauge choice. 
In order to make contact with \cite{WARPED} we choose,
\be
b_5'=\frac{1}{8} e^{2\phi -2 h}\left( 4 e^{2 h + 2 k} + e^{2 g} (-1 + s -a^2 + 2 a b)\right) 
\ee
which gives $b_1=\frac{1}{2} e^{ 2\phi + 2 k} $.

\clearpage 

%\providecommand{\href}[2]{#2}\begingroup\raggedright\begin{thebibliography}{10}

%\end{thebibliography}\endgroup

%\bibliographystyle{JHEP}
\end{document}